\documentclass[
fontsize=12pt,          
numbers=noenddot,    	
parindent=half,        
listof=totoc,        	
bibliography=totoc,  	
headsepline=true,       
footsepline=false, 		
DIV=12                	
]{scrartcl}
\addtokomafont{disposition}{\boldmath}

\usepackage[utf8]{inputenc}
\usepackage[T1]{fontenc}
\usepackage[english]{babel}

\usepackage{graphicx}
\usepackage{color}

\usepackage{amsmath, amsfonts, amssymb, bbm, bm, mathabx}

\usepackage[citestyle = authoryear,
bibstyle = authoryear, dashed = false, giveninits,
backend=bibtex, maxnames = 2, maxbibnames = 10, uniquename = false, uniquelist = false]{biblatex} 
\DefineBibliographyStrings{ngerman}{andothers={et\ al\adddot}}
\renewbibmacro*{volume+number+eid}{%
	\printfield{volume}%
	\setunit*{\addnbthinspace}%
	\printfield{number}%
	\setunit{\addcomma\space}%
	\printfield{eid}}
\DeclareFieldFormat[article]{number}{\mkbibparens{#1}}

\usepackage[defaultlines = 2, all]{nowidow}
\usepackage[a4paper, left = 2.5cm, right = 2.5cm, top = 2.5cm, bottom = 2.5cm]{geometry}

\usepackage{booktabs} 	
\usepackage{longtable} 
\usepackage{tabularx}
\usepackage{xltabular}
\usepackage{lscape}
\usepackage[hidelinks]{hyperref} 
\usepackage{cleveref}

\usepackage{algorithm}
\usepackage{algpseudocode}

\usepackage{todonotes}
\usepackage{lscape}
\usepackage{nowidow}
\usepackage{placeins}
\usepackage[bf, format = plain]{caption}
\usepackage{subcaption}

\newcommand{\dif}{\mathop{}\!\mathrm{d}}
\DeclareMathOperator{\E}{\mathbb{E}}

\title{\Large Simulation study to evaluate when Plasmode simulation is superior to parametric simulation in estimating the mean squared error of the least squares estimator in linear regression}

\author{\normalsize Marieke Stolte$^1$\thanks{Corresponding author, e-mail: \texttt{stolte@statistik.tu-dortmund.de}} \and \normalsize Nicholas Schreck$^2$
	\and\normalsize Alla Slynko$^3$ \and\normalsize Maral Saadati$^2$ \and\normalsize Axel Benner$^2$ \and\normalsize Jörg Rahnenführer$^1$ \and\normalsize Andrea Bommert$^1$}
\date{\small$^1$Department of Statistics, TU Dortmund University\\
	$^2$Division of Biostatistics, German Cancer Research Center\\
	$^3$Department of Statistics and Actuarial Science, University of Waterloo}

\begin{document}
	\maketitle
	\section*{Abstract}	
	Simulation is a crucial tool for the evaluation and comparison of statistical methods. How to design fair and neutral simulation studies is therefore of great interest for both researchers developing new methods and practitioners confronted with the choice of the most suitable method. The term simulation usually refers to parametric simulation, that is, computer experiments using artificial data made up of pseudo-random numbers. 
	Plasmode simulation, that is, computer experiments using the combination of resampling feature data from a real-life dataset and generating the target variable with a known user-selected outcome-generating model, is an alternative that is often claimed to produce more realistic data. We compare parametric and Plasmode simulations for the example of estimating the mean squared error (MSE) of the least squares estimator (LSE) in linear regression. If the true underlying data-generating process (DGP) and the outcome-generating model (OGM) were known, parametric simulation would obviously be the best choice in terms of estimating the MSE well. 
	However, in reality, both are usually unknown, so researchers have to make assumptions: in Plasmode simulation studies for the OGM, in parametric simulation for both DGP and OGM. Most likely, these assumptions do not exactly reflect the truth. Here, we aim to find out how assumptions deviating from the true DGP and the true OGM affect the performance of parametric and Plasmode simulations in the context of MSE estimation for the LSE, and in which situations, which simulation type is preferable. 
	Our results suggest that the preferred simulation method depends on many factors, including the number of features, and on how and to what extent the assumptions of a parametric simulation differ from the true DGP. Also, the resampling strategy used for Plasmode influences the results. In particular, subsampling with a small sampling proportion can be recommended.

	\section{Introduction}
	Simulation studies are usually defined as computer experiments using artificial data generated by a pseudo-random number generator for which some truth about the data-generating process (DGP) and the outcome-generating model (OGM) is known, e.g., the true parameter values of the OGM or the distribution of the features. 
	Well-designed, fair simulation studies are needed both for the evaluation of newly introduced methods and, in particular, for the neutral comparison of existing methods \parencite{boulesteix_plea_2013}. The DGP and OGM are usually chosen to either reflect realistic scenarios or edge cases for the application of the method of interest. We investigate the first case here.
	
	We call the kind of simulation with artificial data, where the DGP and OGM are fully known, ``parametric'' simulations. Non-parametric simulation, where all data are real-life data, is not part of our analysis. Plasmode simulation is a special case of semi-parametric simulation, which is characterized by parts of the data being real-life data and parts of the DGP or OGM being specified. 
	\textcite{morris_using_2019} give a technical introduction to (parametric) simulation studies and focus on guidance for best practices in performing and reporting simulation studies (``ADEMP'' criteria). 
	\textcite{boulesteix_introduction_2020} give a general and more applied introduction to (parametric) simulation studies. 
	
	Parametric simulation studies are a crucial tool in the performance evaluation and comparison of statistical methods since they can offer insights beyond analytical results \parencite{morris_using_2019,boulesteix_introduction_2020} and can be used to evaluate criteria that cannot be assessed on real data where the DGP and OGM are unknown  \parencite{boulesteix_introduction_2020}. An example is the bias of an estimator, which can only be evaluated if the true parameter value can be controlled within the simulation. Therefore, one of the main advantages of parametric simulation studies is the full knowledge of the parameters of the DGP and OGM within the study. Another advantage is the possibility of investigating large numbers of different scenarios, which permits analyzing how the performance of methods depends on the choice of the DGP and OGM. Moreover, it is possible to generate very large numbers of datasets. One of the main disadvantages is the simplification of real-life DGPs. Often, very simple DGPs and OGMs are chosen arbitrarily which then do not reflect the often complex real-life processes. This may lead to wrong conclusions \parencite{boulesteix_introduction_2020}. The over-simplification gets even worse for high-dimensional data, as it gets harder, for example, to specify realistic distributions and correlation structures for an increasing number of variables \parencite{schreck_statistical_2023}.  
	
	A different approach are so-called statistical Plasmodes as first introduced by \textcite{cattell_general_1967}. 
	\textcite{schreck_statistical_2023} distinguish between statistical and biological Plasmodes depending on the procedure used for generating data.
	The motivation of statistical Plasmodes is to preserve a realistic data structure by resampling feature data from real-life datasets instead of using pseudo-random numbers as is usually done in parametric simulation. At the same time, some control over the generated data is given by generating outcome variables for the given resampled feature data according to a known outcome-generating model, like in parametric simulation. So, parametric and Plasmode simulations differ in the generation of features, while outcomes are generated in the same manner. For the feature resampling, different resampling approaches can be utilized \parencite{schreck_statistical_2023}. 
	Biological Plasmodes are generated by natural biological processes, for example, in a wet lab by manipulating biological samples. 
	In this paper, only statistical Plasmodes are considered. 
	
	The main advantage of Plasmode simulations is that the DGP does not have to be specified. The resampling is claimed to ensure the generation of realistic feature data. At the same time, quantities depending on the parameters of the OGM can still be assessed in contrast to fully non-parametric simulations. However, the resampling requires suitable datasets from the true DGP of interest with not too few observations. Depending on the application, this might be a major limitation. 
	A more detailed discussion of the advantages and disadvantages of parametric vs. Plasmode simulation is given in \textcite{schreck_statistical_2023}. The authors especially point out that evidence for the often-made claim of Plasmode simulations producing more realistic data, i.e., data that is closer to the true DGP, is missing. Also, the authors emphasize the lack of studies on the effect of the often arbitrary choice of OGMs, which might affect both Plasmode and parametric simulation studies. 
	
	We aim to compare the ability of Plasmode and parametric simulation to assess the performance of statistical methods, especially concerning how misspecifications affect the results in both cases. We do this in a controlled simulation scenario so that we know both the true DGP and the true OGM, for evaluation purposes. In general, if we knew the truth, parametric simulation using this truth would be best. Since the truth is usually unknown in real-life applications, for parametric simulation researchers instead have to make assumptions about the DGP. These assumptions might deviate from the truth. 
	Without deviations, the parametric simulation will always perform best since it accurately reflects the truth. On the other hand, when the parametric assumptions about the DGP are far from the truth, we expect Plasmode to be superior since the resampling is expected to give results that are rarely very far from the true DGP. Our goal is to determine the extent of deviation for which the parametric simulation gets worse than Plasmode. Therefore, we aim to find out
	\begin{enumerate}
		\item How much the DGP chosen in the parametric simulation can deviate from the truth before the parametric simulation becomes worse than Plasmode. 
		\item How deviations of the chosen OGM from the true OGM affect both parametric and Plasmode simulations.
		\item How the choice of the resampling type affects the Plasmode simulation.
	\end{enumerate}
	Based on the results, we are able to give guidance in which situations to choose parametric or Plasmode simulation and how to perform it.

	We restrict our analysis to a simple scenario and focus on the estimation of the mean squared error (MSE) of the least squares estimator (LSE) in a linear regression model. Therefore, we focus on the explanatory performance of the linear model and do not consider predictive performance. Moreover, we restrict to the low-dimensional setting, i.e., at most $p = 50$ features. We compare how well both parametric simulations with different assumptions about the DGP and the OGM and Plasmode simulations using different resampling strategies, estimate the true MSE. So here, we check how well parametric and Plasmode simulation perform for one particular example of application. We investigate this for different true data-generating processes and outcome-generating models. To compare different methods via simulation, this approach ensures that the simulation studies approximate well the performance of the methods for the true DGP and OGM.
	
	The article is structured as follows. In Section \ref{sec:methods}, we describe parametric and Plasmode simulations in general, and in Section \ref{sec:setup} our specific simulation setup. In Section \ref{sec:results}, we present the results of our simulations, and in Section \ref{sec:conclusions}, we provide recommendations for performing simulations that follow from our results. In Section \ref{sec:discussion}, the results are summarized and discussed.

	\section{Methods}\label{sec:methods}
	In the following, we briefly explain parametric and Plasmode simulations in general, pointing out, in particular, the different options for the resampling strategy in Plasmode simulations.
	
	\subsection{Parametric simulation}
	In parametric simulation studies, the whole data-generating process (DGP) and the outcome-generating model (OGM) have to be specified and are therefore known within the study. They are usually set up to either be as close as possible to a certain kind of data that the researcher is interested in (e.g., gene expression data) or to cover as many situations as possible, possibly including extreme situations. We focus on the first case. Given the specified DGP, a large number of feature datasets are generated using pseudo-random number generators. In all cases that we are investigating, a target variable is then generated from these features by applying the OGM. This yields a large number of datasets, which are then used for applying the methods of interest. This procedure allows the researcher to evaluate the performance of the methods with respect to a metric of interest. The process of generating the datasets can be seen as mimicking the repeated collection of samples from a large population. The results can provide insights into how the methods under study perform on average for datasets that are similar to the chosen DGPs and OGMs \parencite{boulesteix_introduction_2020}. For more details on how to design, perform, analyze, and report parametric simulation studies, refer to \textcite{morris_using_2019}.
	
	\subsection{Plasmode simulation}
	In Plasmode simulation studies, no assumptions on the DGP for the feature data are made. Instead, it is required to have a representative real-life dataset at hand that resulted from the true DGP \parencite{schreck_statistical_2023}. If only real data were used, we would have no control over the DGP and OGM in our simulation. This means that we could not estimate certain quantities (e.g., the bias of an estimator) that directly depend on the true unknown parameters \parencite{boulesteix_introduction_2020}. To enable us to estimate the quantities that directly depend on the true parameters of the OGM (which are important quantities of interest for performance evaluation of models), Plasmode simulation combines the use of real feature data with a known OGM. 
	A Plasmode simulation study then works as follows. In each iteration, a Plasmode dataset is drawn from the real-life dataset at hand. The researcher has to decide on the resampling method. Possible methods include
	\begin{itemize}
		\item $n$ out of $n$ Bootstrap \parencite{efron_bootstrap_1979}, i.e., drawing with replacement a dataset of the same size as the original dataset,
		\item $m$ out of $n$ Bootstrap \parencite{gotze_asymptotic_1993, bickel_resampling_1997, politis_subsampling_1999}, i.e., drawing with replacement $m<n$ observations of the original dataset,
		\item subsampling, i.e., drawing without replacement $m < n$ observations of the original dataset,
	\end{itemize}   
	or other adaptations of Bootstrap, like 
	\begin{itemize}
		\item smoothed Bootstrap \parencite{efron_nonparametric_1981, silverman_bootstrap_1987, hall_smoothing_1989, wang_optimizing_1995}, i.e., applying kernel-estimation to the empirical distribution of the original dataset and resampling from this smoothed empirical distribution, or
		\item wild Bootstrap \parencite{wu_jackknife_1986}, i.e., adding the standardized values of each variable scaled by a random number to the original variable, or
		\item no resampling, i.e., using the whole dataset as it is.
	\end{itemize}  
	A discussion of the first three options in the context of Plasmode simulation can be found in \textcite{schreck_statistical_2023}. In the case of $m$ out of $n$ Bootstrap and subsampling, the researcher also has to decide on the number of observations to draw. For $m$ out of $n$ Bootstrap, there exists a data-dependent algorithm to find the optimal value of $m$ \parencite{bickel_choice_2008}. After resampling a number of Plasmode datasets, the OGM is applied to each of the datasets to generate the outcomes. The resulting datasets can then be used for computing the performance metrics of interest, like in parametric simulation. For more details, see \parencite{schreck_statistical_2023}. 
	
	\section{Setup of the comparison study}\label{sec:setup}
	In the following, we first describe the general approach and then the detailed setup of our comparison study. 
	
	\subsection{General approach}
	We conduct our comparisons with respect to different true DGPs and OGMs. For each scenario, we calculate the true MSE of the least squares estimator (LSE). We then perform a parametric and a Plasmode simulation for estimating the MSE. For these simulations, we choose different DGPs and OGMs. 
	The estimated MSEs resulting from these simulations are then compared to the true MSEs to assess how well the parametric and Plasmode simulations approximate the true values. 
	
	The outcomes are in all cases generated according to a linear model \begin{equation}\label{eq:lm}
		y = X\beta + \varepsilon
	\end{equation}
	for the true scenarios as well as for the parametric and Plasmode simulations. Note that the intercept is included in this model. The true MSE of the LSE $\hat{\beta}$ depends on the number of observations $n$, the residual variance of $\varepsilon$, and the distribution of the features $X$. For fixed $X$, the LSE is unbiased and thus the MSE reduces to the variance, which is given by $\text{Var}(\hat{\beta}| X ) = \sigma^2 (X^TX)^{-1}$.\\
	To define the true DGP and OGM, we have to determine 
	\begin{itemize}
		\item the true distribution of the features,
		\item the true parameter vector $\beta$, and
		\item the true distribution of the error term $\varepsilon$.
	\end{itemize}
	In the context of the parametric simulation, both DGP and OGM have to be chosen, so the distribution of the features, the coefficient vector, and the error distribution have to be specified. Inside the Plasmode simulation, only the OGM has to be chosen, so the coefficient vector and the error distribution have to be specified. 
	
	\subsection{Simulation setup}
	In this section, we describe the true scenarios used in the simulation as well as the deviations from these true scenarios that are assumed for the parametric or Plasmode simulation.
	\subsubsection{True scenarios}
	We use different scenarios as our truth for the comparison. Table \ref{tab:trueDGPOGM} gives an overview of these scenarios. The scenarios differ in the number of features ($p$) and observations ($n$) as well as the true correlation structure. In all scenarios, we assume that our features come from a multivariate normal distribution with mean zero and variances of one, that the true vector of coefficients ($\beta$) consists of all ones, and that the true error distribution is $N(0, 0.3^2)$.\\
	We start with simple scenarios with only two features ($p=2$), which are sampled from a bivariate Gaussian distribution with mean zero, variances of one, and a pairwise correlation of $0.2$ or $0.5$, and $50$ or $100$ observations. We use the same parameter settings for $p = 10$ except that we only look at pairwise correlations of $0.2$. For $p = 50$, we always use $100$ observations for identifiability reasons. Once again, we set all pairwise correlations to $0.2$. Additionally, we use block diagonal correlation matrices with five blocks of ten features each. Within each block, the correlations are once set to $0.2^{|i-j|}$ and once to $0.5^{|i-j|}$ for all $i\ne j$. Features from different blocks are assigned a correlation of 0. \\
	The scenarios are chosen to represent a low, a moderate, and a larger number of features for which the estimation process is still stable for $100$ observations. \\
	Moreover, we use covariance matrices estimated from real datasets to see how the simulations behave with more complicated correlation structures. We chose regression datasets that were available on OpenML \parencite{OpenML_2013}, were used in the benchmark in \textcite{gijsbers_open_2019}, had at most $100$ features, no constant features, no missing values, and pairwise correlations with absolute values of at most $0.95$. With these criteria, we ended up with four datasets: quake \parencite{quake_simonoff_smoothing_1996}, wine\_quality \parencite{wine_quality_cortez_modeling_2009}, pol \parencite{pol_weiss_rule-based_1995}, and Yolanda \parencite{Yolanda_guyon2017analysis}. 
	\begin{table}[!tb]
		\centering
		\resizebox{\textwidth}{!}{
			\begin{tabular}{p{0.19\textwidth} >{\raggedleft}p{0.05\textwidth} >{\raggedleft}p{0.05\textwidth}p{0.7\textwidth}}
				\toprule
				Name & $p$ & $n$ & Distribution of features \\
				\midrule
				$(p2n100\rho0.2)$ & $2$ & $100$ & $(X_1, X_2)^T\sim N_2\left(\boldsymbol{0}_2, \Sigma\right)$ with $\Sigma_{i,j} = 0.2 ~\forall i\ne j$, $\Sigma_{ii} = 1$ \\
				$(p2n50\rho0.2)$ & $2$ & $50$ & $(X_1, X_2)^T\sim N_2\left(\boldsymbol{0}_2, \Sigma\right)$ with  $\Sigma_{i,j} = 0.2 ~\forall i\ne j$, $\Sigma_{ii} = 1$\\
				$(p2n100\rho0.5)$ & $2$ & $100$ & $(X_1, X_2)^T\sim N_2\left(\boldsymbol{0}_2, \Sigma\right)$ with $\Sigma_{i,j} = 0.5 ~\forall i\ne j$, $\Sigma_{ii} = 1$ \\
				$(p10n100\rho0.2)$ & $10$ & $100$ & $(X_1,\dots, X_{10})^T\sim N_{10}\left(\boldsymbol{0}_{10}, \Sigma\right)$ with $\Sigma_{i,j} = 0.2 ~\forall i\ne j$, $\Sigma_{ii} = 1$ \\
				$(p10n50\rho0.2)$ & $10$ & $50$ & $(X_1,\dots, X_{10})^T\sim N_{10}\left(\boldsymbol{0}_{10}, \Sigma\right)$ with $\Sigma_{i,j} = 0.2 ~\forall i\ne j$, $\Sigma_{ii} = 1$ \\
				$(p50n100\rho0.2)$ & $50$ & $100$ & $(X_1,\dots, X_{50})^T\sim N_{50}\left(\boldsymbol{0}_{50}, \Sigma\right)$ with $\Sigma_{i,j} = 0.2 ~\forall i\ne j$, $\Sigma_{ii} = 1$ \\
				$(p50n100\rho0.2^{|i-j|})$ & $50$ & $100$ & $(X_1,\dots, X_{50})^T\sim N_{50}\left(\boldsymbol{0}_{50}, \Sigma\right)$ with covariance matrix $\Sigma$ with blockdiagonal structure where within each of 5 blocks of 10 features the pairwise covariance/ correlation between the $i$th and $j$th feature of the block is given as $0.2^{|i-j|}$ and all variances are equal to 1 \\
				$(p50n100\rho0.5^{|i-j|})$ & 50 & $100$ & $(X_1,\dots, X_{50})^T\sim N_{50}\left(\boldsymbol{0}_{50}, \Sigma\right)$ with covariance matrix $\Sigma$ with block diagonal structure where within each of 5 blocks of 10 features the pairwise covariance/ correlation between the $i$th and $j$th feature of the block is given as $0.5^{|i-j|}$ and all variances are equal to 1  \\
				(quake) & $3$ & $100$ & $(X_1,\dots, X_{3})^T\sim N_{3}\left(\boldsymbol{0}_{3}, \Sigma\right)$ with covariance matrix $\Sigma$ estimated estimated from real dataset quake \parencite{quake_simonoff_smoothing_1996} \\
				(wine\_quality) & $11$ & $100$ & $(X_1,\dots, X_{11})^T\sim N_{11}\left(\boldsymbol{0}_{11}, \Sigma\right)$ with covariance matrix $\Sigma$ estimated from real dataset wine\_quality \parencite{wine_quality_cortez_modeling_2009}\\
				(pol) & $26$ & $100$ & $(X_1,\dots, X_{26})^T\sim N_{26}\left(\boldsymbol{0}_{26}, \Sigma\right)$ with covariance matrix $\Sigma$ estimated from real dataset pol \parencite{pol_weiss_rule-based_1995} \\
				(Yolanda) & $100$ & $200$ & $(X_1,\dots, X_{100})^T\sim N_{100}\left(\boldsymbol{0}_{100}, \Sigma\right)$ with covariance matrix $\Sigma$ estimated from real dataset Yolanda \parencite{Yolanda_guyon2017analysis} \\
				\bottomrule
		\end{tabular}}
		\caption{Parameters for true data-generating processes (DGP) and outcome-generating models (OGM). In all scenarios, the true vector of coefficients is equal to $(1,\dots, 1)^T \in\mathbb{R}^{p + 1}$ and the error distribution is set to $\varepsilon\sim N(0, 0.3^2)$. $\boldsymbol{0}_p$ denotes the $p$-dimensional vector of zeros.}
		\label{tab:trueDGPOGM}
	\end{table}

	
	\subsubsection{Deviations from true scenarios}
	We choose DGPs and OGMs for parametric and Plasmode simulation that present different kinds of deviations from the truth described in the previous section. The general structures of these deviations are listed in Table \ref{tab:deviations}. 
	A complete list of the specific parameter values that were chosen can be found in Table \ref{tab:deviation.list} in the Appendix. As a baseline, we assume the true scenario, which reflects the case that we -- by chance -- correctly specify all parameters in the simulations. Then, we consider choices for each part of the DGP and OGM that reflect increasing deviations from the truth. For the coefficients, we use different values that are either wrong, but of the same order, or that even differ by a large factor. We also included the case of assuming no effect ($\beta = 0$), which is an important special case that might be of interest in many studies. For the distribution of $\varepsilon$, we either only misspecify its standard deviation or misspecify the distribution as either more heavy-tailed (scaled $t$-distribution) or skewed (scaled and shifted $\chi^2$-distribution). As deviations from the true feature distribution, we first still assume a multivariate normal distribution but with wrong correlations, expectations, or variances. We then look at entirely wrong distributions, namely, Gaussian mixture, log-normal, and Bernoulli distributions. The true correlation structure is preserved in those cases. We achieve this by generating Gaussian mixture, log-normal, and Bernoulli variables from multivariate normals and setting the covariance matrix of the underlying normals in a way such that the corresponding variables have the desired variances and covariances. For log-normals and Bernoulli variables with variables of the same distribution, the calculation can be found in \textcite{astivia_population_2017} and \textcite{emrich_method_1991}. The calculation for log-normal and Bernoulli variables in combination with normal variables, as well as all calculations for Gaussian mixture variables, can be found in Appendix \ref{app:calc.cor}.
	
	\FloatBarrier
	\begin{longtable}{p{0.33\textwidth}p{0.66\textwidth}}
		\bottomrule
		\caption{Deviations from true DGP and OGM for parametric and Plasmode simulation.}
		\label{tab:deviations}\\
		\endfoot
		\toprule
		Scenario name & Description \\
		\midrule\endhead
		True model & Assumptions coincide with truth \\
		Coefficients misspecified I & Assumed $\beta$ vector $(0, 1 / p, 2 / p, \dots,1)^T \in\mathbb{R}^{p+1}$ instead of $\mathbf{1}_{p+1}$ \\
		Coefficients misspecified II & Assumed $\beta$ vector $\mathbf{0.05}_{p+1}$ instead of $\mathbf{1}_{p+1}$ \\
		Coefficients misspecified III & Assumed $\beta$ vector $\mathbf{10}_{p+1}$ instead of $\mathbf{1}_{p+1}$ \\
		Coefficients misspecified IV & Assumed $\beta$ vector $\mathbf{0}_{p+1}$ instead of $\mathbf{1}_{p+1}$ \\
		Error sd misspecified $c$ & Assumed $\sigma = c$ instead of $\sigma = 0.3$ for $\varepsilon\sim N(0,\sigma^2)$ \\
		Correlation misspecified $\rho$ & Assumed fixed pairwise correlation of $\rho$ \\
		Correlation misspecified $\rho^{|i-j|}$ & Assumed pairwise correlation of $\rho^{|i-j|}$ for $i$th and $j$th feature for $p = 10$, or for $i$th and $j$th feature within each of 5 blocks of 10 features for $p=50$, respectively\\
		Coefficients (I) and correlation ($\rho$) misspec. & $\mathbf{0.05}_{p+1}$ instead of $\mathbf{1}_{p+1}$ and fixed pairwise correlation of $\rho$ instead of $\rho_{true}$ \\
		Coefficients (II) and correlation ($\rho$) misspec. & Assumed $\beta$ vector $\mathbf{10}_{p+1}$ instead of $\mathbf{1}_{p+1}$ and fixed pairwise correlation of $\rho$ instead of $\rho_{true}$ \\
		Error sd (0.4) and correlation ($\rho$) misspec. & Assumed $\sigma = 0.4$ instead of $\sigma = 0.3$ for $\varepsilon\sim N(0,\sigma^2)$ and fixed pairwise correlation of $\rho$ instead of $\rho_{true}$ \\
		Feature distribution misspecified N(0,1), N($\mu$,1) & Assumed expectation of $\mu$ for second half of features \\
		Feature distribution misspecified N($\mu$,1) & Assumed expectation of $\mu$ for all features \\
		Feature distribution misspecified N(0,1), N(0,$\sigma^2$) & Assumed variance of $\sigma^2$ for second half of features\\
		Feature distribution misspecified N(0,$\sigma^2$) & Assumed variance of $\sigma^2$ for all features\\
		Feature distribution misspecified N(0,$\sigma^2$), $(1-\alpha)$N(0,1)+$\alpha$N(0,10) & Assumed marginal distribution of second half of features as Gaussian mixture with $100\alpha\%$ outliers sampled from $N(0,10)$ and marginal distribution of first half of features misspecified as normal with mean $0$ and variance that matches the variance $\sigma^2$ of the second half of features, $Cor(X_i, X_j) = \rho_{true}, i\ne j$ still holds\\ 
		Feature distribution misspecified N($\mu$,$\sigma^2$), $(1-\alpha)$N(0,1)+$\alpha$N(3,1) & Assumed marginal distribution of second half of features as Gaussian mixture with $100\alpha\%$ of the observations sampled from $N(3,1)$ and marginal distribution of first half of features misspecified as normal with mean $\mu$ and variance $\sigma^2$ chosen such that they match mean and variance of the second half of features, $Cor(X_i, X_j) = \rho_{true}, i\ne j$ still holds\\ 
		Feature distribution misspecified N(1.65,2.83), logN(0,1) &  Assumed marginal distribution of second half of features misspecified as log-normal with parameters 0 and 1 and marginal distribution of first half of features misspecified as normal with matching mean and variance, $Cor(X_i, X_j) = \rho_{true}, i\ne j$ still holds \\
		Feature distribution misspecified Bin($\pi$) & Assumed marginal distribution of second feature misspecified as Bernoulli with a success probability of $\pi$, $Cor(X_i, X_j) = \rho_{true}, i\ne j$ still holds\\
		Error distribution misspecified t(df) scaled & Assumed $\varepsilon\sim t_{\text{df}}$ and scaled $\varepsilon$ to still have sd 0.3  \\
		Error distribution misspecified chisq(df) scaled & Assumed $\varepsilon\sim \chi^2_{\text{df}}$ and shifted and scaled $\varepsilon$ to still have mean 0 and sd 0.3 \\    
	\end{longtable}
	\FloatBarrier
	
	\subsection{Simulation procedure}
	The overall simulation structure is described in Algorithm~\ref{alg:sim}.
	For each true scenario, we first approximate the true MSE by drawing $25\,000\,000$ datasets of size $n$ from the true distribution of $X$ with the first column being a vector of ones, corresponding to the intercept of the model. We then calculate $X\beta$ and add random noise $\varepsilon$ according to the true distribution of $\varepsilon$ and define this as our outcome vector $y$ belonging to the respective dataset. For each pair of data $X$ and corresponding target $y$, we estimate $\hat{\beta}$ using least squares estimation. We then calculate the component-wise means over the replications of the simulation of $(\hat{\beta}_j - \beta_j)^2, j = 0,\dots, p$, with $p$ denoting the number of features, as estimates of the true component-wise MSEs. We refer to these quantities as the ``true'' component-wise MSEs. 
	In each true scenario, we then perform parametric and Plasmode simulations for estimating the component-wise MSEs under the assumption that we do not know the respective true scenario.\\
	The process for data generation for parametric simulation is described in Algorithm~\ref{alg:sim.param}. We make different assumptions on the distribution of the features ($X$), the values of the coefficients $\beta$, and the distribution of $\varepsilon$. We then generate $n.mod = 1000$ datasets according to these assumptions using pseudo-random numbers.
	\begin{algorithm}[!b]
		\caption{Structure of simulation process}\label{alg:sim}
		\begin{algorithmic}[1]
			\Require $n > 0$ (number of observations),
			$0 < p <n$ (number of features), $n.mse>0$ (number of MSE estimations), $n.mod>0$ (number of LSEs, i.e. model estimates, used for estimation of one estimated MSE), true DGP (distribution of features), true OGM ($\beta$, distribution of $\varepsilon$), assumed DGP (assumed distribution of features), assumed OGM ($\beta_{a}$, assumed distribution of $\varepsilon$), $type$ of Bootstrap, proportion $\pi$ for resampling ($=1$ for $n$ out of $n$ Bootstrap, Wild Bootstrap, and Smoothed Bootstrap)
			\Ensure Error in estimated MSE for parametric simulation
			\State $\text{MSE}_{true;j} \gets \mathbb{E}\left[\left(\hat{\beta}_j - \beta_j\right)^2\right], j = 0,\dots,p,$ for the LSE $\hat{\beta}$ in the true model
			\For{$k=1,\dots,n.mse$} 
			\State $X^{(k,i)} \gets$ design matrix generated with Algorithm \ref{alg:sim.param} or \ref{alg:sim.Plasmode} for $i = 1,\dots,n.mod$
			\For{$i=1,\dots,n.mod$} 
			\State $\varepsilon^{(k,i)}\gets$ noise sampled from assumed distribution of $\varepsilon$
			\State $y^{(k,i)} \gets X^{(k,i)}\beta_{a} + \varepsilon^{(k,i)} $
			\State $\hat{\beta}^{(k,i)} \gets \left(\left(X^{(k,i)}\right)^TX^{(k,i)}\right)^{-1}\left(X^{(k,i)}\right)^Ty^{(k,i)}$ \Comment LSE
			\EndFor
			\State $\text{MSE}_{j}^{(k)} \gets \frac{1}{n.mod}\sum_{i=1}^{n.mod}\left(\hat{\beta}^{(k,i)}_{j} - \beta_{a;j}\right)^2, j = 0,\dots,p$
			\State $Err_{j}^{(k)} \gets \text{MSE}_{j}^{(k)}-\text{MSE}_{true;j}, j = 0,\dots,p$
			\EndFor
		\end{algorithmic}
	\end{algorithm}
	
	In some scenarios, we use parametric simulation with estimation of mean and covariance. For this, at the beginning of each simulation, one dataset of size $n = 1000$ is sampled from the true DGP, and the mean and covariance are estimated from this dataset and used as the assumed mean and covariance of the assumed DGP. This corresponds to the case that researchers might have data at hand from which they estimate some characteristics of the DGP to incorporate them into a parametric simulation in order to perform a more realistic simulation.
	
	\begin{algorithm}[!t ]
		\caption{Structure of feature data generation for parametric simulation}\label{alg:sim.param}
		\begin{algorithmic}[1]
			\Require $n > 0$, $0 < p <n$, assumed DGP, $k$ (iteration number of Algorithm 1)
			\Ensure Generated datasets
			\For{$i=1,\dots,n.mod$} \Comment Inner Simulation
			\State $X^{(k,i)} \gets$ design matrix drawn from assumed data generating process using a \par
			\hspace*{2ex} pseudo-random number generator
			\EndFor
		\end{algorithmic}
	\end{algorithm}
	
	The procedure for data generation for Plasmode simulation is described in Algorithm~\ref{alg:sim.Plasmode}. Here, we have to specify the resampling method to use. As a first step for each Plasmode simulation, one dataset is drawn from the true DGP. Note that in each case, the number of observations \textit{after} resampling has to match the number of observations used for parametric simulation and for the true scenario to ensure a fair comparison of methods. For a more detailed discussion of this issue, see Section \ref{sec:wrong.n}. We then draw $n.mod = 1000$ resampled datasets from our dataset according to the chosen resampling method. 
	We utilize the following Bootstrap versions: 
	\begin{itemize}
		\item $m$ out of $n$ Bootstrap \parencite{bickel_resampling_1997, gotze_asymptotic_1993, politis_subsampling_1999} with resampling proportion $\pi \in \{0.01, 0.1, 0.5, 0.632, $ $0.8, 0.9\}$, i.e.\ drawing with replacement $n$ observations out of $\lceil n/\pi\rceil$ observations,
		\item $n$ out of $n$ Bootstrap \parencite{efron_bootstrap_1979}, i.e.\ drawing with replacement $n$ observations out of $n$ (special case of $m$ out of $n$ Bootstrap for $\pi = 1$),
		\item Smoothed Bootstrap \parencite{efron_nonparametric_1981, silverman_bootstrap_1987, hall_smoothing_1989, wang_optimizing_1995}, i.e.\ drawing with replacement $n$ observations out of the smoothed empirical distribution of $n$ observations, 
		\item Wild Bootstrap \parencite{wu_jackknife_1986}, i.e.\ adding the standardized version of each observed feature vector scaled with a noise factor sampled from $N(0,1)$ to the observed feature vectors, and
		\item subsampling with resampling proportion  $\pi \in \{0.01, 0.1, 0.5, 0.632, 0.8, 0.9\}$, i.e.\ drawing without replacement $n$ observations out of $\lceil n/\pi\rceil$ observations, 
		\item no resampling, equivalent to subsampling with resampling proportion $\pi = 1$.
	\end{itemize}  
	We do not determine an optimal resampling proportion, e.g.\ with the algorithm introduced in \textcite{bickel_choice_2008}, since it takes too much time to repeat this for every dataset in the simulation. Instead, we try a range of resampling proportions. 
	\begin{algorithm}[!t ]
		\caption{Structure of feature data generation for Plasmode simulation}\label{alg:sim.Plasmode}
		\begin{algorithmic}[1]
			\Require $n > 0$, $0 < p <n$, true DGP, $type$ of Bootstrap, proportion $\pi$ for resampling ($=1$ for $n$ out of $n$ Bootstrap, Wild Bootstrap and Smoothed Bootstrap), $k$ (iteration number of Algorithm 1)
			\Ensure Plasmode datasets
			\State $X^{(k)}_{Plasm}\gets$ design matrix $\in\mathbb{R}^{\lceil n/\pi\rceil\times (p + 1)}$ drawn from true DGP
			\For{$i=1,\dots,n.mod$} \Comment Inner Simulation
			\If{$type$ == ``$m$ out of $n$ Bootstrap'' \textbf{or} $type$ == ``$n$ out of $n$ Bootstrap''}
			\State $X^{(k,i)} \gets $ $n$ rows sampled from $X^{(k)}_{Plasm}$ with replacement
			\Else 
			\If{$type$ == ``Subsampling''}
			\State $X^{(k,i)} \gets $ $n$ rows sampled from $X^{(k)}_{Plasm}$ without replacement
			\Else 
			\If{$type$ == ``Wild Bootstrap''}
			\State $a \gets $ vector of $p$ numbers sampled from $N(0,1)$
			\State $X^{(k,i)}_{1} \gets \mathbf{1}_n$ 
			\State $X^{(k,i)}_{j} \gets $ $X^{(k,i)}_{j} + a_j \cdot (X^{(k,i)}_{j} - \bar{X}^{(k,i)}_{j}) / SD(X^{(k,i)}_{j}), \: j = 2,\dots, p+1$
			\Else 
			\If{$type$ == ``Smoothed Bootstrap''}
			\State $X^{(k,i)} \gets $ $n$ rows sampled from $X^{(k)}_{Plasm}$ with replacement + random \par
			\hspace*{16ex} noise from a multivariate normal distribution centered at the \par
			\hspace*{16ex} data points and parameterized by corresponding bandwidth \par
			\hspace*{16ex} matrix estimated by Silverman`s rule \parencite{silverman_bw_density_1998}); 
			\EndIf
			\EndIf
			\EndIf
			\EndIf
			\EndFor
		\end{algorithmic}
	\end{algorithm}

	On each dataset generated either according to the parametric or Plasmode approach, the linear model (\ref{eq:lm}) using the chosen parameters for the OGM is then applied to generate the outcome variable. From these, $\hat{\beta}$ is estimated for each dataset. The MSE is estimated as the average component-wise squared difference of the estimated and assumed coefficient vectors. One estimated MSE value corresponds to the result of one parametric or Plasmode simulation study. The whole process is repeated $100$ times so we can see how much variation exists in the MSE estimation when repeating the parametric or Plasmode simulation study.
	
	\subsection{Performance evaluation}
	To compare the performance of parametric and Plasmode simulations, we look at their errors in MSE estimation. For each type of simulation (parametric, parametric with estimation of mean and variance, Plasmode with different resampling methods and proportions), we obtain 100 estimated MSEs for each deviation from the true DGP and OGM. We calculate the component-wise absolute errors as the differences between estimated MSEs and corresponding true MSEs in each case. Additionally, we calculate the relative errors by dividing the absolute errors by the corresponding true MSEs. We aggregate the absolute and relative errors per simulation over the coefficients by taking the arithmetic mean over the absolute component-wise values. We aggregate over the repetitions of the simulation studies by taking the median of the aggregated values. With this strategy, runs with large errors in single coefficients obtain large aggregated values, while the overall aggregated value across simulation repetitions is robust against single simulations with large aggregated errors. 
	
	We examine the errors graphically using boxplots. An example with a corresponding explanation will be shown in Section~\ref{sec:Example}. Additionally, we analyze how much the assumptions in parametric simulation can deviate from the truth until the results are worse than with Plasmode simulation. Therefore, we sort the parametric deviations within each subgroup (e.g.\ deviation from the variance of the multivariate normal) in increasing order of the magnitude of the deviation (e.g.\ if the true variance is $1$ and the tested values are $0.1, \dots, 0.99$, these are ordered decreasingly) and identify the first value in this order for which the fully aggregated error for parametric simulation is larger than that for the considered type of Plasmode simulation. In this way, we can quantitatively compare the parametric simulation to the different Plasmode variants. If the first value, where parametric is worse than Plasmode, is close to the true value, it follows that for deviations of this type, the parametric simulation is very sensitive to small deviations, and we have to be very confident in our parameter settings for the DGP if we want to use parametric simulation. These are the cases where Plasmode might be superior to parametric simulation.
	
	\subsection{Software}
	All analyses are performed using R 4.2.2 \parencite{R}. We use the \texttt{mvtnorm} package \parencite{mvtnorm1,mvtnorm2} to simulate data from multivariate normal distributions. For smoothed Bootstrap, the R package \texttt{kernelboot} \parencite{kernelboot} is used. For visualization of the results, we use the \texttt{ggplot2} package \parencite{ggplot2} and \texttt{ggh4x} \parencite{ggh4x}. The R code and results for the simulation are available on Zenodo (\url{https://doi.org/10.5281/zenodo.10567144}, \url{https://doi.org/10.5281/zenodo.10567059}).
	
	\section{Results}\label{sec:results}
	In this section, we evaluate the results of the simulations. First, we explain the plots for one simple scenario and type of deviation (Section \ref{sec:Example}). 
	Then, the different resampling strategies for Plasmode are compared (Section \ref{sec:Plasmode.comp}). Afterward, we discuss the results for the different types of deviations (Section \ref{sec:feat.dist} to \ref{sec:err.dist}). 
	Last, we consider the results for correlation structures estimated from real data (Section \ref{sec:est.cor}) as well as the effect of the size of the resampled dataset (Section \ref{sec:wrong.n}). 
	
	\subsection{Example}\label{sec:Example}
\begin{figure}[!tb ]
		\centering
		\begin{subfigure}{\textwidth}
			\centering
			\caption{$p = 2,\: n = 100, \:\beta = (1, 1, 1)^T, \:\sigma = 0.3,\: Cor(X_i, X_j) = 0.2~\forall i\ne j$.}\label{fig:ex.true.rel.var0.3}
			\includegraphics[width = \textwidth, page = 2]{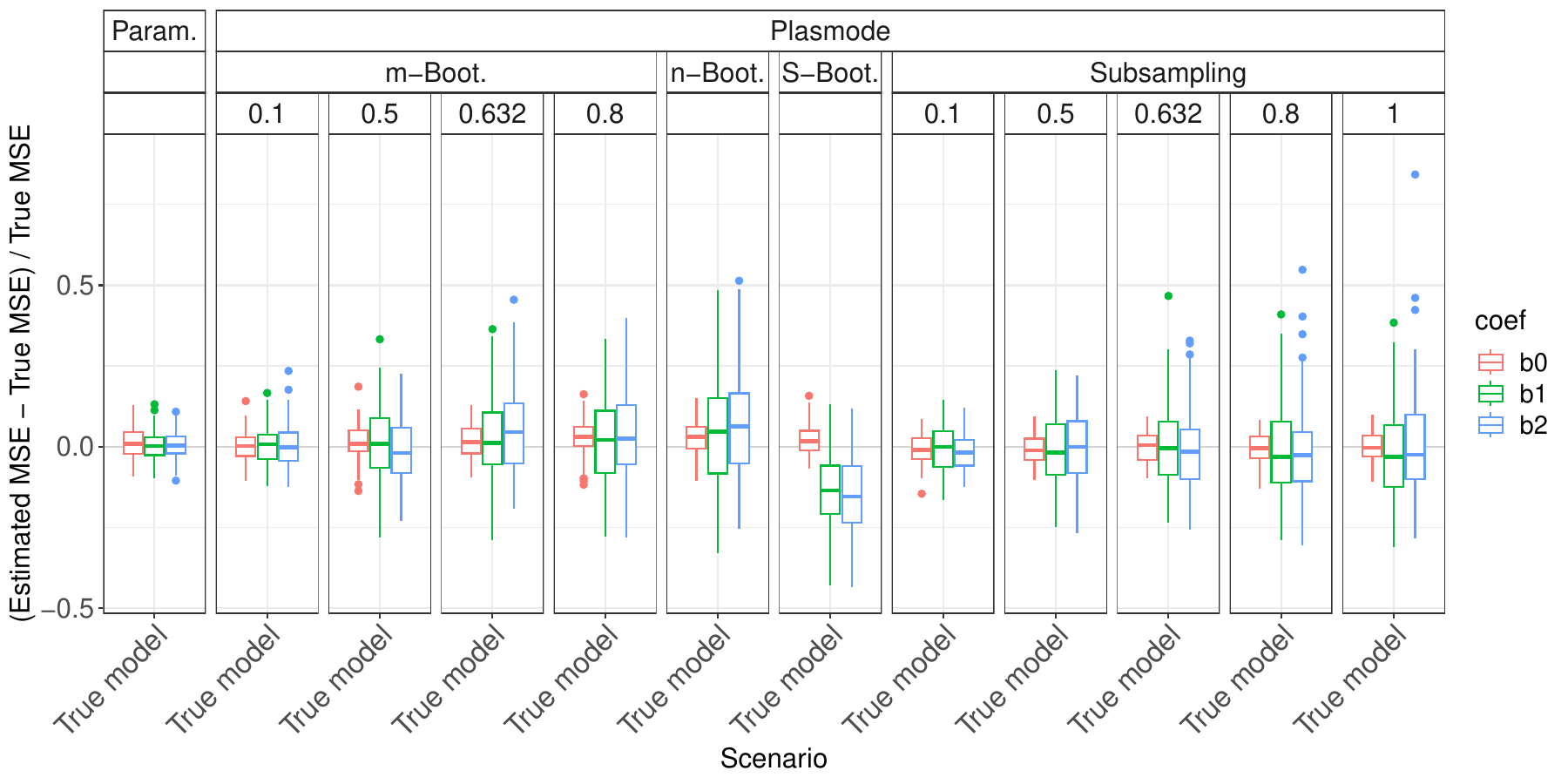}
		\end{subfigure}
		\begin{subfigure}{\textwidth}
			\centering
			\caption{$p = 2,\: n = 100, \:\beta = (1, 1, 1)^T, \:\sigma = 3,\: Cor(X_i, X_j) = 0.2~\forall i\ne j$.} \label{fig:ex.true.rel.var3}
			\includegraphics[width = \textwidth, page = 2]{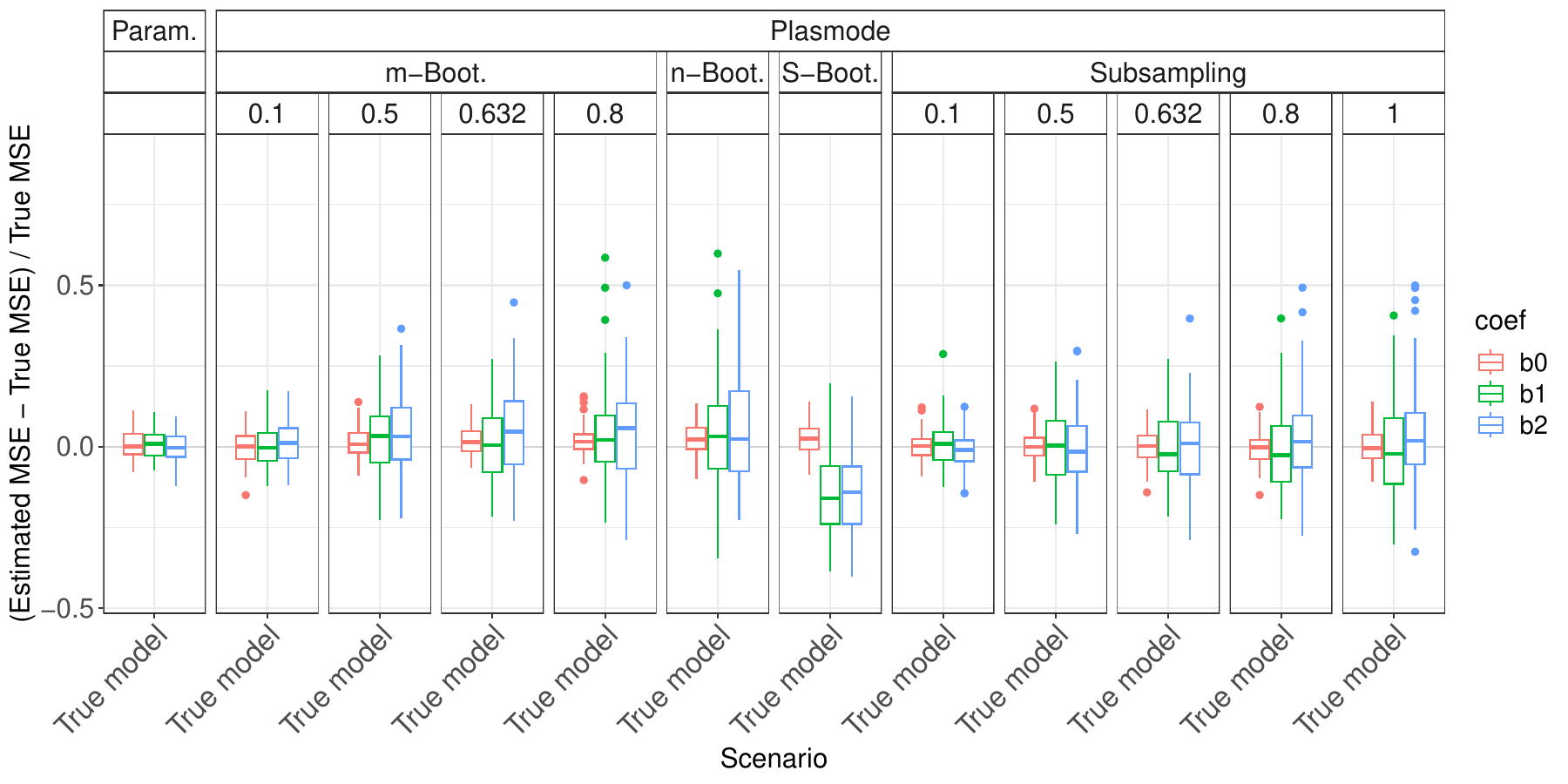}
		\end{subfigure}
		
		\caption{Absolute (A) and relative (B) error in MSE estimation for individual coefficients for different types of Plasmode simulation compared to parametric simulation under assumption of true DGP and OGM.}\label{fig:ex.true.rel}
	\end{figure}

\begin{figure}[!tb ]
		\centering
		\centering
		\includegraphics[width = \textwidth, page = 4]{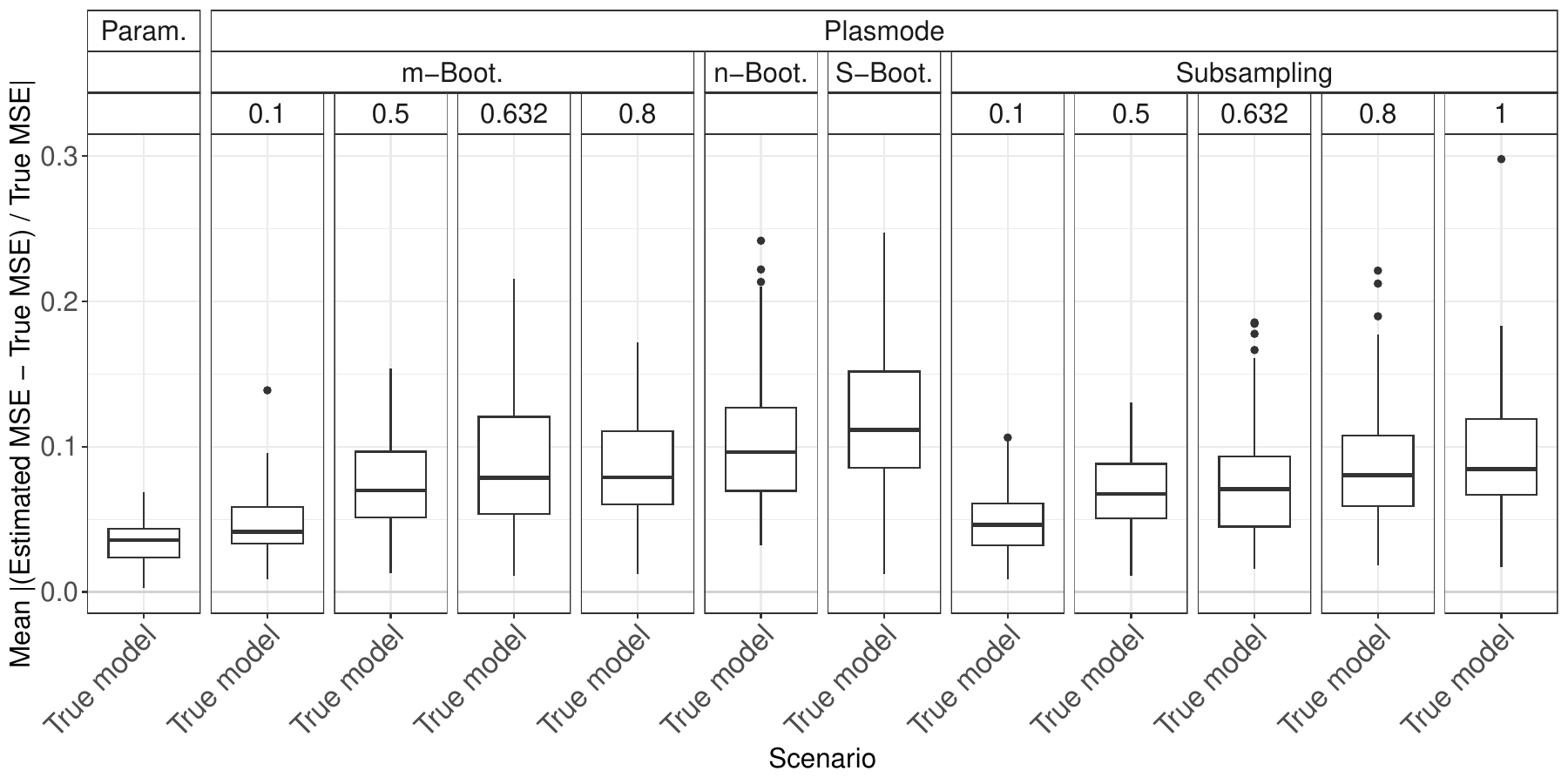}
		\caption{Absolute value of the relative error in MSE estimation averaged over individual coefficients, for different types of Plasmode simulation compared to parametric simulation under the assumption of the true data generating process and outcome generating model, for $p = 2,\: n = 100, \:\beta = (1, 1, 1)^T, \:\sigma = 0.3,\: Cor(X_i, X_j) = 0.2~\forall i\ne j$.}
		\label{fig:ex.true.rel.agg}
	\end{figure}

\begin{figure}[!tb ]
		\centering
		\centering
		\includegraphics[width = \textwidth, page = 2]{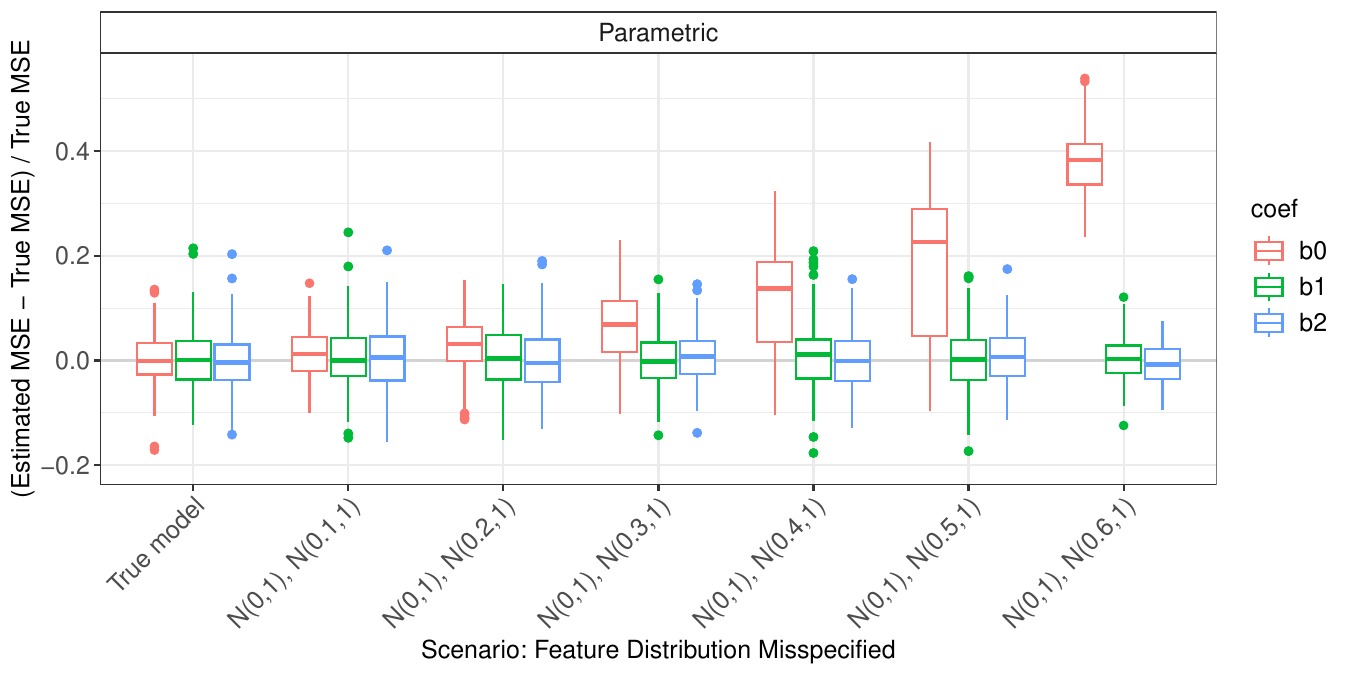}
		\caption{Absolute value of relative error in MSE estimation for individual coefficients when the assumed feature distribution in parametric simulation deviates from the true distribution, for $p = 2,\: n = 100, \:\beta = (1, 1, 1)^T, \:\sigma = 0.3,\: Cor(X_i, X_j) = 0.2~\forall i\ne j$.}\label{fig:ex.mean.rel}
	\end{figure}

	\FloatBarrier
	In the following, we explain the displays that we use in the subsequent sections using one concrete example. 
	We again consider the two scenarios with $p = 2$, $n = 100$, pairwise correlation of $0.2$, $\beta = \boldsymbol{1}_{3}$, and $\varepsilon\sim N(0, 0.3^2)$ or $\varepsilon\sim N(0, 3^2)$. 
	We calculated the errors in the MSE estimation using parametric and Plasmode simulations as described in the previous section. 
	We display the errors in different ways using boxplots. We display the absolute or relative errors for each coefficient individually, like in Figure \ref{fig:ex.true.rel}, Figure \ref{fig:ex.mean.rel}, and Figure \ref{fig:ex.true}, or aggregated over the coefficients like in Figure \ref{fig:ex.true.rel.agg}. 
	We use the unaggregated version in cases where the error for different coefficients might behave differently. 
	This is, for example, the case for deviations in the feature distribution of the second half of features. 
	Otherwise, if all coefficients behave similarly, we use the aggregated version.
	
	For the individual coefficients, the absolute or relative errors of the $100$ repetitions of each type of simulation (parametric, different types of Plasmode) are displayed in one box per coefficient. 
	This is done separately for the true model and each deviation. 
	The deviations are described on the $x$-axis, and coefficients are distinguished by differently colored boxes. 
	The headers give information about the type of simulation used. 
	The first row is the distinction between parametric and Plasmode simulation. 
	The second row gives the type of Plasmode simulation. 
	The third row gives the resampling proportion. 
	For example, in Figure \ref{fig:ex.true.rel} in the third facet, the relative errors per coefficient for Plasmode using $m$ out of $n$ Bootstrap with a resampling proportion of $0.5$ are displayed. 
	This corresponds to sampling with replacement $100$ observations from a dataset of $200$ observations for each simulation. 
	For parametric simulation, $n$ out of $n$ Bootstrap and Smoothed Bootstrap, there is no subsampling proportion so this field is left empty. 
	We leave out Wild Bootstrap in the following analyses since it produces very large outliers and is consistently outperformed by all other Bootstrap types (see e.g.\ Figure \ref{fig:comp.plasmode}). 
	We abbreviate $m$ out of $n$ Bootstrap as $m$-Bootstrap, $n$ out of $n$ Bootstrap as $n$-Bootstrap and Smoothed Bootstrap as $S$-Bootstrap. 
	If necessary we further abbreviate Bootstrap as Boot.\ or B., Parametric as Param.\ or Prm.\ and Subsampling as Sub..\\
	For the aggregated errors, we display the mean over the absolute values of the errors of the individual coefficients per deviation, i.e.\ the mean error per coefficient of one simulation, for the $100$ repetitions of each simulation type in one box. 
	Apart from the aggregation, the figures are constructed in the same way as for the individual coefficients. \\
	There are two types of comparisons: we can compare the performance of different types of simulation for the true model, like in Figures \ref{fig:ex.true.rel}, \ref{fig:ex.true.rel.agg}, and \ref{fig:ex.true} to see how well each simulation type would perform if we knew the truth.
	Or we can compare the performance for differently strong deviations like in Figure \ref{fig:ex.mean.rel}. 
	This allows us to assess the impact of different deviations on the performance.
	We can also combine both displays and show the performance for one kind of deviation for all those types of simulation that are affected by it, and the performance for all other simulation types for the true model only. 
	For example, in the case of deviation from the mean of the second feature distribution, we can show the errors for parametric simulation for different amounts of deviation as in Figure \ref{fig:ex.mean.rel} along with the performance of the Plasmode types under the true model as in Figure \ref{fig:ex.true.rel} or \ref{fig:ex.true}. 
	We cannot misspecify the feature distribution in Plasmode simulation, so only the true model is shown.
	This combined version is the display that we will use for the rest of our analysis. \\
	In general, we might be interested in both absolute and relative errors. 
	As can be seen in Figure \ref{fig:ex.true}, the absolute errors for our specific problem are directly dependent on the chosen parameter for the error standard deviation: 
	if the standard deviation changes by a factor of $10$, e.g.\ here from $\sigma = 0.3$ to $\sigma = 3$, the errors change by a factor of approximately $10^2 = 100$, which can easily be checked by the theoretical relation $\text{Var}(\hat{\beta}|X) = \sigma^2 X^TX$, using that the LSE is unbiased for fixed $X$. 
	The relative errors, on the other hand, are independent of $\sigma$ since the factor affects both the absolute error and the true MSE, by which the absolute error is divided, in the same way. This is, for example, demonstrated in Figure \ref{fig:ex.true.rel}. 
	Therefore, we will only display the relative version for the rest of this analysis since the absolute values could be scaled to be arbitrarily small or large by choosing the error variance accordingly. 
	We will also restrict our analysis to the case $\sigma = 0.3$, since this leads to more stable simulations than $\sigma = 3$, as the latter corresponds to an extremely low signal-to-noise ratio.
	
	\subsection{Comparison of different Plasmode types and resampling proportions}\label{sec:Plasmode.comp}
\begin{figure}[!tb]
		\centering
		\includegraphics[width = \textwidth, page = 2]{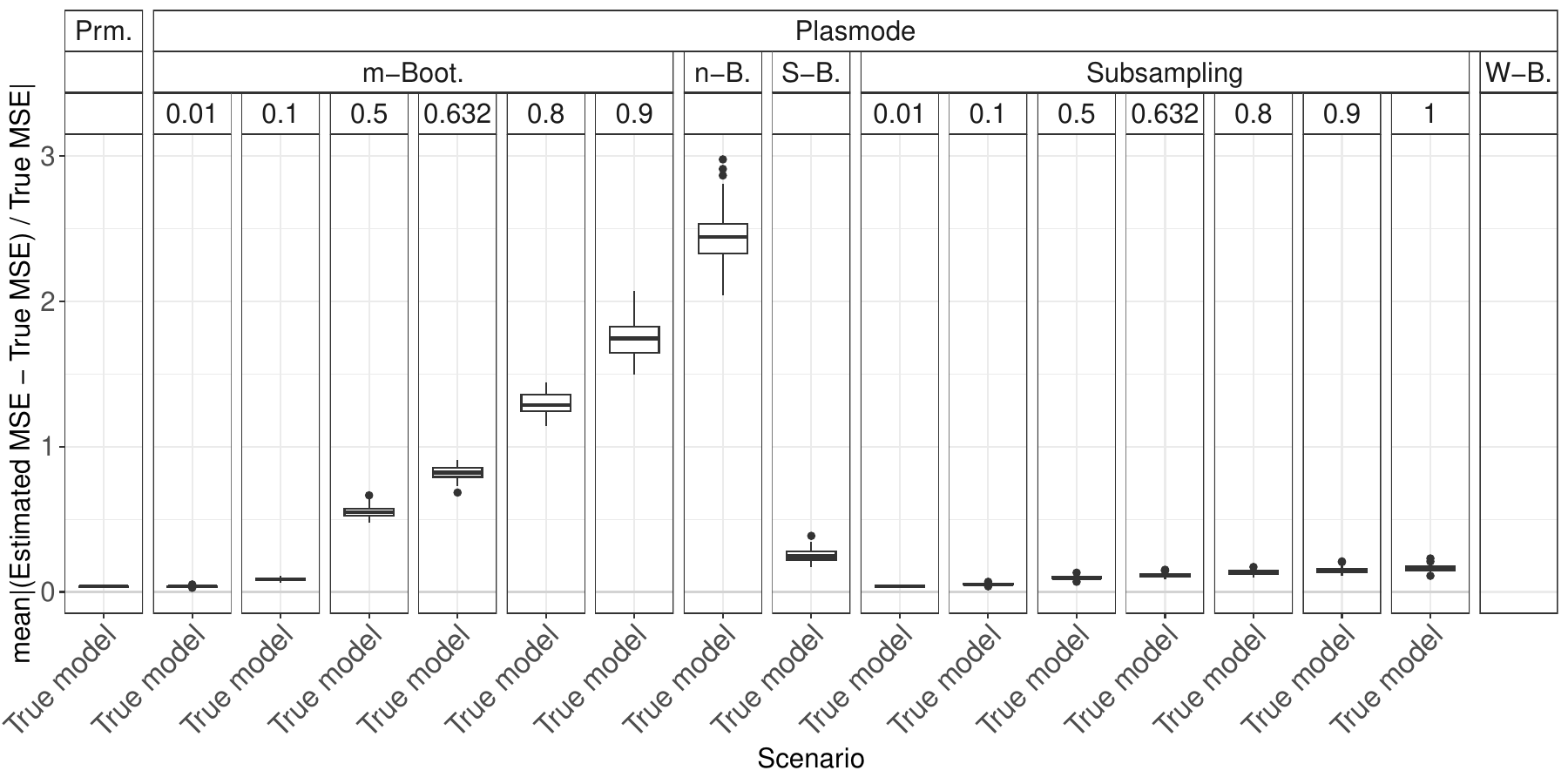}
		\caption{Absolute value of relative error in the MSE estimation averaged over individual coefficients for different types of Plasmode simulation compared to parametric simulation, under the assumption of the true data generating process and outcome generating model, for $p = 50,\: n = 100, \:\beta = \boldsymbol{1}_{51}, \:\sigma = 0.3,\: Cor(X_i, X_j) = 0.2~\forall i\ne j$.}
		\label{fig:comp.plasmode}
	\end{figure}

	Figure \ref{fig:comp.plasmode} shows the aggregated relative errors for $p = 50$ with fixed pairwise correlations of $0.2$ for the true model for all types of simulation. 
	This example confirms that overall, Plasmode using Wild Bootstrap performs the worst. 
	All values for its relative errors lie outside the range of all other resampling types. 
	This is similar for other scenarios, such that we do not show the results for Wild Bootstrap in any other plot.
	Within the other simulation types, Plasmode using the $n$ out of $n$ Bootstrap performs worst with relative mean errors of around $2.5$ and also relatively high variation. 
	$m$ out of $n$ Bootstrap and subsampling perform better both in terms of the median aggregated error and in terms of smaller variability with decreasing resampling proportion, i.e., the larger the dataset from which the $100$ observations are sampled, the lower the variability. 
	$m$ out of $n$ Bootstrap converges towards $n$ out of $n$ Bootstrap for increasing subsampling proportions. 
	Except for very low subsampling proportions ($0.1$ and $0.01$), Bootstrap performs worse than subsampling both with regard to median aggregated error and variability. 
	It is interesting to note that no resampling (i.e.\ subsampling with a subsampling rate of one), which means using the same feature data for the whole simulation and only sampling new observations of the target, still outperforms $m$ out of $n$ Bootstrap with subsampling proportions from $0.5$ on as well as the smoothed, $n$ out of $n$, and wild bootstrap. 
	Smoothed Bootstrap performs worse than all subsampling versions, but better than the $m$ out of $n$ Bootstrap for subsampling proportions from $0.5$ on. 
	With Smoothed Bootstrap, the true MSE of the slope coefficients gets consistently underestimated under the true model (see e.g. Figure \ref{fig:ex.true.rel}). 
	Subsampling and $m$ out of $n$ Bootstrap are indistinguishable for very low subsampling proportions since the impact of duplicate observations decreases with increasing size of the dataset from which we resample. 
	For a proportion of $0.01$, both these approaches perform as well as the parametric simulation. 
	These results reflect what we have also seen in all other scenarios, although for lower $p$, the differences between the simulation types become very small. 
	It should be noted that in these simulations, subsampling and $m$ out of $n$ Bootstrap require a larger dataset to resample from for lower resampling proportions. This might give them an advantage.
	Due to its very poor performance, we will exclude the wild Bootstrap from now on. 
	We will also reduce the values of resampling proportions to 0.1 and 0.632 for $m$ out of $n$ Bootstrap and to 0.1, 0.632, and 1 for subsampling for more clarity. 
	The numbers were chosen to represent a relatively low and a relatively high resampling proportion. 
	Moreover, 0.632 has been used in Plasmode simulations, motivated by the expected proportion of non-duplicated observations for $n$ out of $n$ Bootstrap \parencite{de_bin_subsampling_2016}.
	
	\subsection{Deviations from true feature distribution} \label{sec:feat.dist}
	We will now take a look at the different deviations from the true feature distribution. 
	These only affect the parametric simulation. 
	Since in all cases, different coefficients are affected differently, we always show the individual errors per coefficient. 
	We focus on the case $p = 2$ and $n = 100$ since for this we can still display the errors for individual coefficients in a clear manner. 
	The results can be transferred to higher numbers of features or lower numbers of observations. 
	As expected, the absolute values of the errors are larger for higher values of $p$ or smaller values of $n$, but the qualitative results are the same. 
	In all cases, we only display the range of deviations that is relevant to the comparison of parametric and Plasmode simulations. 
	
	\subsubsection{Gaussian with wrong expectation}
\begin{figure}[!b]
		\centering
		\includegraphics[width = \textwidth, page = 2]{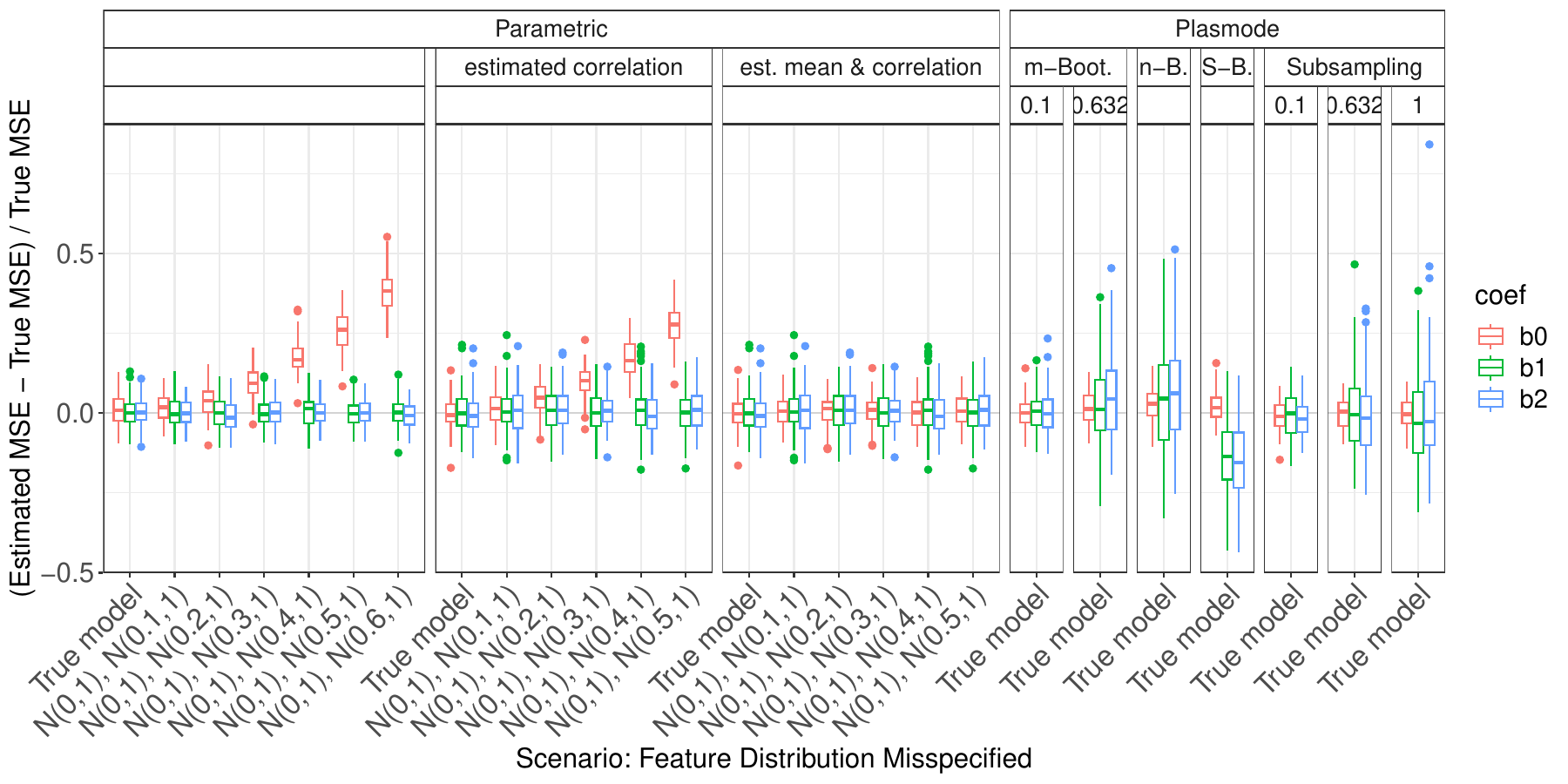}
		\caption{Relative error in MSE estimation for individual coefficients when the assumed mean of the marginal distribution of the second feature in parametric simulation deviates from the true mean, for $p = 2,\: n = 100, \:\beta = (1, 1, 1)^T, \:\sigma = 0.3,\: Cor(X_i, X_j) = 0.2~\forall i\ne j$. N(0,1), N($\mu$,1) denotes that the first feature is generated from a standard normal (truth), and the second feature is generated from a normal distribution with mean $\mu$ instead (deviation).}
		\label{fig:feat.dist.mean}
	\end{figure}

	Figure \ref{fig:feat.dist.mean} shows the relative errors in case of deviations from the expectation of the second feature (Feature distribution misspecified N(0,1), N($\mu$,1), cf.\ Table \ref{tab:deviations}). 
	We can observe that the second coefficient stays unaffected while the errors for the intercept increase with increasing deviations from the true mean. 
	This result is to be expected, as can be seen by reparametrization. 
	If the truth is $X_2\sim N(0,1)$ and we assume $X_2^a\sim N(\mu, 1),\,\mu>0$, we can rewrite the resulting linear model using $X_2^a$ instead of $X_2$ as 
	\begin{align*}
		Y &= \beta_0 + X_1 \beta_1 +X_2^a \beta_2 + \varepsilon\\
		&= \beta_0 + X_1 \beta_1 + (X_2 + \mu) \beta_2 + \varepsilon\\
		&= \underbrace{\beta_0  + \mu \beta_2}_{=: \beta_0^{new}} + X_1 \beta_1 + X_2 \beta_2  + \varepsilon.
	\end{align*}
	As $\mu>0$ and $\beta_2>0$ in our case, it holds $\beta_0^{new} > \beta_0$.
	
	The errors in the intercept can be prevented by estimating the mean using a dataset sampled from the true DGP. 
	This leads to slightly increased variance in the errors of the parametric simulation, but stable median errors that are close to zero for all coefficients. 
	
	\subsubsection{Gaussian with wrong variance}
\begin{figure}[!tb]
		\centering
		\includegraphics[width = \textwidth, page = 4]{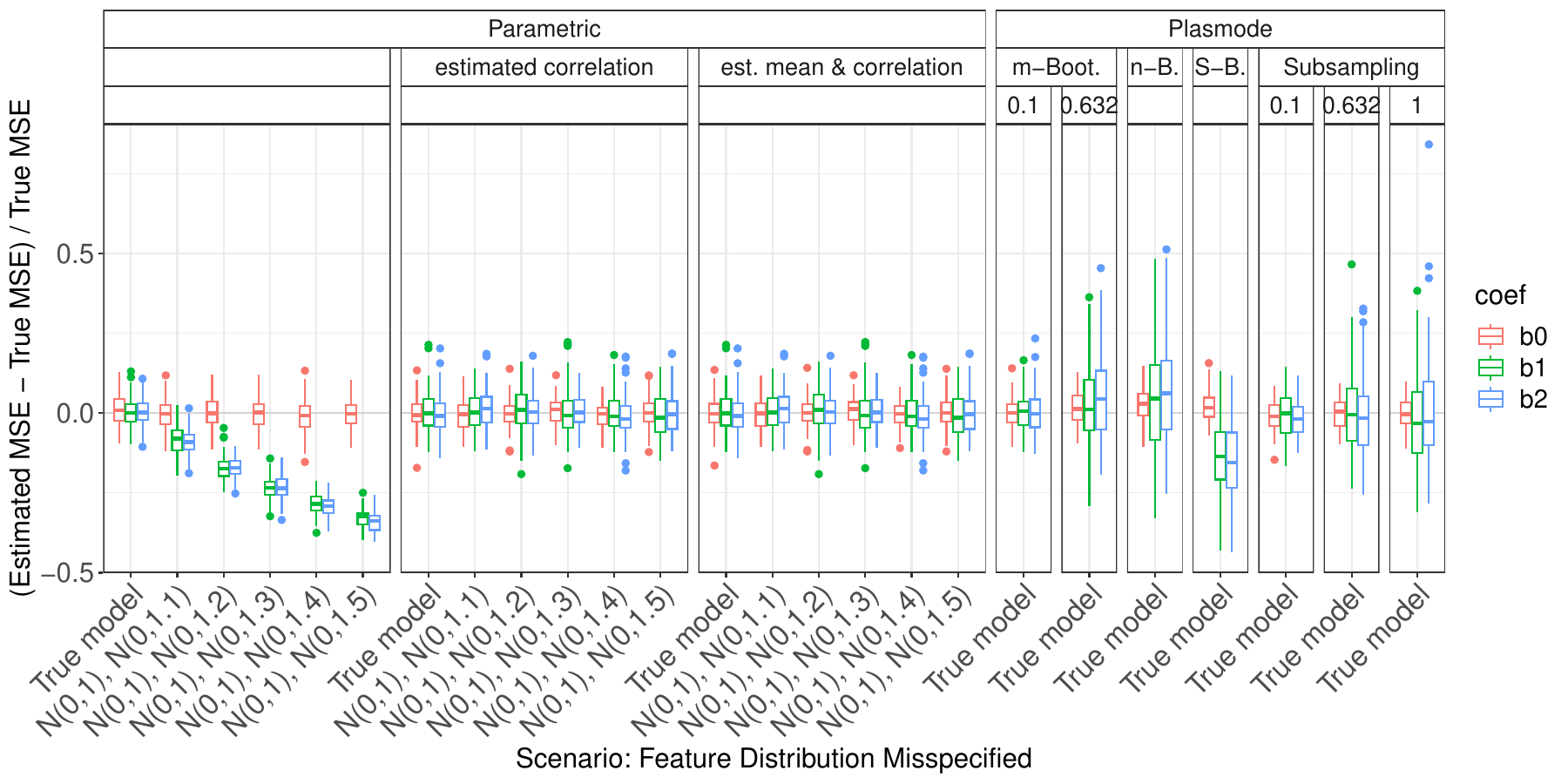}
		\caption{Relative error in MSE estimation for individual coefficients when the assumed variance of the marginal distribution of the second feature in parametric simulation deviates from the true variance, for $p = 2,\: n = 100, \:\beta = (1, 1, 1)^T, \:\sigma = 0.3,\: Cor(X_i, X_j) = 0.2~\forall i\ne j$. N(0,1), N(0,$\sigma^2$) denotes that the first feature is generated from a standard normal (truth), and the second feature is generated from a normal distribution with variance $\sigma^2$ instead (deviation).}
		\label{fig:feat.dist.var}
	\end{figure}

	Figure \ref{fig:feat.dist.var} shows the relative errors for deviations from the true variance of the second feature. 
	We see that both slope coefficients are affected. 
	The true MSE is underestimated by the simulation, and this underestimation gets worse as the (misspecified) variance of the second feature increases. 
	Again, this behavior can be prevented by estimating the covariance matrix from a dataset from the true DGP at the cost of slightly increased variation. 
	For estimation of the covariance matrix, the dataset from the true DGP must have sufficiently many observations. 
	Here, we used $1000$ observations, which is sufficient for $p = 2$, as well as for $p = 50$. 
	Smaller numbers of observations are insufficient for $p = 50$ as can be seen in Figure \ref{fig:no.est.var} in the Appendix. 
	When increasing the variance of the second feature, the error in MSE converges to an upper bound corresponding to the true MSE, since the estimated MSE converges to zero for increasing variances.
	This can lead to problems later on, when we look for the first deviation where the aggregated error for parametric simulation exceeds the error for Plasmode simulation. 
	For low $p$, the upper bound of the error for increasing the feature variance in parametric simulation is still larger than the errors obtained with Plasmode simulation. 
	However, for large $p$, where Plasmode performs worse, the error reached even with very high values for the variance of the second half of features is smaller than that of some Plasmode types. 
	This is demonstrated in Figure \ref{fig:50.inc.var} in the Appendix for the case of $p = 50$. 
	There, we show the mean of the relative errors of the coefficients per simulation run. 
	In that case, we do not take the absolute values before averaging, to demonstrate the direction of the errors. 
	In the present case, this is no problem since either the MSEs for all coefficients are overestimated or all are underestimated, so there is no risk of the errors of different coefficients cancelling out in the mean.
	Decreasing instead of increasing the variance of the second half of features leads to an overestimation of the true MSE, and this overestimation is unbounded. 
	Therefore, in settings where the upper bound does not exceed the errors of all Plasmode types, we use decreasing instead of increasing variances, see e.g. Figure \ref{fig:50.dec.var} in the Appendix.

	\subsubsection{Gaussian with wrong correlations}
\begin{figure}[!b]
		\centering
		\includegraphics[width = \textwidth, page = 2]{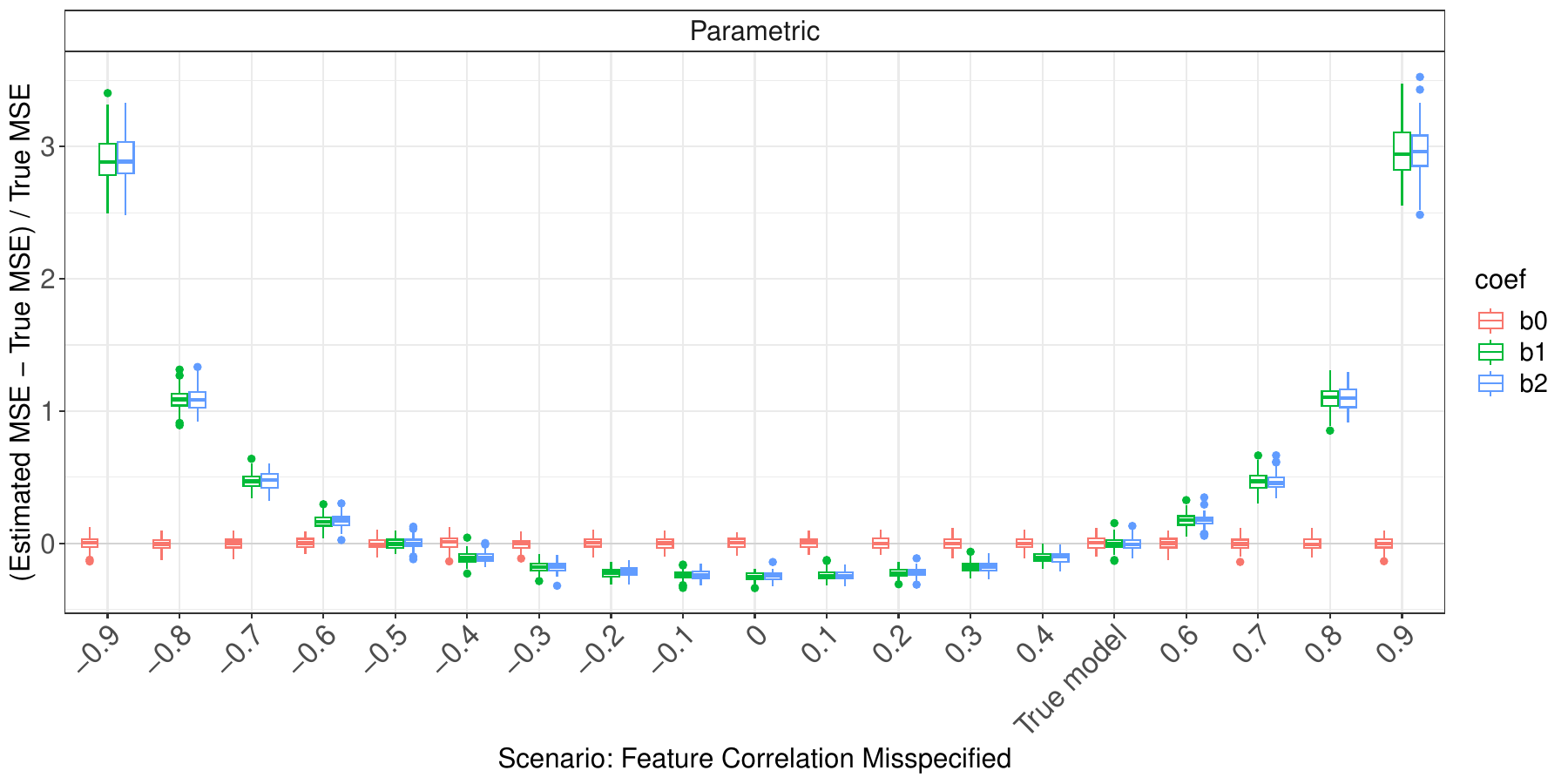}
		\caption{Relative error in MSE estimation for individual coefficients when the assumed correlation of the features in parametric simulation deviates from true correlation, for $p = 2,\: n = 100, \:\beta = (1, 1, 1)^T, \:\sigma = 0.3,\: Cor(X_i, X_j) = 0.5~\forall i\ne j$.}
		\label{fig:feat.corr.0.5.only.param}
	\end{figure}

	The overall influence of misspecifying the pairwise correlations of the features is more easily demonstrated when the true pairwise correlations are $0.5$ instead of $0.2$. 
	The relative errors in this case for parametric simulation are shown in Figure \ref{fig:feat.corr.0.5.only.param}. \\
	The intercept is unaffected when misspecifying the correlation. 
	For the errors in the slopes, we observe a parabolic shape that intersects with zero at the true correlation of $0.5$ and at $-0.5$. 
	For the MSE estimation, the sign of the correlation does not seem to have any influence, only the absolute value, as the parabolic shape is symmetrical around zero. 
	When overestimating the absolute value of the true correlation, the true MSE is overestimated. 
	For underestimating the absolute value of the true correlation, the true MSE is underestimated. 
	
\begin{figure}[!tb]
		\centering
		\includegraphics[width = \textwidth, page = 2]{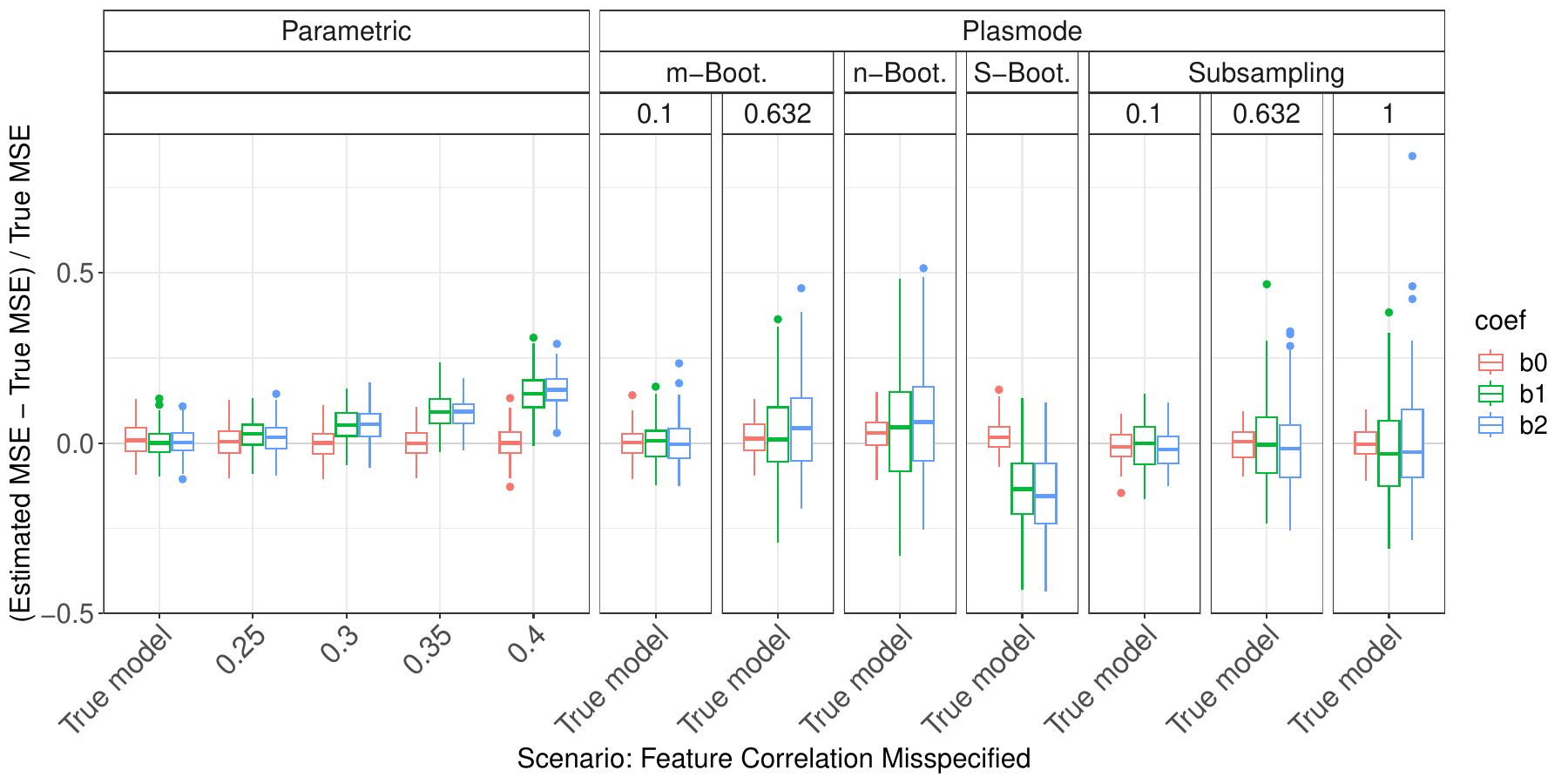}
		\caption{Relative error in MSE estimation for individual coefficients when the assumed correlation of the features in parametric simulation deviates from true correlation, for $p = 2,\: n = 100, \:\beta = (1, 1, 1)^T, \:\sigma = 0.3,\: Cor(X_i, X_j) = 0.2~\forall i\ne j$.}
		\label{fig:feat.corr.0.2}
	\end{figure}

	This pattern is also observed for a true correlation of $0.2$ (Figure \ref{fig:feat.corr.0.2}). 
	For the comparison of parametric and Plasmode, we concentrate on assuming a correlation that is higher than the true correlation, since for these deviations, the errors are monotonously increasing. 
	This can, for example, be seen in the comparison for true fixed pairwise correlations of $0.2$ and $p =2,\: n = 100$ as shown in Figure \ref{fig:feat.corr.0.2}.
	
	The observed shape is plausible from a theoretical point of view. 
	The MSE of the LSE given $X$ is equal to its variance, as it is unbiased. This variance is given as the diagonal of $\sigma^2 (X^T X)^{-1}$.
	For $X$ drawn from a multivariate normal distribution, i.e.\ ignoring the intercept term, $(X^T X)^{-1}$ follows an inverse Wishart distribution. 
	Its expectation is given by the inverse covariance matrix $\Sigma^{-1}$ of this multivariate normal. 
	When explicitly calculating the diagonal values of $\Sigma^{-1}$ in the case of pairwise fixed correlations of $\rho$, we can see that this expectation depends quadratically on $\rho$, which matches the observed form. 
	
\begin{figure}[!tb ]
		\centering
		\includegraphics[width = \textwidth, page = 2]{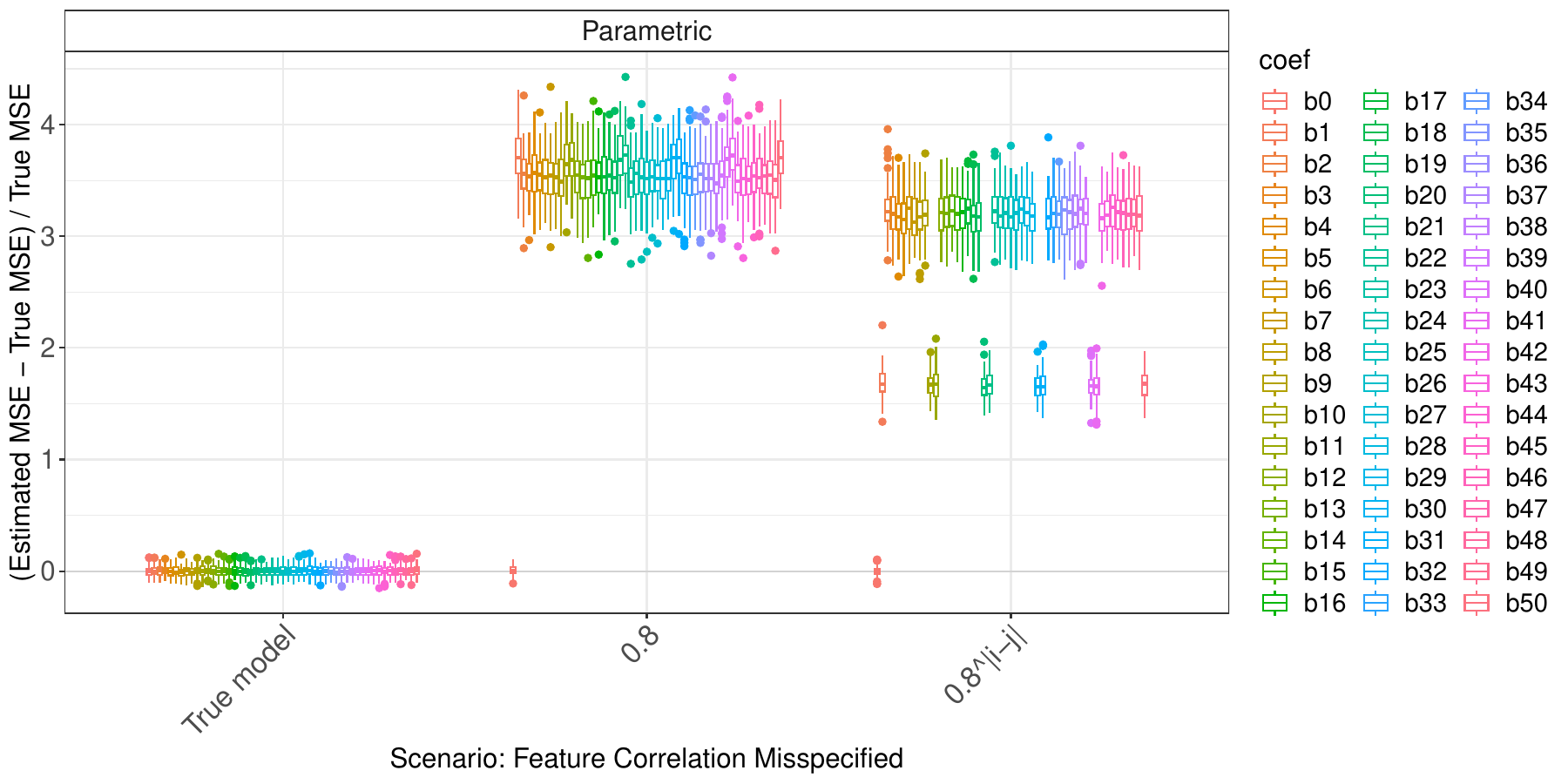}
		\caption{Relative error in MSE estimation for individual coefficients when the assumed correlation of the features in parametric simulation deviates from true correlation, for $p = 50,\: n = 100, \:\beta = (1, 1, 1)^T, \:\sigma = 0.3,\: Cor(X_i, X_j) = 0.2^{|i-j|}$ for $i$th and $j$th feature within each of the 5 blocks.}
		\label{fig:feat.corr.block}
	\end{figure}

	When the true correlation matrix has a block structure, we observe lower errors for the coefficients at the margins of the blocks if the value of the correlations but not their structure is misspecified (Figure \ref{fig:feat.corr.block}). 
	Again, this can be derived theoretically for the very simple case described above when inserting the block diagonal structure for $\Sigma$. 
	
	\subsubsection{Gaussian mixture}
\begin{figure}[!b]
		\centering
		\includegraphics[width = \textwidth, page = 6]{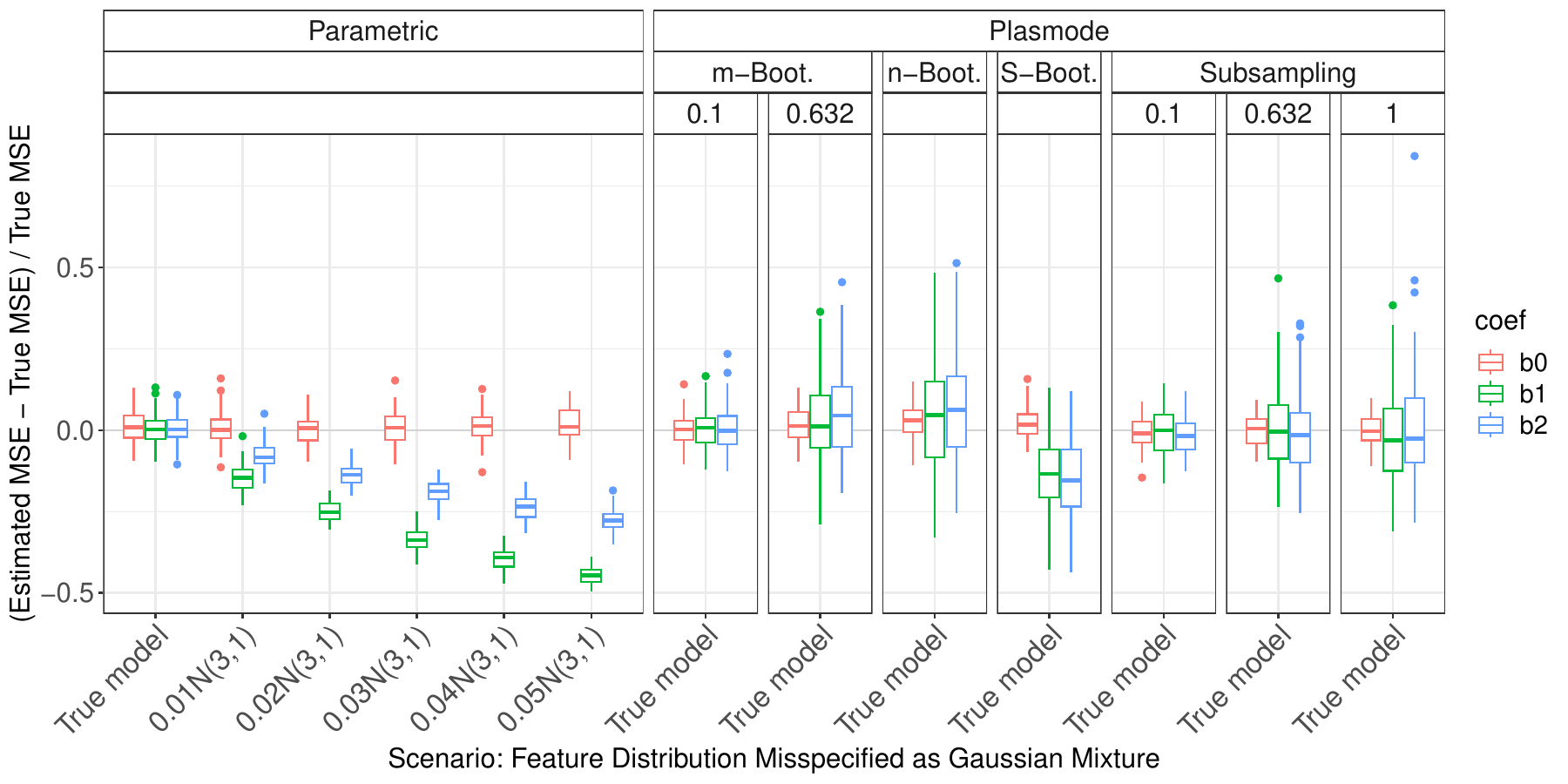}
		\caption{Relative error in MSE estimation for individual coefficients when the assumed marginal distribution of the second feature in parametric simulation is misspecified as Gaussian mixture with increasing proportion of data drawn from Gaussian with different expectations (bimodal distribution), for $p = 2,\: n = 100, \:\beta = (1, 1, 1)^T, \:\sigma = 0.3,\: Cor(X_i, X_j) = 0.2~\forall i\ne j$. The mean and the variance of the marginal normal distribution of the first feature are set to match those of the second. The mixing proportion is given on the $x$-axis.}
		\label{fig:feat.dist.bimodal}
	\end{figure}

\begin{figure}[!tb ]
		\centering
		\includegraphics[width = \textwidth, page = 8]{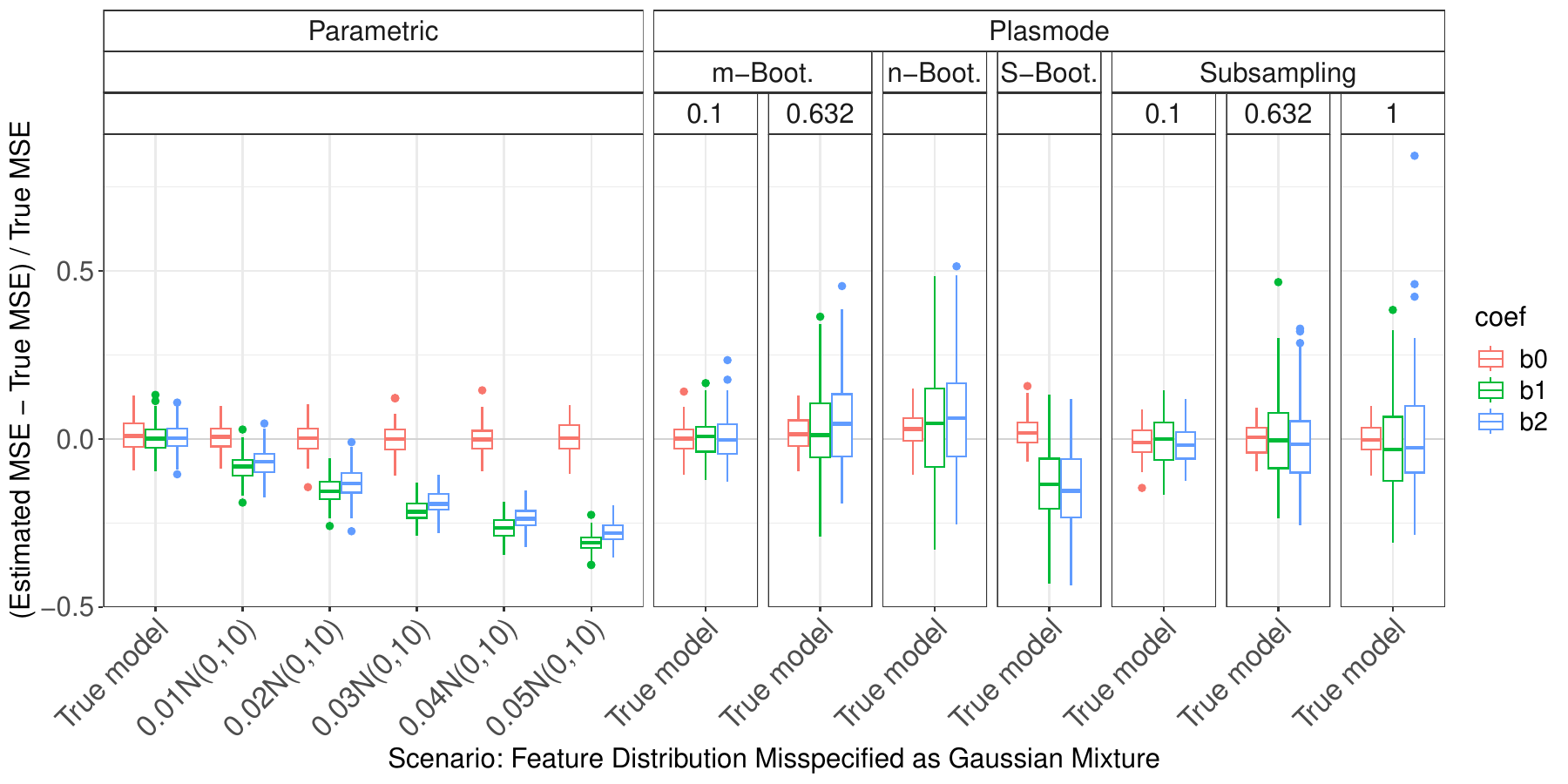}
		\caption{Relative error in MSE estimation for individual coefficients when the assumed marginal distribution of the second feature in parametric simulation is misspecified as Gaussian mixture with increasing proportion of data drawn from Gaussian with different variance (contaminated distribution), for $p = 2,\: n = 100, \:\beta = (1, 1, 1)^T, \:\sigma = 0.3,\: Cor(X_i, X_j) = 0.2~\forall i\ne j$. The mean and the variance of the marginal normal distribution of the first feature are set to match those of the second. The mixing proportion is given on the $x$-axis.}
		\label{fig:feat.dist.outlier}
	\end{figure}

	Next, we use two different versions of Gaussian mixtures as feature distributions for the second half of the features. 
	With this, not only the parameter but the whole shape of the distribution is altered. 
	For the first type of Gaussian mixture, a proportion of $\alpha$ of the observations stems from a normal distribution with mean $3$ and variance $1$. 
	This yields a bimodal distribution.
	For the second type of Gaussian mixture, a proportion of $\alpha$ of the observations stems from a normal distribution with mean $0$ and variance $10$. 
	This represents a contamination model with outliers. 
	In both cases, the remaining proportion of $1-\alpha$ stems from the standard normal, in agreement with the true distribution. 
	We always set the marginal distribution of the first feature to a normal that has the same mean and variance as the Gaussian mixture for the second marginal distribution, and successively increase the proportion in the mixing distribution. 
	This enables us to separate the influence of the change in expectation and variance of the distribution from the effect of the bimodality and outliers.\\
	In the bimodal case (Figure \ref{fig:feat.dist.bimodal}), we see that with an increasing proportion of observations from the $N(3,1)$ distribution, the underestimation of the MSE for the corresponding second coefficient also increases. 
	It is still less pronounced than for the first coefficient, which corresponds to the normal with wrong expectation and variance. 
	This might be due to the fact that most of the observations in the mixture distribution belong to the true distribution. 
	In the case of a normal distribution with wrong expectation and variance, all observations come from a distribution that differs from the true one.
	For the contamination model (Figure \ref{fig:feat.dist.outlier}), we observe the same behavior, but the differences between the coefficients are smaller there. 
	
	\subsubsection{Log-normal}
\begin{figure}[!tb ]
		\centering
		\includegraphics[width = \textwidth, page = 12]{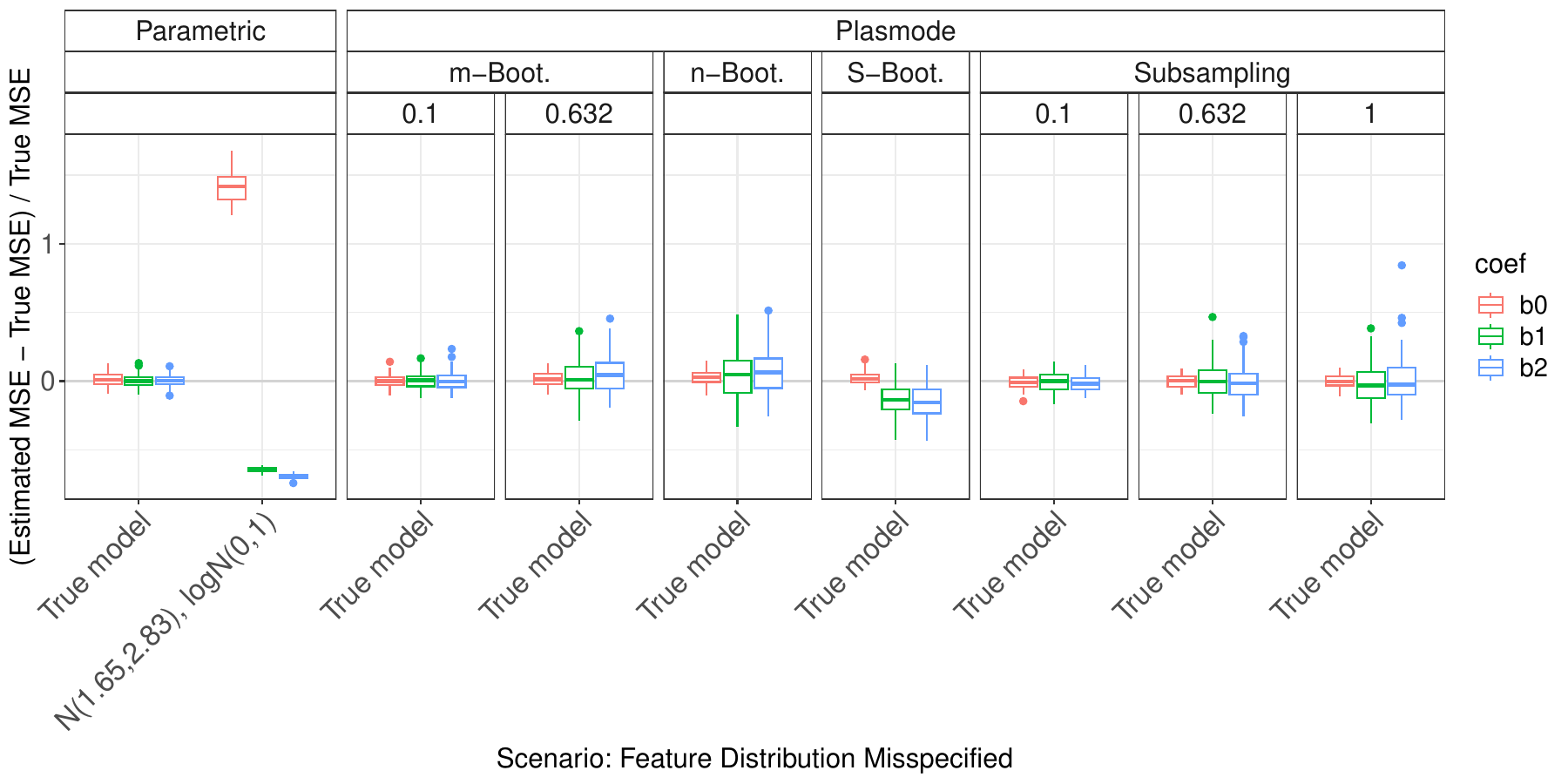}
		\caption{Relative error in MSE estimation for individual coefficients when the assumed marginal distribution of the second feature in parametric simulation is misspecified as log-normal, for $p = 2,\: n = 100, \:\beta = (1, 1, 1)^T, \:\sigma = 0.3,\: Cor(X_i, X_j) = 0.2~\forall i\ne j$. The mean and the variance of the marginal normal distribution of the first feature are set to match those of the second.}
		\label{fig:feat.dist.log}
	\end{figure}

	Figure \ref{fig:feat.dist.log} shows the relative errors for the individual coefficients when the distribution of the second feature is misspecified as log-normal and the distribution of the first feature is misspecified as a normal with matching mean and variance. 
	There is a large overestimation of the MSE for the intercept, while the MSEs for the other coefficients are underestimated. 
	The underestimation is slightly worse for the second coefficient than for the first, so the additional skewness of the log-normal leads to worse MSE estimation compared to a normal with the same mean and variance. 
	The errors in all coefficients for this deviation are considerably higher than those of any Plasmode variant that is compared here.
	
	\subsubsection{Bernoulli}
\begin{figure}[!tb ]
		\centering
		\includegraphics[width = \textwidth, page = 10]{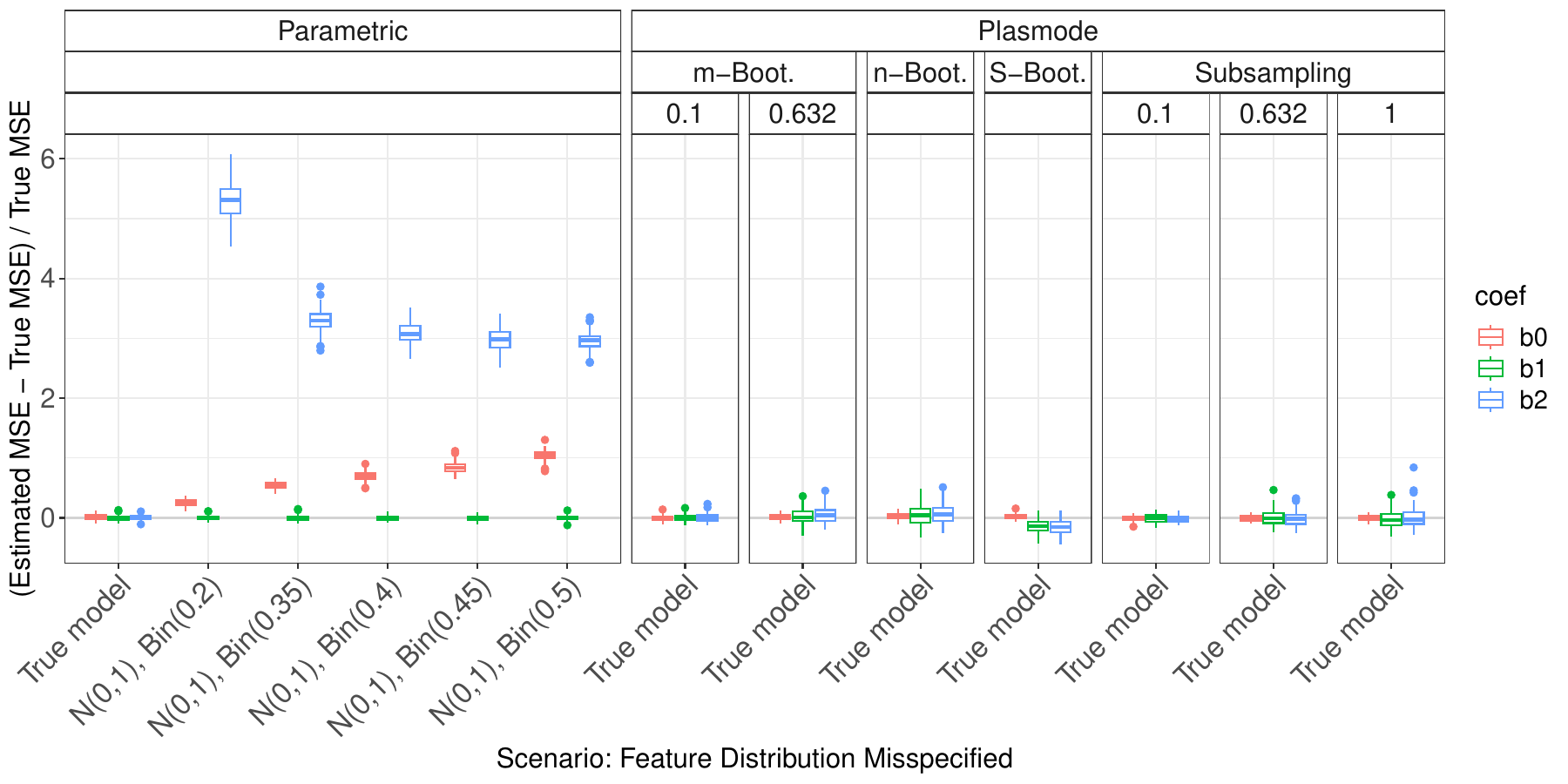}
		\caption{Relative error in MSE estimation for individual coefficients when the assumed marginal distribution of the second feature in parametric simulation is misspecified as Bernoulli with different success probabilities, for $p = 2,\: n = 100, \:\beta = (1, 1, 1)^T, \:\sigma = 0.3,\: Cor(X_i, X_j) = 0.2~\forall i\ne j$.}
		\label{fig:feat.dist.bin}
	\end{figure}

	Figure \ref{fig:feat.dist.bin} shows the relative errors for the individual coefficients when the distribution of the second feature is misspecified as Bernoulli and the distribution of the first feature is correctly specified as a standard normal. 
	We observe an increasing overestimation of the MSE for the intercept with increasing success probabilities. 
	The MSE for the first coefficient is unaffected. 
	The MSE for the coefficient belonging to the binary feature is also clearly overestimated, where the overestimation decreases towards success probabilities of $0.5$. 
	The errors for the intercept and the second coefficient for this deviation are considerably higher than those of any Plasmode variant that is compared here.
	
	\subsection{Deviations from true coefficients}\label{sec:coef}
\begin{figure}[!t ]
		\centering
		\includegraphics[width = \textwidth, page = 2]{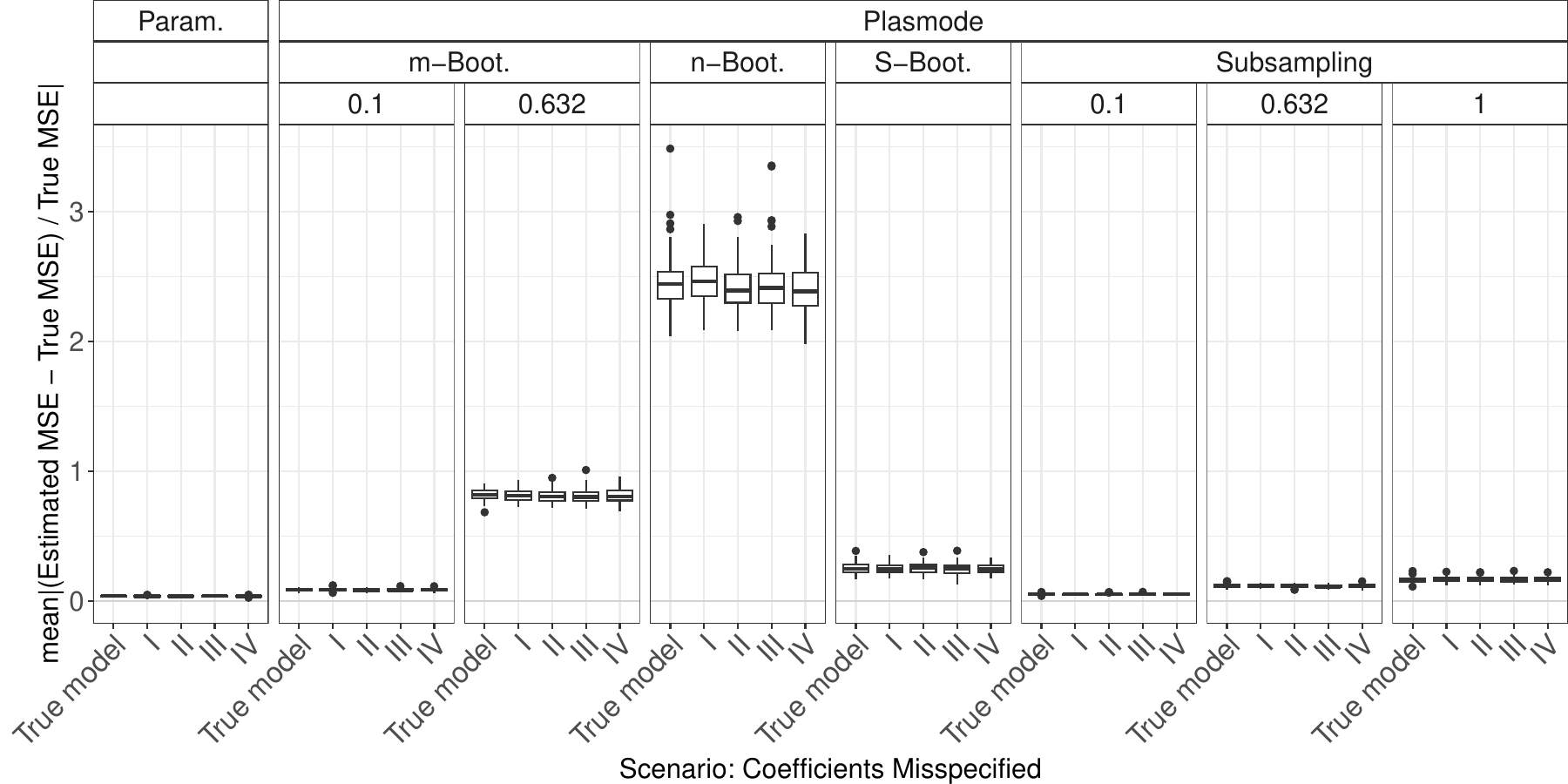}
		\caption{Absolute value of relative error in MSE estimation averaged over individual coefficients when the assumed coefficients in parametric and Plasmode simulation are misspecified, for $p = 50,\: n = 100, \:\beta = \boldsymbol{1}_{51}, \:\sigma = 0.3,\: Cor(X_i, X_j) = 0.2~\forall i\ne j, \:\beta_I = (0, 0.02,\dots,1)^T, \:\beta_{II} = \boldsymbol{0.05}_{51}, \:\beta_{III} = \boldsymbol{10}_{51}, \:\beta_{IV} = \boldsymbol{0}_{51}$. Large outliers for $n$ out of $n$ Bootstrap are not displayed.}
		\label{fig:coef}
	\end{figure}

	Figure \ref{fig:coef} shows the aggregated relative errors in MSE estimation for $p = 50$ and fixed correlations of $0.2$ for misspecifications of the coefficient vector $\beta$. 
	Since the specification of the coefficient vector is part of the OGM, this concerns all types of simulations. 
	For each simulation type, the errors for the misspecified coefficients do not differ from the errors for the true model. 
	Therefore, we conclude that the assumed values for the coefficients do not affect the simulation results. 
	The theoretical MSE formula for given $X$ is also only dependent on $\sigma$ and $X$, so independent of $\beta$.
	
	\subsection{Deviations from true error variance}\label{sec:err.var}
\begin{figure}[!tb ]
		\centering
		\includegraphics[width = \textwidth, page = 2]{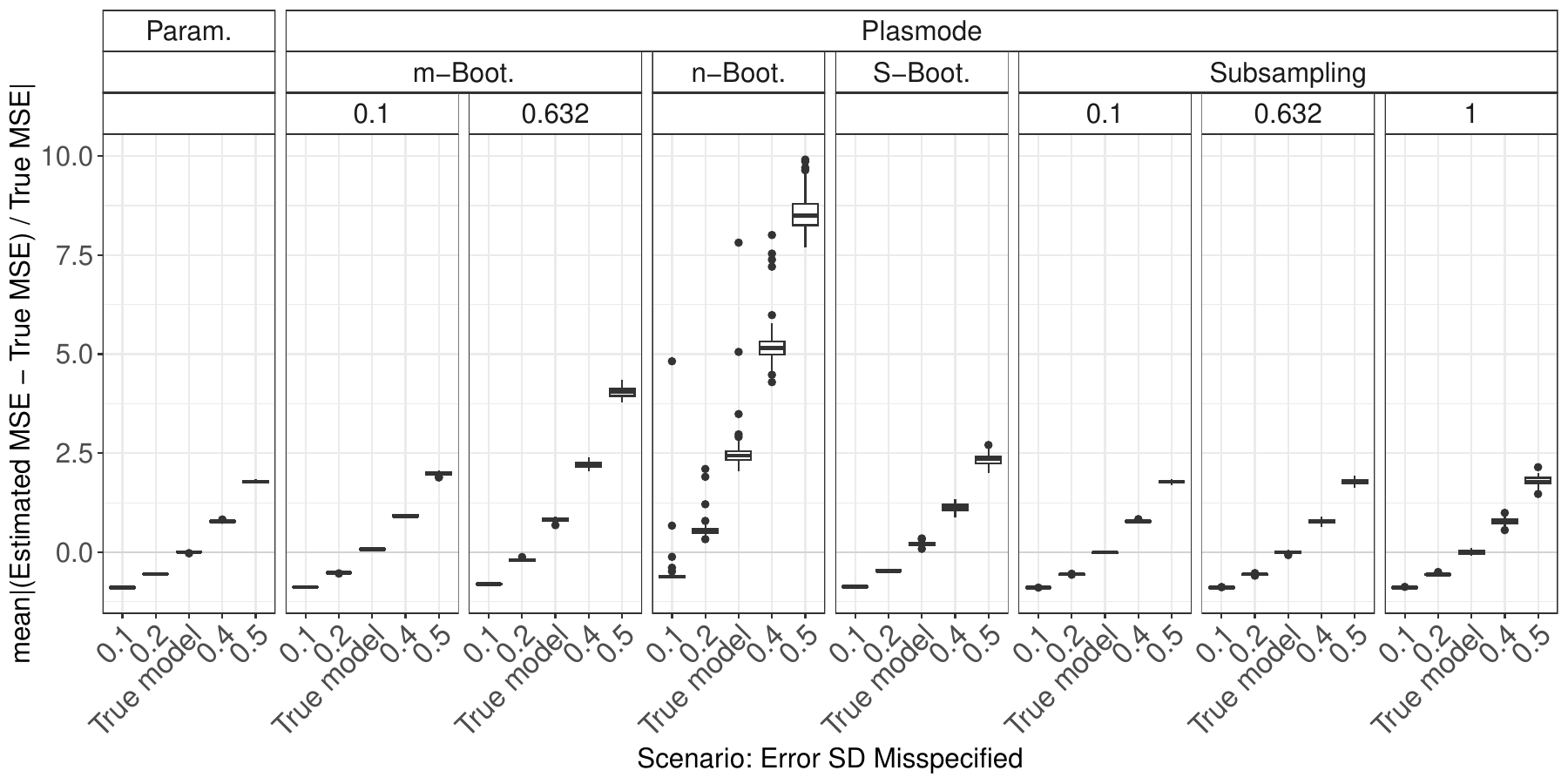}
		\caption{Absolute value of relative error in MSE estimation averaged over individual coefficients when the assumed error variance in parametric and Plasmode simulation are misspecified for $p = 50,\: n = 100, \:\beta = \boldsymbol{1}_{51}, \:\sigma = 0.3,\: Cor(X_i, X_j) = 0.2~\forall i\ne j$. Large outliers for $n$ out of $n$ Bootstrap are not displayed.}
		\label{fig:err.sd}
	\end{figure}

	In Figure \ref{fig:err.sd}, the aggregated relative errors in MSE estimation for $p = 50$ and fixed correlations of $0.2$ for misspecifications of the standard deviation of the error term $\varepsilon$ are shown. 
	Here, we use the relative errors directly without taking the absolute value to demonstrate under- and over-estimation.
	This again concerns all types of simulation. 
	In general, for too small error standard deviations, the true MSE is underestimated, and for too large error standard deviations, the true MSE is overestimated. 
	This pattern is visible for nearly all types of simulations. 
	For $m$ out of $n$ Bootstrap with large resampling proportions as well as for $n$ out of $n$ Bootstrap, the MSE is overestimated even for the true model, and the errors for other values of the error standard deviation are shifted up accordingly. 
	This leads to values closest to zero for too small error standard deviations. 
	In all cases, the variability of the errors increases with increasing error standard deviation. 
	We observe the same ordering that has already resulted for the true model (see Figure \ref{fig:comp.plasmode}) when comparing the errors from different simulation types for misspecified error standard deviations. 
	
	\subsection{Deviations from true error distribution}\label{sec:err.dist}
\begin{figure}[!tb ]
		\centering
		\includegraphics[width = \textwidth, page = 2]{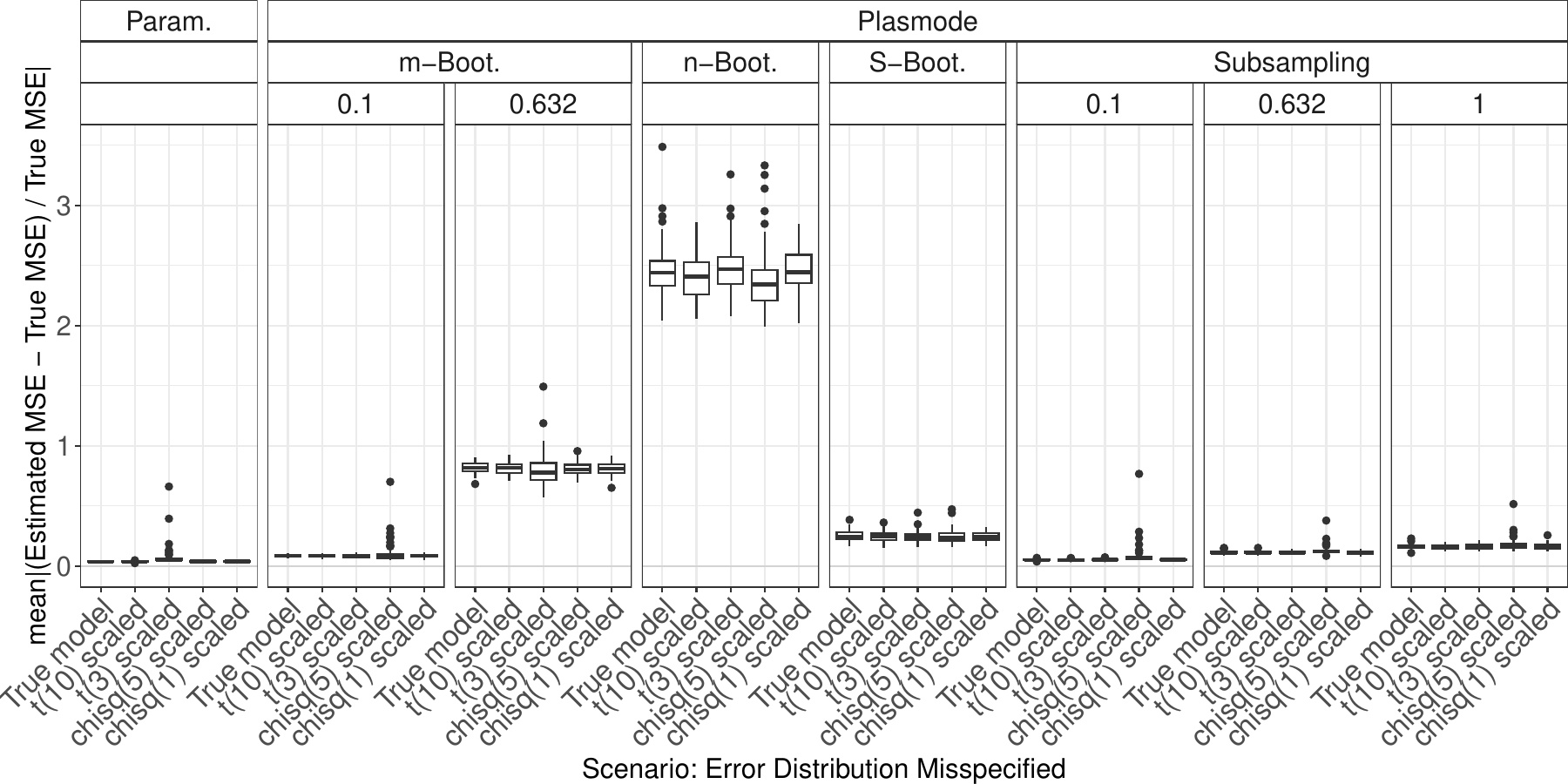}
		\caption{Absolute value of relative error in MSE estimation averaged over individual coefficients when the assumed error distributions in parametric and Plasmode simulation are misspecified, for $p = 50,\: n = 100, \:\beta = \boldsymbol{1}_{51}, \:\sigma = 0.3,\: Cor(X_i, X_j) = 0.2~\forall i\ne j$. Large outliers for $n$ out of $n$ Bootstrap are not displayed.}
		\label{fig:err.dist}
	\end{figure}

	In Figure \ref{fig:err.dist}, the aggregated relative errors in MSE estimation for $p = 50$ and fixed correlations of $0.2$ for misspecifications of the distribution of the error term are shown. 
	There are two types of misspecifications that we compare. 
	We use $t$-distributed errors as an example of a heavier-tailed distribution and $\chi^2$-distributed errors as an example of a skewed distribution. 
	Both are scaled and shifted in a way that the errors still have zero expectation and a standard deviation of $0.3$. 
	Overall, the distribution of the errors does not seem to have any influence on the error in MSE estimation as long as the error standard deviation and zero mean are preserved. 
	
	\subsection{True DGP: Correlation estimated from real data}\label{sec:est.cor}
\begin{figure}[!b ]
		\centering
		\includegraphics[width = \textwidth, page = 2]{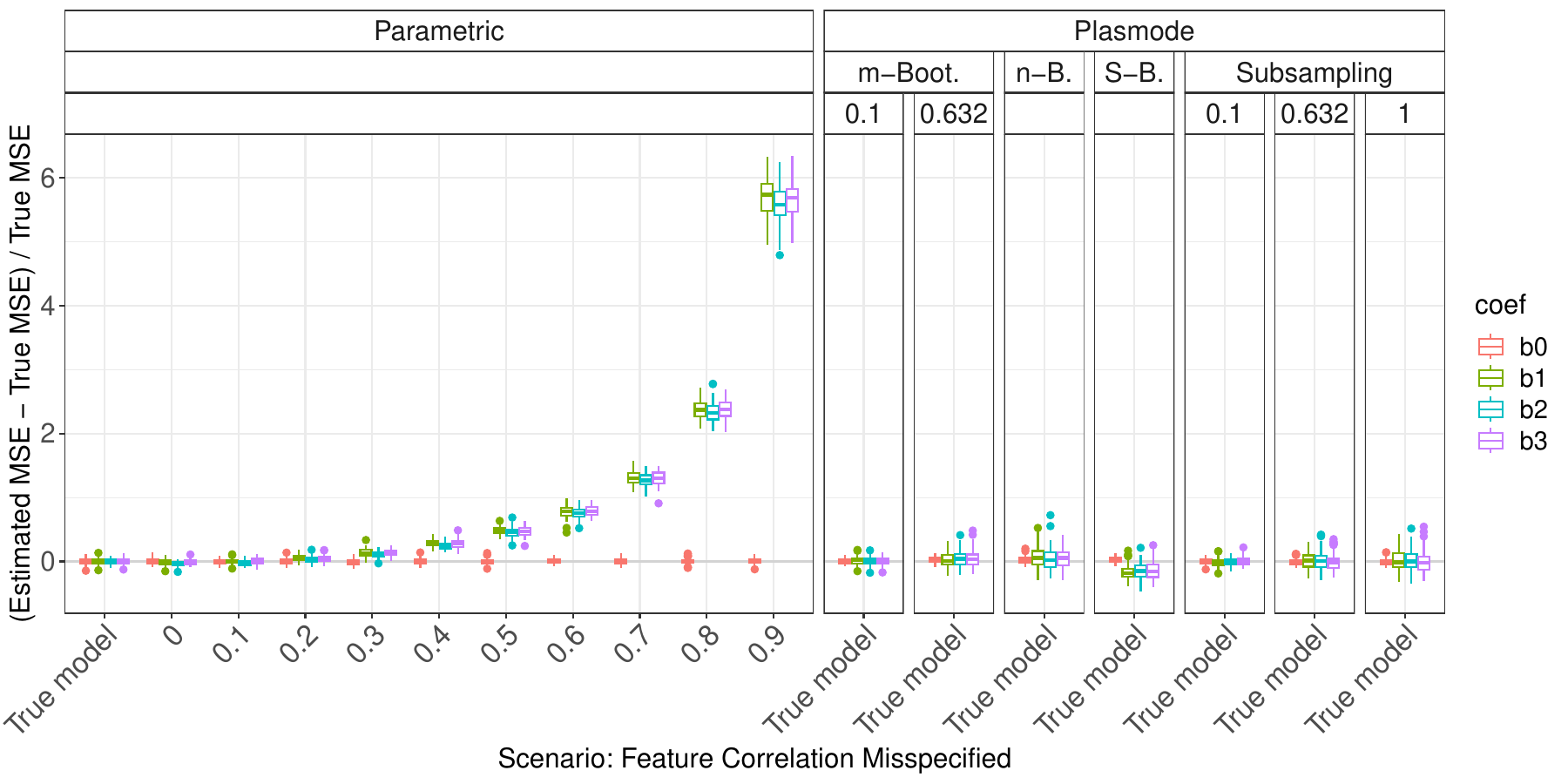}
		\caption{Absolute value of relative error in MSE estimation for individual coefficients when the assumed feature correlation matrix in parametric simulation is misspecified. True correlation matrix is estimated from the benchmark dataset quake ($p = 3,\: n = 100, \:\beta = \boldsymbol{1}_4, \:\sigma = 0.3$).}
		\label{fig:cor.quake}
	\end{figure}

\begin{figure}[!tb ]
		\centering
		\includegraphics[width = \textwidth, page = 4]{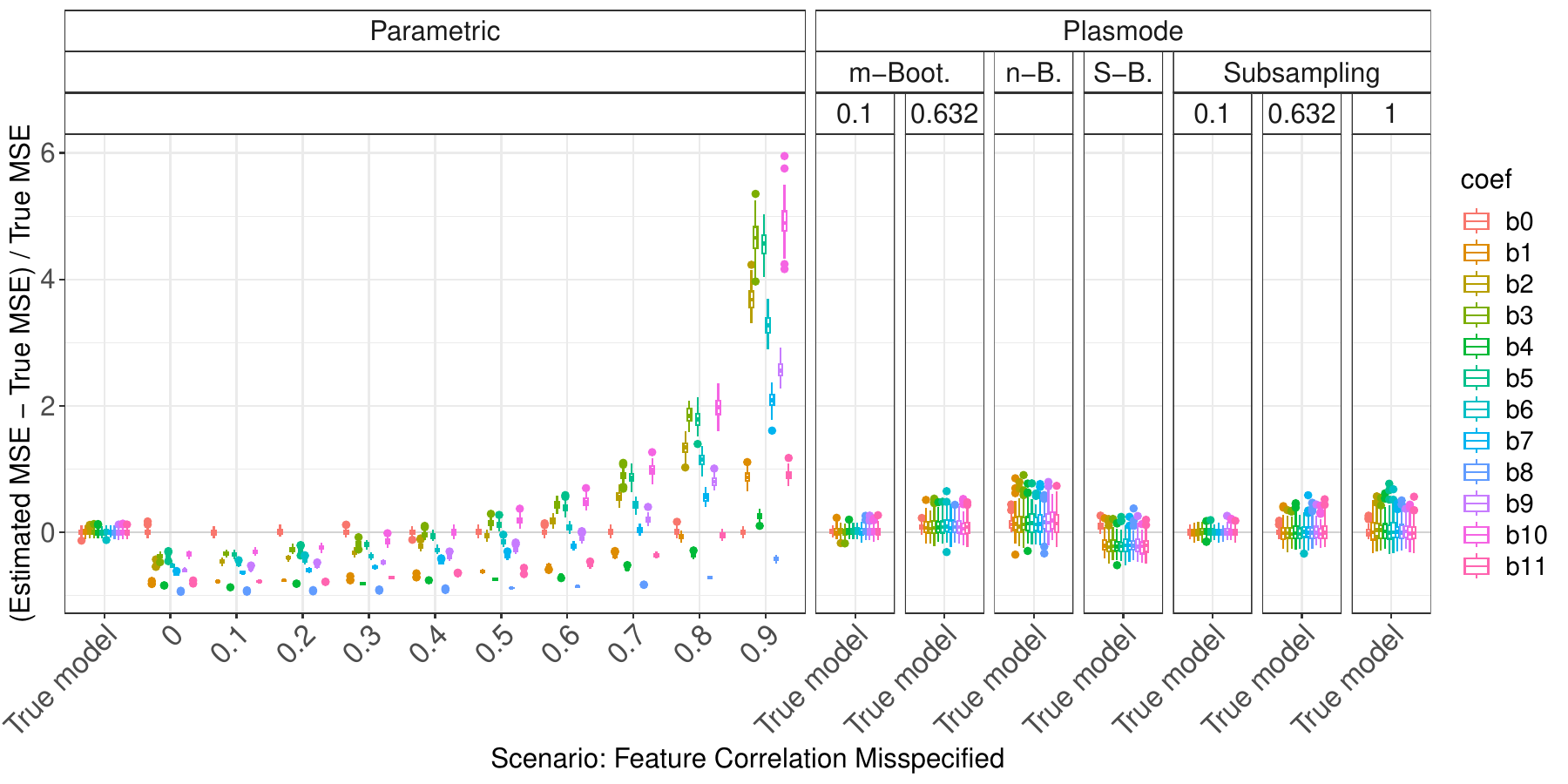}
		\caption{Absolute value of relative error in MSE estimation for individual coefficients when the assumed feature correlation matrix in parametric simulation is misspecified. True correlation matrix is estimated from benchmark dataset wine\_quality ($p = 11,\: n = 100, \:\beta = \boldsymbol{1}_{12}, \:\sigma = 0.3$).}
		\label{fig:wine.quality}
	\end{figure}

\begin{figure}[!tb ]
		\centering
		\includegraphics[width = \textwidth, page = 6]{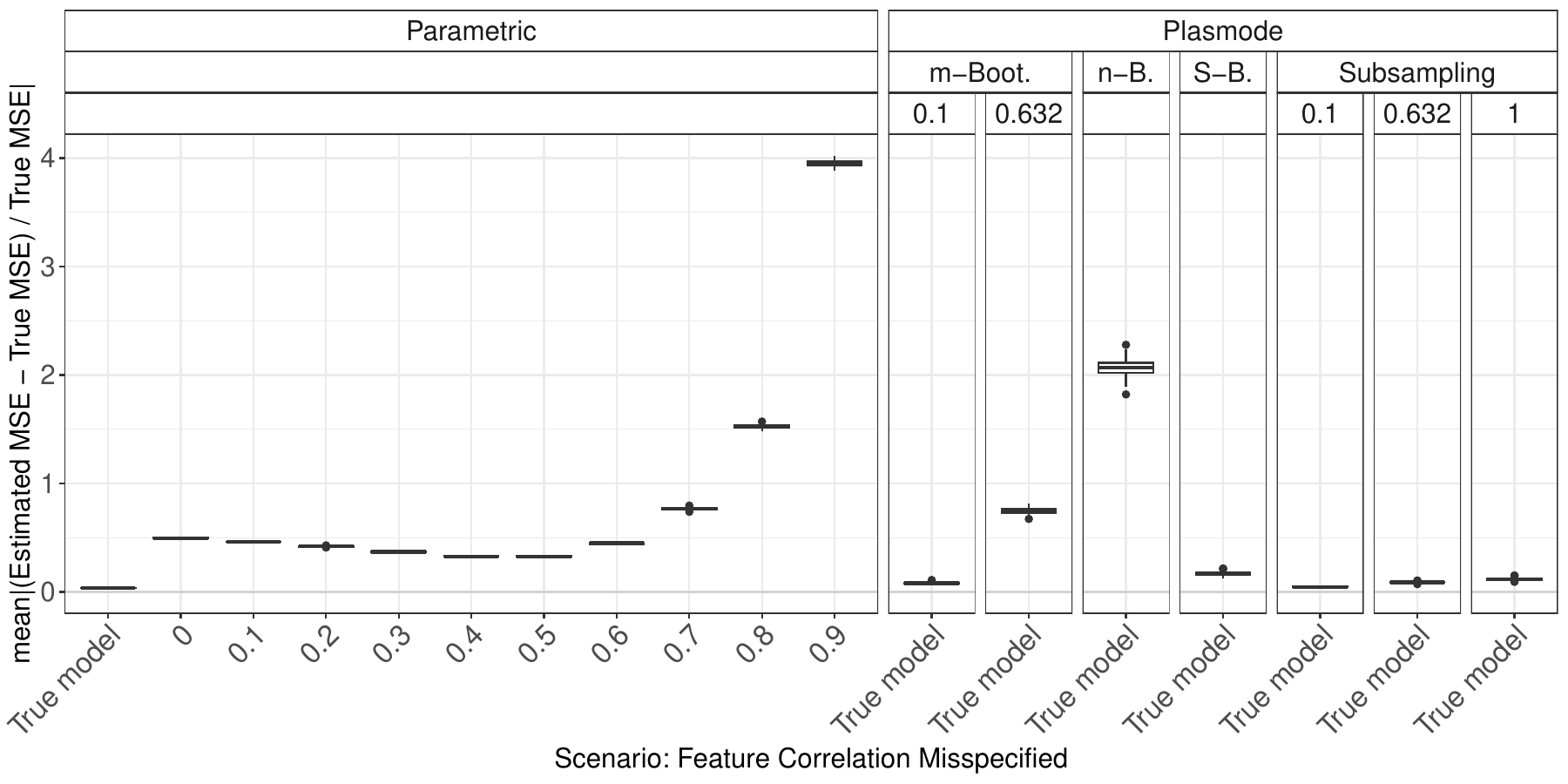}
		\caption{Absolute value of relative error in MSE estimation averaged over individual coefficients when the assumed feature correlation matrix in parametric simulation is misspecified. True correlation matrix is estimated from benchmark dataset Yolanda ($p = 100,\: n = 200, \:\beta = \boldsymbol{1}_{101}, \:\sigma = 0.3$).}
		\label{fig:Yolanda}
	\end{figure}

	We now analyze the results for the scenarios where the true correlation matrix is estimated from a real dataset. 
	In the following, we only discuss the results that differ from those for the simpler correlation structures we looked at before.
	These are all deviations that do not alter the correlation matrix. 
	For deviations from the true correlations, it gets more complicated. 
	In the case of small correlations that differ little, the results are still similar to those that we saw before. 
	For example, Figure \ref{fig:cor.quake} shows the results for the correlation estimated from the dataset quake. 
	The true pairwise correlations are $\text{Cor}(X_1, X_2) = -0.1286,\ \text{Cor}(X_1, X_3) = -0.0151$, and $\text{Cor}(X_2, X_3) = 0.1353$. 
	The results look similar to those we saw before for fixed correlations of $0.2$. 
	On the other hand, for the other datasets, the estimated pairwise correlations show higher variation, which means that no fixed value can be used to approximate all correlations simultaneously in a good way. 
	This is, for example, clearly visible in Figure \ref{fig:wine.quality} for the correlation matrix estimated from the dataset wine\_quality. 
	For each choice of fixed pairwise correlation, there are some coefficients with very large relative errors. 
	This can also lead to errors showing a pattern that differs from the parabolic shape we observed before (Figure \ref{fig:feat.corr.0.5.only.param}), as can be seen in Figure \ref{fig:Yolanda} for the dataset Yolanda. 
	For those cases where no constant correlation approximates all real correlations well, many of the Plasmode variants outperform parametric simulation for all assumed oversimplified correlation structures. 
	A possible cure for parametric simulation would be to estimate the correlation structure from real data, which -- in this case -- corresponds to the true model. 
	Overall, assuming some simple correlation structure, like often done in parametric simulations, might lead to high errors in the estimation of the MSE in cases where the true correlation structure is more complicated. 
	To correctly guess this correlation structure is highly unlikely, and it might even be impossible to specify complicated correlation structures in high-dimensional settings.

	\subsection{Size of resampled datasets}\label{sec:wrong.n}
\begin{figure}[!b]
		\centering
		\includegraphics[width = \textwidth, page = 1]{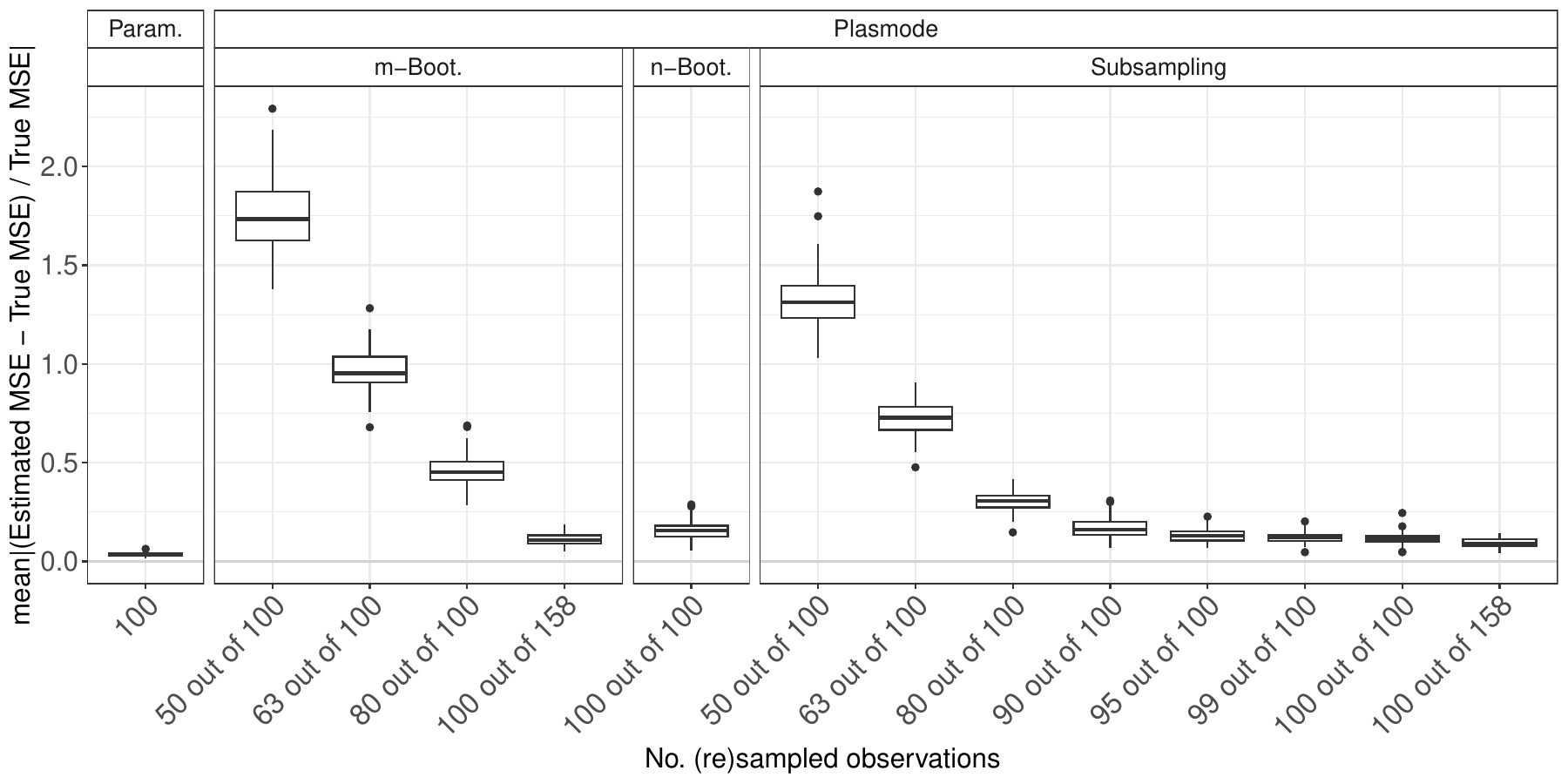}
		\caption{Comparison of different resampling types for different numbers of observations resampled from a dataset with 100 observations. Absolute value of relative error in MSE estimation averaged over individual coefficients when the true model is assumed in parametric and Plasmode simulation, for $p = 10,\: n = 100, \:\beta = \boldsymbol{1}_{11}, \:\sigma = 0.3,\: Cor(X_i, X_j) = 0.2~\forall i\ne j$.}
		\label{fig:wrong.n}
	\end{figure}

	Until now, we have always compared simulations that use the same number of observations, which leads to differently sized datasets from which the Plasmode data is resampled. 
	This might seem unintuitive, but it is necessary to ensure a fairer comparison of the simulation methods since the true MSE that the estimations are compared to is monotonously decreasing in the number of observations in the dataset. 
	Therefore, if we set the size of the dataset that we are resampling from to $100$ and resample smaller datasets from this, the MSE will always be overestimated, even for the true model. 
	This means that if we want to estimate the MSE for datasets of a certain size $n$, we have to use datasets of that exact size in our simulations. 
	However, it might be unrealistic that we have a dataset of the correct size at hand to resample from for our simulation. 
	For example, if we use simulation to estimate a quantity that cannot be estimated directly from the data since it depends on unknown parameters (e.g. the bias of an estimator), we might have a concrete dataset at hand for which we want to estimate this quantity. 
	In this case, Plasmode would be a natural choice, and since the number of observations is limited, we might use resampled datasets of smaller size to estimate the quantity for the whole dataset. 
	We now discuss the results for this case for $p = 10$ for the true model. 
	For $p = 2$, differences between the resampling methods are very small anyway. 
	For $p = 50$, it will be hard to differentiate between the errors occurring due to the differently sized datasets and the errors caused by approaching the boundary of identifiability. 
	Figure \ref{fig:wrong.n} shows the results for the different Bootstrap methods compared to parametric simulation for differing sizes of datasets resampled from a dataset of size $100$.  
	For comparison, the case of resampling $100$ out of $158$ observations that have been used in the analysis so far for a resampling proportion of $0.632$ is also included. 
	The estimated MSEs are compared to the true MSE for $n = 100$ in all cases. 
	Higher errors are observed for smaller sizes of the resampled dataset. 
	The smallest errors are observed for subsampling with the subsampling proportion approaching the number of observations in the dataset. 
	So in the case where the number of observations is limited to the number of observations that we are interested in, it might even be the best choice to do no resampling at all and just generate different responses for the MSE estimation. 
	It should be noted that when fixing the size of the dataset to resample from, the $n$ out of $n$ Bootstrap performs comparably well. 
	A reason for this might be that it uses a dataset of size $100$ for estimating the MSE. 
	Therefore, no errors occur due to the dependency of the MSE on $n$. 
	Moreover, the $n$ out of $n$ Bootstrap can use the dataset more efficiently since it uses more samples for the MSE estimation than subsampling or the $m$ out of $n$ Bootstrap with lower resampling proportions.

	\section{Conclusions and recommendations}\label{sec:conclusions}
	In the following, we summarize what we have learned from the comparisons that we performed. 
	First, we provide some general insights. 
	Then, we present detailed comparisons for which types of deviations from the data-generating process Plasmode was superior to parametric simulation in our analyses.
	
	\subsection{General insights}
	We looked at different true data-generating processes (DGP) and deviations from those for the estimation of the MSE of the least squares estimator (LSE) in linear regression to compare how well different simulation types perform in this case. 
	Overall, we saw that if there is no deviation from the true scenario, parametric simulation outperforms all Plasmode simulations. 
	The same holds for deviations that affect parametric as well as Plasmode, i.e.\ deviations from the outcome generating model (OGM), given that the DGP used for parametric simulation is close to the truth. 
	We saw that the misspecification of the coefficients and of the error distribution (as long as expectation and variance are kept) does not have any effect on the quality of the MSE estimation, while the misspecification of the error standard deviation does have an effect. \\
	Misspecifications of the DGP only affect parametric simulation. 
	For all kinds of misspecifications of the DGP in parametric simulation (misspecification of expectation, variance, correlation, whole distribution), parametric simulation can become worse than Plasmode. 
	The degree of misspecification needed for Plasmode to be superior depends on the type of misspecification, the resampling method used in the context of Plasmode that we compare with, and on the number of observations $n$ and the number of features $p$.
	A detailed analysis of the degree of misspecification that is needed for Plasmode to be superior is given in Subsection \ref{sec:detailed.comp}.\\
	Within the different resampling strategies for Plasmode simulations, we observed that in general, Wild Bootstrap performed worst, followed by $n$ out of $n$ Bootstrap. 
	$m$ out of $n$ Bootstrap performed better than $n$ out of $n$ and subsampling usually performed best. 
	For both $m$ out of $n$ Bootstrap and subsampling, smaller resampling proportions are favorable.
	This means that for a fixed number of subsampled observations $n$ of interest, larger datasets to resample from are required. 
	Smoothed Bootstrap usually performs worse than subsampling and even than no resampling (subsampling proportion of one), but better than $m$ out of $n$ Bootstrap with moderate resampling rates, i.e.\ rates larger than $0.5$. 
	When the number of observations for resampling is limited to the number of observations that we are interested in, we are restricted to $n$ out of $n$ Bootstrap, Smoothed Bootstrap, Wild Bootstrap, no resampling at all (i.e. subsampling with the proportion of one), or resampling a dataset of smaller size for Plasmode. 
	Our analyses suggest that no resampling at all or subsampling with a subsampling proportion very close to one might be the best choice in this case. 
	This is due to the dependence of the MSE on the number of observations, which leads to biased estimates of the MSE if the number of observations used for the simulation differs from the number of observations of interest.
	
	\subsection{Detailed comparisons}\label{sec:detailed.comp}
	Table \ref{tab:comparison} presents the values for each scenario and deviation at which certain types of Plasmode simulation are superior to parametric simulation.
	As discussed before, this is only applicable to deviations regarding the data-generating process (DGP). 
	The numbers given in the Plasmode columns are calculated as follows. 
	For the given scenario, deviation and Plasmode type, the deviations are ordered increasingly. 
	Then, the first deviation for which the median aggregated relative error of the parametric is higher than that of the Plasmode type is identified. 
	These values correspond to the medians in the aggregated boxplots.
	For example, in the first row, the case of $p = 2,\: n = 100$ and fixed pairwise correlations of $0.2$ is analyzed for deviations of the assumed expected value for the second feature. 
	The true expectation is 0. 
	Plasmode with $m$ out of $n$ Bootstrap or subsampling with a resampling proportion of $0.1$ is superior to parametric simulation for assumed expectations of $0.25$ and higher. 
	Plasmode with $m$ out of $n$ Bootstrap or subsampling with a resampling proportion of $0.632$ is only superior for assumed expectations of $0.4$ and higher, $n$ out of $n$ Bootstrap for values of $0.5$ and higher, Smoothed Bootstrap for values of $0.55$ and higher, and Plasmode without resampling (subsampling with proportion of 1) for values of $0.45$ and higher.\\
	When using correlation matrices estimated from real datasets, the order for the deviations in the correlations is unclear, as discussed in Section \ref{sec:est.cor}. 
	Therefore, they are excluded from the comparison. 
	Also, in all cases, assuming log-normal or binary data instead of normal data is worse than all Plasmode variants and, therefore, also excluded.\\
	For these analyses, in the parametric simulations, the expectations and high variances were increased in steps of $0.05$, and the low variances were decreased in steps of $0.1$. 
	The mixing proportion for Gaussian mixtures and the pairwise correlations were increased in steps of $0.01$. \\
	For $p = 50$ and assuming Gaussian mixtures, in some cases, even a proportion of 100\% data for the second half of features coming from the wrong distribution is not sufficient for Plasmode to be superior, as can be concluded from the values found for deviating expectations and variances. 
	The corresponding entries in Table \ref{tab:comparison} are left empty in these cases. 
	\begin{landscape}
		\centering
		\begin{xltabular}{\linewidth}{ >{\raggedleft\arraybackslash} X >{\raggedleft\arraybackslash}X >{\raggedleft\arraybackslash}X p{6cm} >{\raggedleft\arraybackslash}X  >{\raggedleft\arraybackslash}X >{\raggedleft\arraybackslash}X >{\raggedleft\arraybackslash}X >{\raggedleft\arraybackslash}X >{\raggedleft\arraybackslash}X >{\raggedleft\arraybackslash}X >{\raggedleft\arraybackslash}X}
			\toprule
			$p$ & $n$ & True $\rho$ & Scenario type & True value & \multicolumn{2}{c}{ $m$-Bootstrap} &  $n$-Bootstrap &  Smoothed Bootstrap & \multicolumn{2}{c}{ Subsampling} &  No resampling \\
			& & & & & 0.1 & 0.632 & & & 0.1 & 0.632 & \\ 
			\midrule
			\endhead
			\bottomrule
			\caption{Smallest deviations in parametric simulations for which Plasmode simulation is superior to parametric simulation. $p$ denotes the number of features, $n$ the number of observations. True $\rho$ gives the true correlation structure, scenario type the type of deviation and true value the true parameter value that the deviation refers to.}\\
			\label{tab:comparison}\\
			\endfoot
			2 & 100 & 0.2 & Expectation of 2nd feature misspecified & 0 & 0.25 & 0.4 & 0.5 & 0.55 & 0.25 & 0.4 & 0.45 \\ 
			2 & 100 & 0.2 & Variance of 2nd feature misspecified & 1 & 1.05 & 1.15 & 1.15 & 1.2 & 1.1 & 1.1 & 1.15 \\ 
			2 & 100 & 0.2 & Distribution misspecified: Gaussian mixture with N(0,10) & 0 & 0.01 & 0.02 & 0.02 & 0.03 & 0.01 & 0.02 & 0.02 \\ 
			2 & 100 & 0.2 & Distribution misspecified: Gaussian mixture with N(3,1) & 0 & 0.01 & 0.01 & 0.02 & 0.02 & 0.01 & 0.01 & 0.01 \\ 
			2 & 100 & 0.2 & Feature correlation misspecified & N(0,1) & 0.28 & 0.35 & 0.39 & 0.4 & 0.29 & 0.35 & 0.36 \\ 
			2 & 50 & 0.2 & Expectation of 2nd feature misspecified & 0 & 0.3 & 0.5 & 0.55 & 0.6 & 0.3 & 0.5 & 0.55 \\ 
			2 & 50 & 0.2 & Variance of 2nd feature misspecified & 1 & 1.1 & 1.2 & 1.25 & 1.3 & 1.1 & 1.2 & 1.2 \\ 
			2 & 50 & 0.2 & Distribution misspecified: Gaussian mixture with N(0,10) & 0 & 0.01 & 0.03 & 0.03 & 0.04 & 0.01 & 0.03 & 0.03 \\ 
			2 & 50 & 0.2 & Distribution misspecified: Gaussian mixture with N(3,1) & 0 & 0.01 & 0.02 & 0.02 & 0.03 & 0.01 & 0.02 & 0.02 \\ 
			2 & 50 & 0.2 & Feature correlation misspecified & N(0,1) & 0.33 & 0.41 & 0.41 & 0.41 & 0.29 & 0.39 & 0.41 \\ 
			2 & 50 & 0.5 & Expectation of 2nd feature misspecified & 0 & 0.25 & 0.35 & 0.4 & 0.5 & 0.25 & 0.35 & 0.4 \\ 
			2 & 50 & 0.5 & Variance of 2nd feature misspecified & 1 & 1.05 & 1.15 & 1.15 & 1.25 & 1.05 & 1.1 & 1.15 \\ 
			2 & 50 & 0.5 & Distribution misspecified: Gaussian mixture with N(0,10) & 0 & 0.01 & 0.02 & 0.02 & 0.03 & 0.01 & 0.02 & 0.02 \\ 
			2 & 50 & 0.5 & Distribution misspecified: Gaussian mixture with N(3,1) & 0 & 0.01 & 0.02 & 0.02 & 0.03 & 0.01 & 0.01 & 0.02 \\ 
			2 & 50 & 0.5 & Feature correlation misspecified & N(0,1) & 0.54 & 0.57 & 0.57 & 0.61 & 0.54 & 0.56 & 0.57 \\ 
			10 & 100 & 0.2 & Expectation of 2nd half of features misspecified & 0 & 0.25 & 0.45 & 0.6 & 0.7 & 0.25 & 0.4 & 0.5 \\ 
			10 & 100 & 0.2 & Variance of 2nd half of features misspecified & 1 & 1.1 & 1.25 & 1.4 & 1.55 & 1.1 & 1.2 & 1.25 \\ 
			10 & 100 & 0.2 & Distribution misspecified: Gaussian mixture with N(0,10) & 0 & 0.01 & 0.02 & 0.03 & 0.04 & 0.01 & 0.02 & 0.02 \\ 
			10 & 100 & 0.2 & Distribution misspecified: Gaussian mixture with N(3,1) & 0 & 0.01 & 0.02 & 0.02 & 0.03 & 0.01 & 0.01 & 0.02 \\ 
			10 & 100 & 0.2 & Feature correlation misspecified & N(0,1) & 0.24 & 0.29 & 0.33 & 0.36 & 0.24 & 0.28 & 0.3 \\ 
			10 & 100 & 0.2 & Feature correlation misspecified $\rho^{|i-j|}$ & N(0,1) & 0.24 & 0.38 & 0.41 & 0.44 & 0.24 & 0.36 & 0.39 \\ 
			10 & 50 & 0.2 & Expectation of 2nd half of features misspecified & 0 & 0.3 & 0.7 & 0.9 & 0.7 & 0.3 & 0.55 & 0.65 \\ 
			10 & 50 & 0.2 & Variance of 2nd half of features misspecified & 1 & 1.1 & 1.55 & 2.45 & 1.55 & 1.1 & 1.3 & 1.45 \\ 
			10 & 50 & 0.2 & Distribution misspecified: Gaussian mixture with N(0,10) & 0 & 0.01 & 0.05 & 0.08 & 0.04 & 0.01 & 0.03 & 0.04 \\ 
			10 & 50 & 0.2 & Distribution misspecified: Gaussian mixture with N(3,1) & 0 & 0.01 & 0.03 & 0.06 & 0.03 & 0.01 & 0.02 & 0.03 \\ 
			10 & 50 & 0.2 & Feature correlation misspecified & N(0,1) & 0.26 & 0.36 & 0.43 & 0.36 & 0.25 & 0.31 & 0.34 \\ 
			10 & 50 & 0.2 & Feature correlation misspecified $\rho^{|i-j|}$ & N(0,1) & 0.33 & 0.44 & 0.5 & 0.44 & 0.22 & 0.39 & 0.42 \\ 
			50 & 100 & 0.2 & Expectation of 2nd half of features misspecified & 0 & 0.4 & 1.55 & 2.7 & 0.8 & 0.25 & 0.5 & 0.65 \\ 
			50 & 100 & 0.2 & Variance of 2nd half of features misspecified (too small) & 1 & 0.88 & 0.38 & 0.17 & 0.69 & 0.94 & 0.84 & 0.77 \\ 
			50 & 100 & 0.2 & Distribution misspecified: Gaussian mixture with N(0,10) & 0 & 0.02 & 0.57 &  & 0.05 & 0.01 & 0.02 & 0.03 \\ 
			50 & 100 & 0.2 & Distribution misspecified: Gaussian mixture with N(3,1) & 0 & 0.01 & 0.98 &  & 0.04 & 0.01 & 0.02 & 0.02 \\ 
			50 & 100 & 0.2 & Feature correlation misspecified & N(0,1) & 0.27 & 0.57 & 0.78 & 0.37 & 0.24 & 0.29 & 0.32 \\ 
			50 & 100 & 0.2 & Feature correlation misspecified $\rho^{|i-j|}$ & N(0,1) & 0.25 & 0.62 & 0.79 & 0.46 & 0.34 & 0.21 & 0.42 \\ 
			50 & 100 & $0.2^{|i-j|}$ in 5 blocks & Expectation of 2nd half of features misspecified & 0 & 0.4 & 2.05 & 2.6 & 0.8 & 0.25 & 0.5 & 0.6 \\ 
			50 & 100 & $0.2^{|i-j|}$ in 5 blocks & Variance of 2nd half of features misspecified (too small) & 1 & 0.88 & 0.39 & 0.17 & 0.68 & 0.94 & 0.84 & 0.77 \\ 
			50 & 100 & $0.2^{|i-j|}$ in 5 blocks & Distribution misspecified: Gaussian mixture with N(0,10) & 0 & 0.02 & 0.51 &  & 0.05 & 0.01 & 0.02 & 0.03 \\ 
			50 & 100 & $0.2^{|i-j|}$ in 5 blocks & Distribution misspecified: Gaussian mixture with N(3,1) & 0 & 0.01 & 0.26 &  & 0.03 & 0.01 & 0.02 & 0.02 \\ 
			50 & 100 & $0.2^{|i-j|}$ in 5 blocks & Feature correlation misspecified & N(0,1) & 0.2 & 0.5 & 0.74 & 0.28 & 0.2 & 0.2 & 0.22 \\ 
			50 & 100 & $0.2^{|i-j|}$ in 5 blocks & Feature correlation misspecified $\rho^{|i-j|}$ & $0.2^{|i-j|}$ & 0.3 & 0.59 & 0.78 & 0.41 & 0.25 & 0.32 & 0.36 \\ 
			50 & 100 & $0.5^{|i-j|}$ in 5 blocks & Expectation of 2nd half of features misspecified & 0 & 0.5 & 2.05 & 3.4 & 2.05 & 0.3 & 0.6 & 0.8 \\ 
			50 & 100 & $0.5^{|i-j|}$ in 5 blocks & Variance of 2nd half of features misspecified (too small) & 1 & 0.88 & 0.39 & 0.17 & 0.68 & 0.94 & 0.84 & 0.78 \\ 
			50 & 100 & $0.5^{|i-j|}$ in 5 blocks & Distribution misspecified: Gaussian mixture with N(0,10) & 0 & 0.02 & 0.43 &  & 0.05 & 0.01 & 0.02 & 0.04 \\ 
			50 & 100 & $0.5^{|i-j|}$ in 5 blocks & Distribution misspecified: Gaussian mixture with N(3,1) & 0 & 0.01 & 0.26 &  & 0.02 & 0.01 & 0.01 & 0.01 \\ 
			50 & 100 & $0.5^{|i-j|}$ in 5 blocks & Feature correlation misspecified & N(0,1) & 0.5 & 0.67 & 0.83 & 0.51 & 0.5 & 0.5 & 0.5 \\ 
			50 & 100 & $0.5^{|i-j|}$ in 5 blocks & Feature correlation misspecified $\rho^{|i-j|}$ & $0.5^{|i-j|}$ & 0.54 & 0.72 & 0.85 & 0.6 & 0.53 & 0.55 & 0.58 \\ 
			3 & 100 & quake & Expectation of 2nd half of features misspecified & 0 & 0.3 & 0.5 & 1 & 1 & 0.3 & 0.45 & 1 \\ 
			3 & 100 & quake & Variance of 2nd half of features misspecified (too small) & 1 & 0.99 & 0.99 & 0.99 & 0.99 & 0.99 & 0.99 & 0.99 \\ 
			3 & 100 & quake & Distribution misspecified: Gaussian mixture with N(0,10) & 0 & 0.01 & 0.02 & 0.02 & 0.03 & 0.01 & 0.02 & 0.02 \\ 
			3 & 100 & quake & Distribution misspecified: Gaussian mixture with N(3,1) & 0 & 0.01 & 0.01 & 0.02 & 0.02 & 0.01 & 0.01 & 0.01 \\ 
			
			\bottomrule
		\end{xltabular}
	\end{landscape}
	
	\section{Summary and Discussion}\label{sec:discussion}
	We performed a simulation study to compare the performance of parametric and Plasmode simulations in the context of MSE estimation for the least squares estimator (LSE) in the linear regression model. 
	For parametric simulation, artificial data is generated according to a fully user-specified data-generating process (DGP) for generating the feature data and outcome-generating model (OGM) for generating the outcome variable.
	In contrast to that, in Plasmode simulation, the feature data is generated by resampling from a real-life dataset, and only the OGM has to be specified. 
	For comparing the two approaches, we need control of the true underlying DGP and OGM. 
	We used different true DGPs and OGMs. 
	Since the true DGP and OGM are unknown in practice, they must be specified when conducting a simulation study. 
	For the Plasmode simulation, the DGP is implicitly given by the chosen dataset.
	This specification is likely a deviation from the truth. 
	Therefore, we examined the influence of different deviations on both types of simulation studies. 
	Note that for Plasmode, there is no explicit deviation from the DGP.
	When resampling from a dataset, one samples from the empirical DGP, which ideally converges to the true DGP.
	
	Within Plasmode simulations, we compared different resampling strategies, namely $n$ out of $n$ Bootstrap, $m$ out of $n$ Bootstrap, subsampling, smoothed Bootstrap, and wild Bootstrap, and, where applicable, also different resampling proportions. 
	Each simulation strategy was evaluated based on the differences between the MSEs estimated using the respective method and the true MSEs. 
	If the true DGP and OGM are known, it is obvious that parametric simulation is the optimal choice as long as drawing from the true DGP and OGM is feasible. 
	However, in reality, the true DGP and OGM are unknown and can at best be approximated using expert knowledge. 
	In Plasmode simulations, as long as a dataset from the DGP of interest is given, only the OGM has to be specified. 
	Therefore, our aim was to find out 
	\begin{enumerate}
		\item How much the DGP chosen in the parametric simulation can deviate from the truth before the parametric simulation becomes worse than the Plasmode simulation. 
		\item How deviations of the chosen OGM from the true OGM affect both parametric and Plasmode simulations.
		\item How the choice of the resampling type affects the Plasmode simulation.
	\end{enumerate} 
	In general, we observed that parametric simulation is superior to Plasmode in all situations where the DGP is specified correctly, i.e., for the true situation or deviations from the true OGM only. 
	For deviations from the DGP in parametric simulation, it depends on the kind of deviation, the degree of deviation, the number of observations, and especially on the number of features in the dataset and the type of resampling used for the Plasmode simulation. 
	For very small deviations, parametric simulations usually remain superior. 
	For low numbers of observations, or especially for higher numbers of features, the performance of Plasmode simulation decreases more drastically than that of parametric simulation, both in terms of the median difference between the estimated and true MSE and the variation of the estimated MSE. 
	This means that the deviations from the true DGP in the assumptions of parametric simulation have to be larger for Plasmode to be superior.
	The effect is more pronounced when using resampling strategies with replacement and a high resampling proportion. 
	A reason for this might be that in these cases, the number of unique observations is lower. 
	Therefore, less information is contained in the data, so the variance and consequently the MSE of the estimator are inflated.
	On the other hand, there are certain settings where Plasmode was always superior to parametric simulation in our study, such as when the DGP was severely misspecified, e.g., when using binary instead of standard normal features.

	The effect that Plasmode notably overestimates the true MSE for increasing $p$, especially for resampling with replacement and high resampling proportions, might be a property of the chosen simulation setup. 
	It is known that using Bootstrap to estimate the variance of the LSE in linear regression models for $p / n \to \kappa \in (0, 1)$ can lead to severe overestimation of the true variance. 
	For $n$ out of $n$ Bootstrap and features sampled from a multivariate standard Gaussian distribution, this property was formally shown and additionally demonstrated via simulation in \textcite{karoui_can_2016}.
	Overestimation of the variance implies overestimation of the MSE, so the arguments made in \textcite{karoui_can_2016} might in part explain the bad performance of Plasmode simulations that we observed. The authors also derived an overestimation of the variance by Jackknifing, which is similar to subsampling with resampling proportions very close to one.
	
	If the distribution class of the features was misspecified as log-normal or even Bernoulli instead of normal or if the true correlation matrix of the features is more complex and parametric simulation uses an oversimplified approximation for it, all types of resampling used for Plasmode simulations were superior. 
	
	Regarding the resampling strategy used for Plasmode simulations, we observed that wild Bootstrap performed by far the worst with respect to MSE estimation. 
	For the remaining types, $n$ out of $n$ Bootstrap was usually inferior to the other resampling strategies. 
	The performance of $m$ out of $n$ Bootstrap and subsampling depends on the chosen subsampling proportion. 
	Generally, smaller proportions are beneficial. 
	Note that the size of the resulting dataset after resampling has to be fixed, so a smaller proportion corresponds to a larger dataset from which to sample. 
	For small resampling proportions, $m$ out of $n$ Bootstrap and subsampling behave very similarly; for larger proportions, subsampling performs better. 
	Smoothed Bootstrap performs similarly to $m$ out of $n$ Bootstrap and subsampling with moderate resampling proportions. 
	For the resampling proportion approaching one, $m$ out of $n$ Bootstrap converges to $n$ out of $n$ Bootstrap. 
	No resampling (subsampling with a resampling proportion of one), i.e., using the whole dataset for the features and only generating new responses in each iteration of the simulation, performed better than $n$ out of $n$ Bootstrap.
	The differences between the resampling types, except for wild Bootstrap, are negligible for small numbers of features ($p = 2$). 
	In that case, Plasmode using any resampling strategy might be a good option since even very small deviations from the DGP lead to parametric simulation being inferior. 
	In general, we suggest using subsampling with a small resampling proportion if feasible.
	
	For larger numbers of features, the performance of Plasmode simulations gets worse in general. 
	Nevertheless, there might still be good reasons for Plasmode simulations in this case. 
	For example, the specification of the DGP gets more and more complicated with an increasing number of features. 
	Especially, the specification of the correlation structure is non-trivial as the number of pairwise correlations increases quadratically with the number of features. 
	This might lead to the choice of oversimplified correlation structures for which we observed a clearly inferior performance of parametric compared to Plasmode simulation. 
	A remedy could be to at least estimate key parameters like the mean and the covariance matrix from a real dataset for the parametric simulation. 
	We observed good results for that strategy at least as long as the dataset from which the parameters are estimated is large enough.
	
	The availability of datasets might be a major limitation for the application of Plasmode simulations. 
	In general, at least one suitable dataset from the DGP of interest is required, see Section 3.2 in \textcite{schreck_statistical_2023} for a discussion. 
	Ideally, this dataset is considerably larger than the sample size $n$ of interest, to allow for a low resampling proportion. 
	In practice, this might often not be given. 
	If the dataset size is limited to the sample size one is interested in, our comparison suggests that no resampling (subsampling with a resampling proportion of one) might be a reasonable variant since the error made by using a dataset of the wrong sample size might outweigh the advantage of lower resampling proportions.
	
	Overall, the choice of the simulation type should be carefully considered for each application. 
	A combination of parametric and Plasmode simulation within a simulation study might be a solution to use both the flexibility of parametric simulation and the ability of Plasmode to preserve characteristics of real-life data.
	
	So far, the comparison of parametric and Plasmode simulation is limited to the specific example of estimating the MSE of the LSE. Therefore, further studies on other endpoints as well as a comparison for high-dimensional data, which brings additional challenges, might be interesting extensions of the analysis at hand.
	
	\section*{Acknowledgement}
	We thank Markus Pauly (TU Dortmund University \& UA Ruhr, Research Center Trustworthy Data Science and Security) for helpful discussions.
	
	\section*{Funding}
	This work has been supported (in part) by the Research Training Group ”Biostatistical Methods for High-Dimensional Data in Toxicology” (RTG 2624, Project P1) funded by the Deutsche Forschungsgemeinschaft (DFG, German Research Foundation - Project Number 427806116). 
	\section*{Competing Interests}
	The authors declare that they have no conflict of interest.
	
	\begin{sloppypar}
		\printbibliography[title = References]
	\end{sloppypar}
	
	\newpage
	\appendix
	\section*{Appendix}
	\section{Additional Figures and Tables}
	\begin{figure}[!htb]
		\centering
		\begin{subfigure}{\textwidth}
			\caption{$p = 2,\: n = 100, \:\beta = (1, 1, 1)^T, \:\sigma = 0.3,\: Cor(X_i, X_j) = 0.2~\forall i\ne j$.}\label{fig:ex.true.var.0.3}
			\centering
			\includegraphics[width = \textwidth, page = 1]{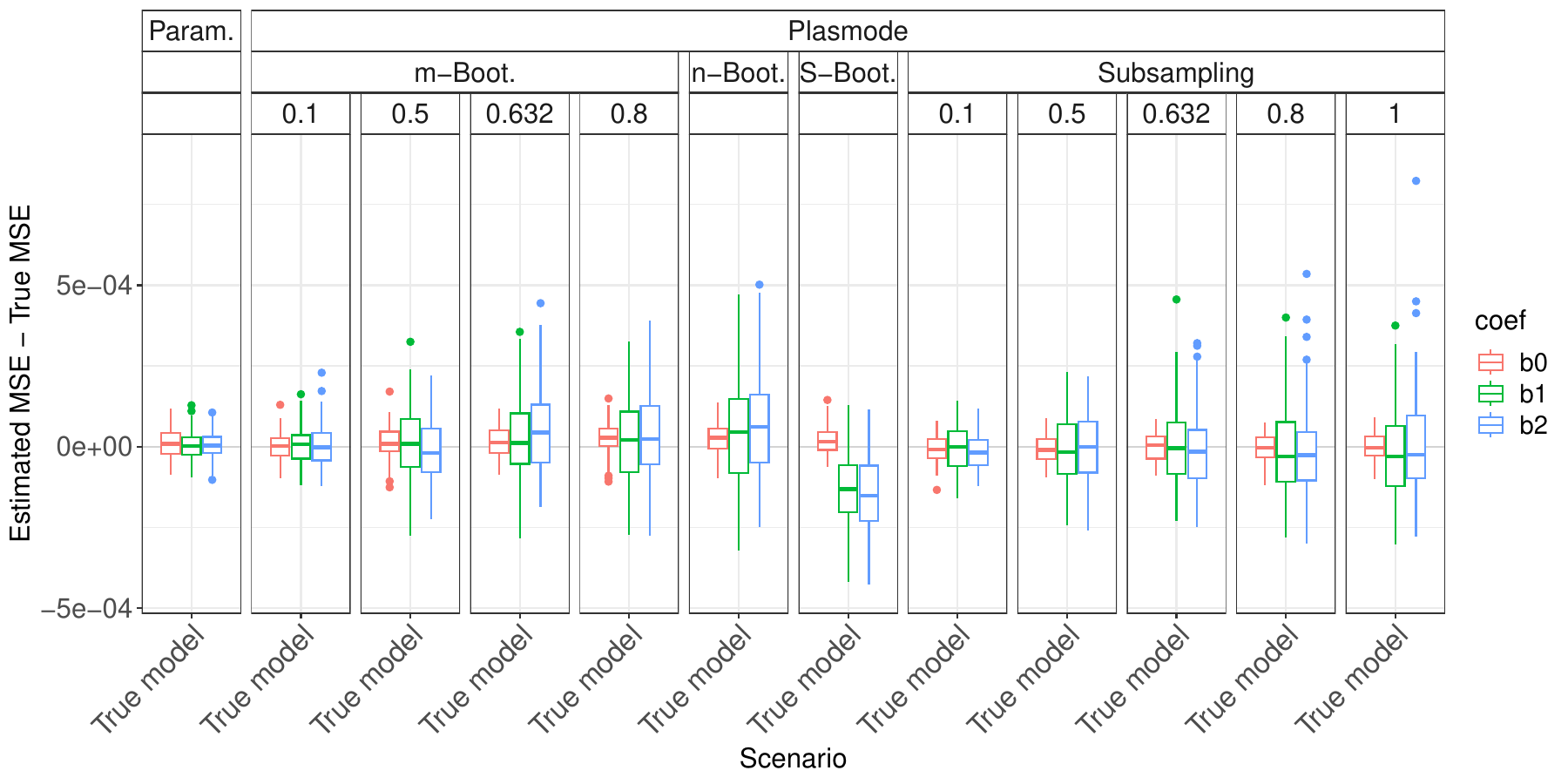}
		\end{subfigure}
		\begin{subfigure}{\textwidth}
			\caption{$p = 2,\: n = 100, \:\beta = (1, 1, 1)^T, \:\sigma = 3,\: Cor(X_i, X_j) = 0.2~\forall i\ne j$.}\label{fig:ex.true.var3}
			\centering
			\includegraphics[width = \textwidth, page = 1]{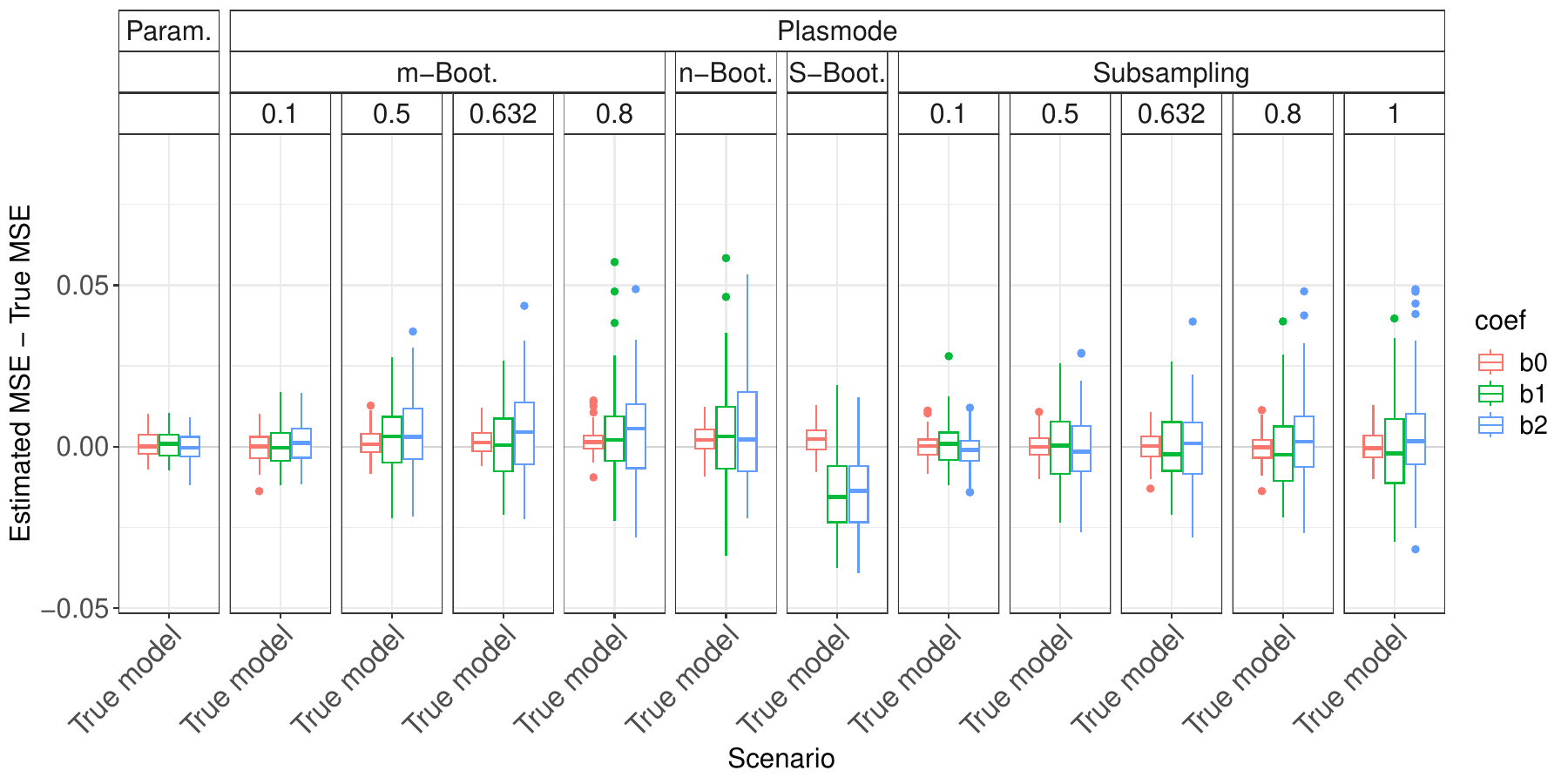}
		\end{subfigure}
		
		\caption{Absolute error in MSE estimation for individual coefficients for different types of Plasmode simulation compared to parametric simulation under assumption of true data generating process and outcome generating model.}\label{fig:ex.true}
	\end{figure}
	
	\begin{figure}[!tb]
		\centering
		\includegraphics[width = \textwidth, page = 3]{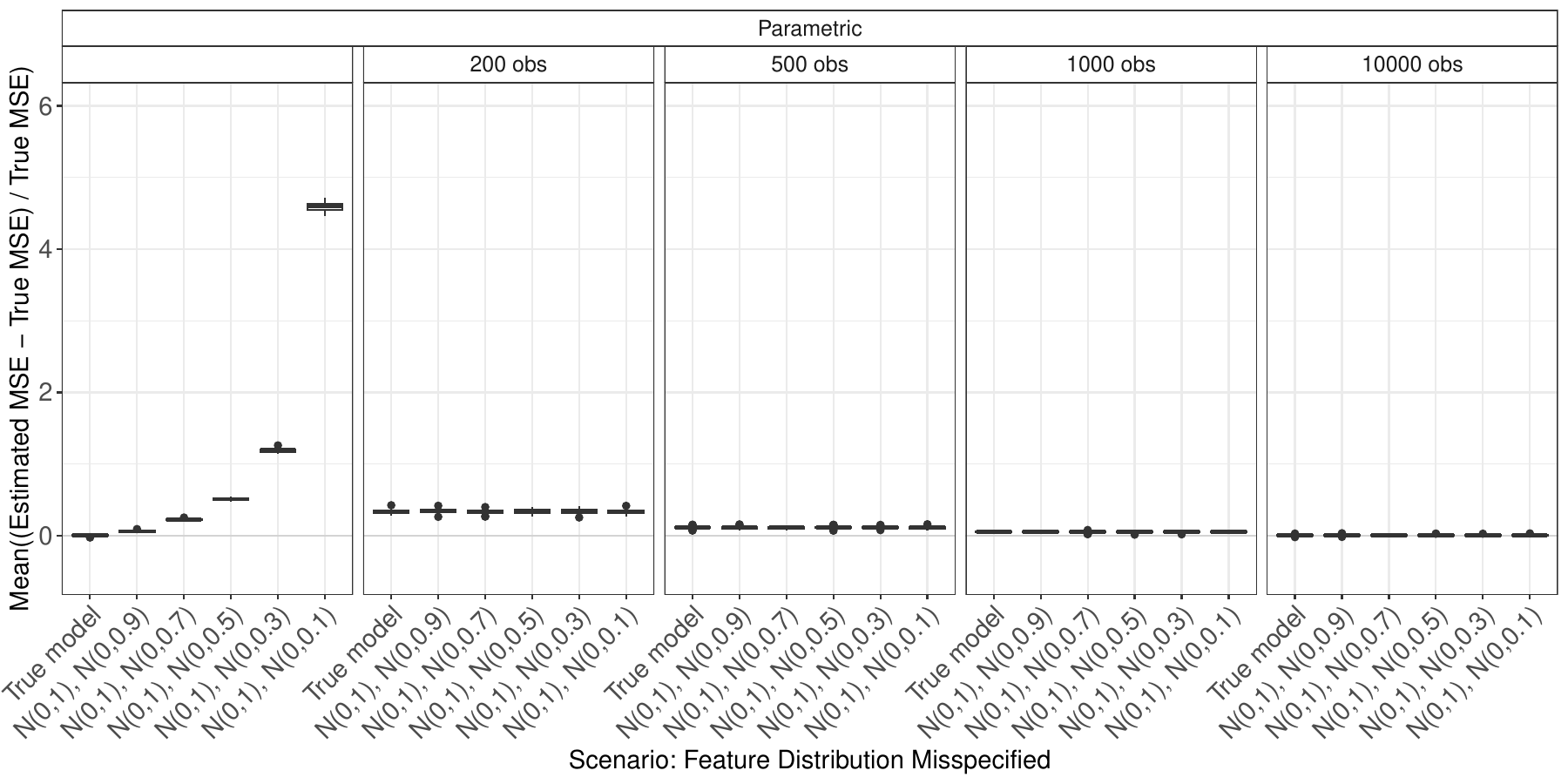}
		\caption{Relative error in MSE estimation averaged over individual coefficients when the variance of the second half of features is misspecified for $p = 50,\: n = 100, \:\beta = \boldsymbol{1}_{51}, \:\sigma = 0.3,\: Cor(X_i, X_j) = 0.2~\forall i\ne j$. The first facet displays the errors in case the misspecified variances are used in the simulation. The remaining facets display the errors for using a variance that is estimated using datasets of different sizes from the true DGP for parametric simulation instead.}\label{fig:no.est.var}
	\end{figure}
	
	\begin{figure}[!tb]
		\centering
		\includegraphics[width = \textwidth, page = 1]{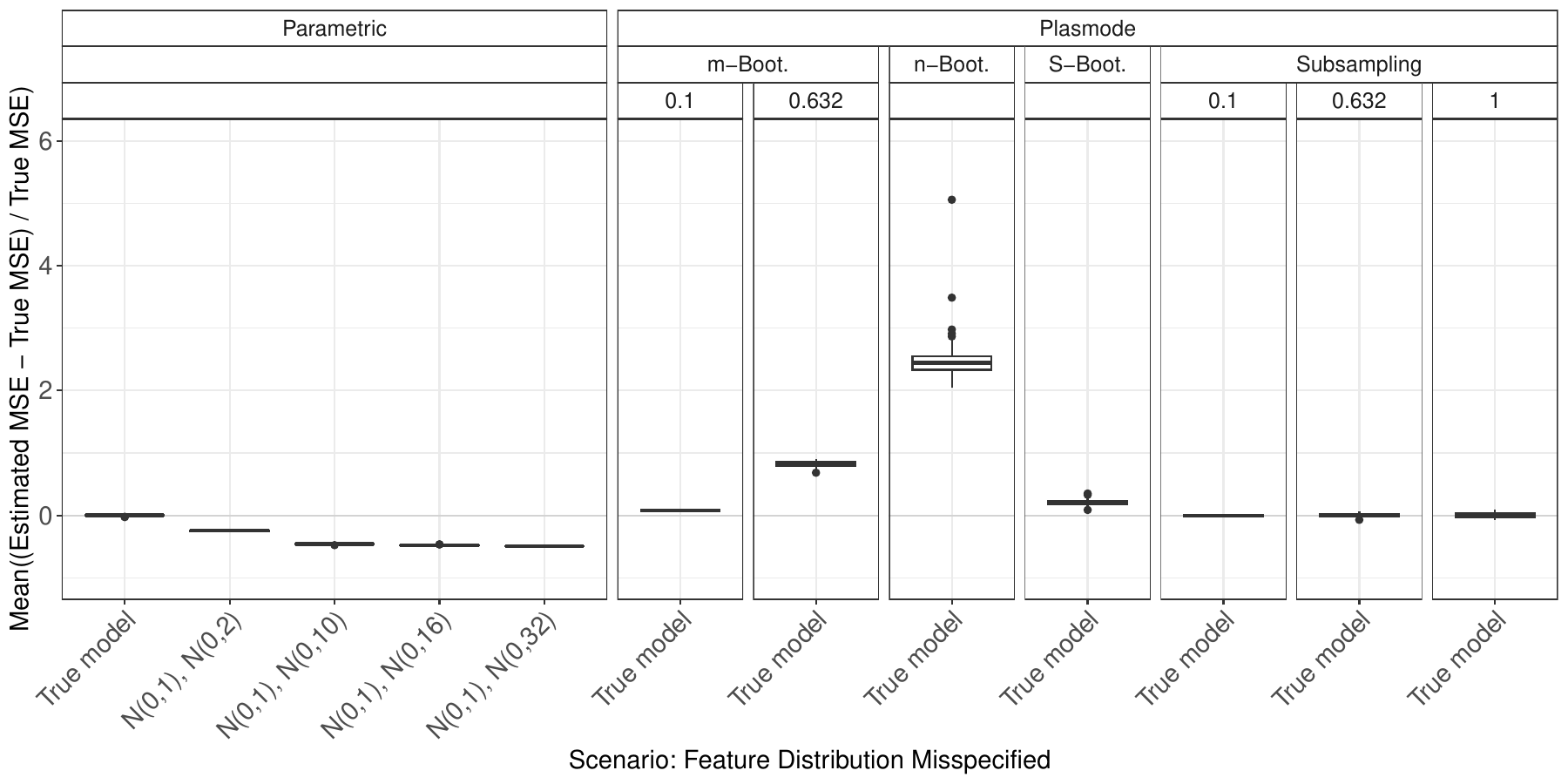}
		\caption{Relative error in MSE estimation averaged over individual coefficients when the assumed variance of the second half of features exceeds the true variance for $p = 50,\: n = 100, \:\beta = \boldsymbol{1}_{51}, \:\sigma = 0.3,\: Cor(X_i, X_j) = 0.2~\forall i\ne j$. Large outliers for $n$ out of $n$ Bootstrap not displayed.}\label{fig:50.inc.var}
	\end{figure}
	
	\begin{figure}[!tb ]
		\centering
		\includegraphics[width = \textwidth, page = 2]{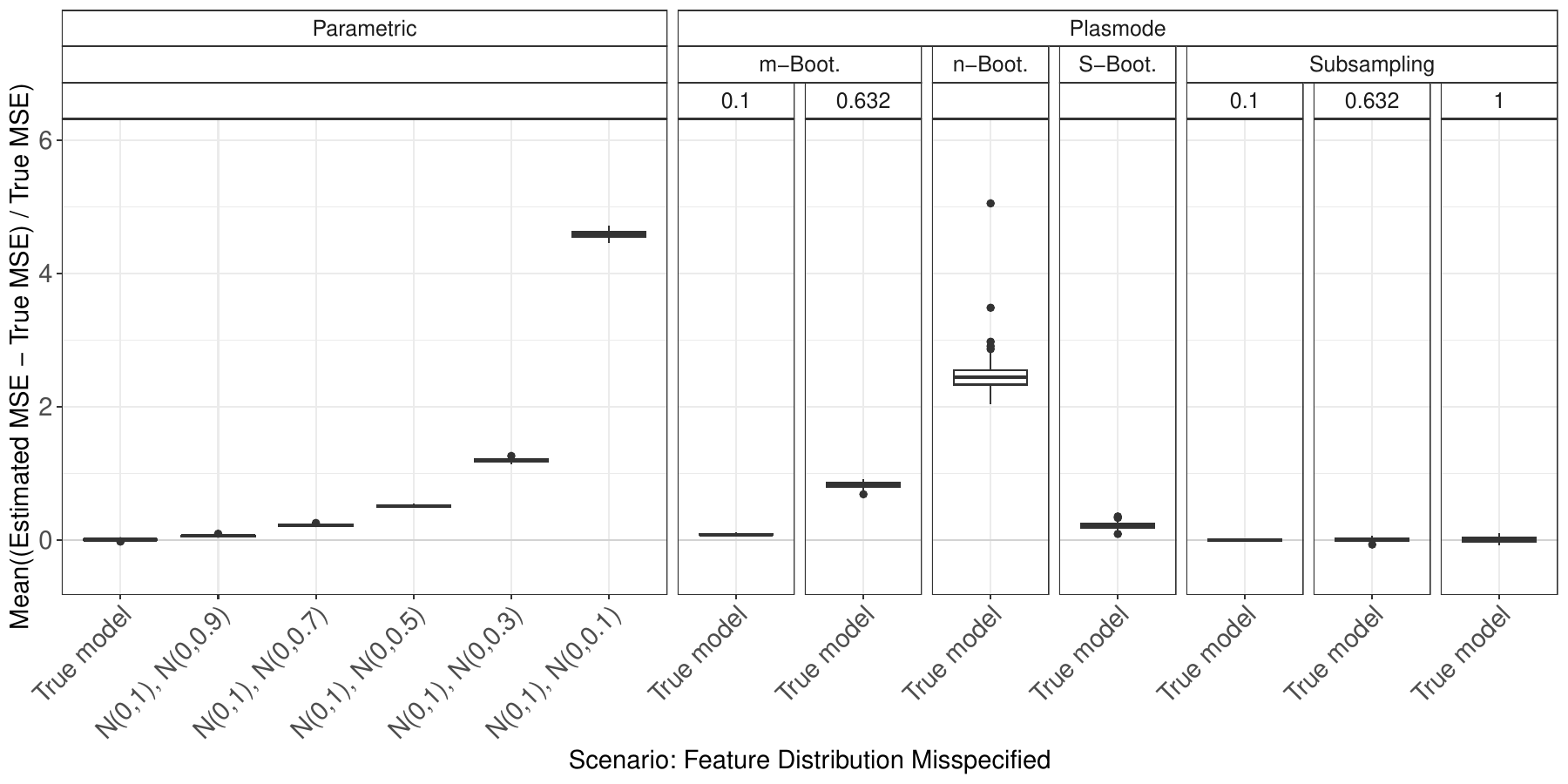}
		\caption{Relative error in MSE estimation averaged over individual coefficients when the assumed variance of the second half of features underestimates the true variance for $p = 50,\: n = 100, \:\beta = \boldsymbol{1}_{51}, \:\sigma = 0.3,\: Cor(X_i, X_j) = 0.2~\forall i\ne j$. Large outliers for $n$ out of $n$ Bootstrap not displayed.}\label{fig:50.dec.var}
	\end{figure}
	
	\clearpage
	\begin{xltabular}{\linewidth}{p{2.8cm}p{7cm}X}
		\toprule
		True scenario & Type of deviation & Values \\ 
		\midrule
		\endhead
		\bottomrule
		\caption{Complete list of deviations from true scenarios}\label{tab:deviation.list}
		\endfoot
		$(p2n100\rho0.2)$ & Error sd misspecified & $\sigma\in \{0.1, 0.2, 0.4, 0.5\}$ \\
		$(p2n100\rho0.2)$ & Correlation misspecified & $\rho\in \{-0.9, -0.8, \allowbreak\dots,\allowbreak 0,\allowbreak 0.01,\allowbreak \dots,\allowbreak 0.19,\allowbreak 0.21,\allowbreak 0.22,\allowbreak \dots,\allowbreak 0.5,\allowbreak 0.6,\allowbreak \dots,\allowbreak 0.9\}$\\  
		$(p2n100\rho0.2)$ & Coefficients and correlation misspecified & $\beta_I$ and $\rho= -0.5$ or $\beta_I$ and $\rho= 0.5$ or $\beta_{II}$ and $\rho= -0.5$\\
		$(p2n100\rho0.2)$ & Error sd and correlation misspecified & $\sigma= 0.4$ and $\rho= -0.5$ or $\sigma= 0.4$ and $\rho= 0.5$\\
		$(p2n100\rho0.2)$ & Expectation of second feature misspecified & $\mu\in\{0.05,\allowbreak 0.1,\allowbreak \dots,\allowbreak 1,\allowbreak 2,\allowbreak 3\}$\\
		$(p2n100\rho0.2)$ & Expectation of both features misspecified & $\mu = 10$\\
		$(p2n100\rho0.2)$ & Variance of second feature misspecified & $\sigma^2\in\{1.05,\allowbreak 1.1,\allowbreak \dots,\allowbreak 1.5,\allowbreak 2,\allowbreak 3,\allowbreak 5\}$\\
		$(p2n100\rho0.2)$ & Variance of both features misspecified & $\sigma^2 =  5$\\
		$(p2n100\rho0.2)$ & Mean and variance of both features misspecified & $\mu= 10$, $\sigma^2 =  5$\\
		$(p2n100\rho0.2)$ & Distribution of second feature misspecified as Gaussian mixture with $N(0,10)$ & $\alpha \in \{0.01, 0.02,\dots, 0.05\}$\\
		$(p2n100\rho0.2)$ & Distribution of second feature misspecified as Gaussian mixture with $N(3,1)$ & $\alpha \in \{0.01, 0.02,\dots, 0.05\}$\\
		$(p2n100\rho0.2)$ & Distribution of second feature misspecified as log-normal & $logN(0,1)$\\
		$(p2n100\rho0.2)$ & Distribution of second feature misspecified as Bernoulli & $\pi\in\{0.2, 0.35, 0.4, 0.45, 0.5\}$\\
		$(p2n100\rho0.2)$ & Error distribution misspecified & $t(10)$, $t(3)$, $\chi^2(1)$, $\chi^2(5)$\\
		$(p2n50\rho0.2)$ & Error sd misspecified & $\sigma\in \{0.1, 0.2, 0.4, 0.5\}$ \\
		$(p2n50\rho0.2)$ & Correlation misspecified & $\rho\in \{-0.9, -0.8,\allowbreak\dots,\allowbreak 0,\allowbreak 0.01,\allowbreak \dots,\allowbreak 0.19,\allowbreak 0.21,\allowbreak 0.22,\allowbreak \dots,\allowbreak 0.4,\allowbreak 0.5,\allowbreak \dots,\allowbreak 0.9\}$\\  
		$(p2n50\rho0.2)$ & Coefficients and correlation misspecified & $\beta_I$ and $\rho= -0.5$ or $\beta_I$ and $\rho= 0.5$ or $\beta_{II}$ and $\rho= -0.5$\\
		$(p2n50\rho0.2)$ & Error sd and correlation misspecified & $\sigma= 0.4$ and $\rho= -0.5$ \\
		$(p2n50\rho0.2)$ & Expectation of second feature misspecified & $\mu\in\{0.05,\allowbreak 0.1,\allowbreak \dots,\allowbreak 1,\allowbreak 2,\allowbreak 3\}$\\
		$(p2n50\rho0.2)$ & Expectation of both features misspecified & $\mu = 10$\\
		$(p2n50\rho0.2)$ & Variance of second feature misspecified & $\sigma^2\in\{1.05,\allowbreak 1.1,\allowbreak \dots,\allowbreak 1.5,\allowbreak 2,\allowbreak 3\}$\\
		$(p2n50\rho0.2)$ & Variance of both features misspecified & $\sigma^2 =  5$\\
		$(p2n50\rho0.2)$ & Mean and variance of both features misspecified & $\mu= 10$, $\sigma^2 =  5$\\
		$(p2n50\rho0.2)$ & Distribution of second feature misspecified as Gaussian mixture with $N(0,10)$ & $\alpha \in \{0.01, 0.02,\dots, 0.05\}$\\
		$(p2n50\rho0.2)$ & Distribution of second feature misspecified as Gaussian mixture with $N(3,1)$ & $\alpha \in \{0.01, 0.02,\dots, 0.05\}$\\
		$(p2n50\rho0.2)$ & Distribution of second feature misspecified as log-normal & $logN(0,1)$\\
		$(p2n50\rho0.2)$ & Distribution of second feature misspecified as Bernoulli & $\pi\in\{0.2, 0.35, 0.4, 0.45, 0.5\}$\\
		$(p2n50\rho0.2)$ & Error distribution misspecified & $t(10)$, $t(3)$, $\chi^2(1)$, $\chi^2(5)$\\
		$(p2n100\rho0.5)$ & Error sd misspecified & $\sigma\in \{0.1, 0.2, 0.4, 0.5\}$ \\
		$(p2n100\rho0.5)$ & Correlation misspecified & $\rho\in \{-0.9, -0.8,\allowbreak\dots,\allowbreak 0.4,\allowbreak 0.51,\allowbreak 0.52,\allowbreak \dots,\allowbreak 0.9\}$\\  
		$(p2n100\rho0.5)$ & Coefficients and correlation misspecified & $\beta_I$ and $\rho= -0.2$ or $\beta_{II}$ and $\rho= -0.2$\\
		$(p2n100\rho0.5)$ & Error sd and correlation misspecified & $\sigma= 0.4$ and $\rho= -0.2$ \\
		$(p2n100\rho0.5)$ & Expectation of second feature misspecified & $\mu\in\{0.05,\allowbreak 0.1,\allowbreak \dots,\allowbreak 2,\allowbreak 3\}$\\
		$(p2n100\rho0.5)$ & Expectation of both features misspecified & $\mu = 10$\\
		$(p2n100\rho0.5)$ & Variance of second feature misspecified & $\sigma^2\in\{1.05,\allowbreak 1.1,\allowbreak \dots,\allowbreak 1.5,\allowbreak 2,\allowbreak 3\}$\\
		$(p2n100\rho0.5)$ & Variance of both features misspecified & $\sigma^2 =  5$\\
		$(p2n100\rho0.5)$ & Mean and variance of both features misspecified & $\mu= 10$, $\sigma^2 =  5$\\
		$(p2n100\rho0.5)$ & Distribution of second feature misspecified as Gaussian mixture with $N(0,10)$ & $\alpha \in \{0.01, 0.02,\dots, 0.05\}$\\
		$(p2n100\rho0.5)$ & Distribution of second feature misspecified as Gaussian mixture with $N(3,1)$ & $\alpha \in \{0.01, 0.02,\dots, 0.05\}$\\
		$(p2n100\rho0.5)$ & Distribution of second feature misspecified as log-normal & $logN(0,1)$\\
		$(p2n100\rho0.5)$ & Distribution of second feature misspecified as Bernoulli & $\pi\in\{0.2, 0.35, 0.4, 0.45, 0.5\}$\\
		$(p2n100\rho0.5)$ & Error distribution misspecified & $t(10)$, $t(3)$, $\chi^2(1)$, $\chi^2(5)$\\
		$(p10n100\rho0.2)$ & Error sd misspecified & $\sigma\in \{0.1, 0.2, 0.4, 0.5\}$ \\
		$(p10n100\rho0.2)$ & Correlation misspecified & $\rho\in \{0,\allowbreak 0.1,\allowbreak 0.21,\allowbreak 0.22,\allowbreak \dots,\allowbreak 0.5,\allowbreak 0.6,\allowbreak \dots,\allowbreak 0.9\}$,
		
		$\rho\in \{(-0.9)^{|i-j|},\allowbreak \dots,\allowbreak (-0.1)^{|i-j|},\allowbreak 0.1^{|i-j|},\allowbreak 0.2^{|i-j|},\allowbreak 0.21^{|i-j|},\allowbreak \dots,\allowbreak 0.5^{|i-j|},\allowbreak 0.6^{|i-j|},\allowbreak \dots,\allowbreak 0.9^{|i-j|}\}$  (single block)\\  
		$(p10n100\rho0.2)$ & Coefficients and correlation misspecified & $\beta_I$ and $\rho= 0.5$ \\
		$(p10n100\rho0.2)$ & Error sd and correlation misspecified & $\sigma= 0.4$ and $\rho= 0.5$\\
		$(p10n100\rho0.2)$ & Expectation of second half of features misspecified & $\mu\in\{0.05,\allowbreak 0.1,\allowbreak \dots,\allowbreak 2,\allowbreak\}$\\
		$(p10n100\rho0.2)$ & Variance of second half of features misspecified & $\sigma^2\in\{1.05,\allowbreak 1.1,\allowbreak \dots,\allowbreak 2, 5\}$\\
		$(p10n100\rho0.2)$ & Distribution of second half of features misspecified as Gaussian mixture with $N(0,10)$ & $\alpha \in \{0.01, 0.02,\dots, 0.05\}$\\
		$(p10n100\rho0.2)$ & Distribution of second half of features misspecified as Gaussian mixture with $N(3,1)$ & $\alpha \in \{0.01, 0.02,\dots, 0.05\}$\\
		$(p10n100\rho0.2)$ & Distribution of second half of features misspecified as log-normal & $logN(0,1)$\\
		$(p10n100\rho0.2)$ & Distribution of second half of features misspecified as Bernoulli & $\pi\in\{0.3, 0.35, 0.4, 0.45, 0.5\}$\\
		$(p10n100\rho0.2)$ & Error distribution misspecified & $t(10)$, $t(3)$, $\chi^2(1)$, $\chi^2(5)$\\
		$(p10n50\rho0.2)$ & Error sd misspecified & $\sigma\in \{0.1, 0.2, 0.4, 0.5\}$ \\
		$(p10n50\rho0.2)$ & Correlation misspecified & $\rho\in \{0,\allowbreak 0.1,\allowbreak 0.21,\allowbreak 0.22,\allowbreak \dots,\allowbreak 0.9\}$,
		
		$\rho\in \{(-0.9)^{|i-j|},\allowbreak \dots,\allowbreak (-0.1)^{|i-j|},\allowbreak 0.1^{|i-j|},\allowbreak 0.2^{|i-j|},\allowbreak 0.21^{|i-j|},\allowbreak \dots,\allowbreak 0.5^{|i-j|},\allowbreak 0.6^{|i-j|},\allowbreak \dots,\allowbreak 0.9^{|i-j|}\}$ (single block)\\  
		$(p10n50\rho0.2)$ & Coefficients and correlation misspecified & $\beta_I$ and $\rho= 0.5$ \\
		$(p10n50\rho0.2)$ & Error sd and correlation misspecified & $\sigma= 0.4$ and $\rho= 0.5$\\
		$(p10n50\rho0.2)$ & Expectation of second half of features misspecified & $\mu\in\{0.05,\allowbreak 0.1,\allowbreak \dots,\allowbreak 3,\allowbreak\}$\\
		$(p10n50\rho0.2)$ & Variance of second half of features misspecified & $\sigma^2\in\{1.05,\allowbreak 1.1,\allowbreak \dots,\allowbreak 5\}$\\
		$(p10n50\rho0.2)$ & Distribution of second half of features misspecified as Gaussian mixture with $N(0,10)$ & $\alpha \in \{0.01, 0.02,\dots, 0.1\}$\\
		$(p10n50\rho0.2)$ & Distribution of second half of features misspecified as Gaussian mixture with $N(3,1)$ & $\alpha \in \{0.01, 0.02,\dots, 0.1\}$\\
		$(p10n50\rho0.2)$ & Distribution of second half of features misspecified as log-normal & $logN(0,1)$\\
		$(p10n50\rho0.2)$ & Distribution of second half of features misspecified as Bernoulli & $\pi\in\{0.3, 0.35, 0.4, 0.45, 0.5\}$\\
		$(p10n50\rho0.2)$ & Error distribution misspecified & $t(10)$, $t(3)$, $\chi^2(1)$, $\chi^2(5)$\\
		$(p50n100\rho0.2)$ & Error sd misspecified & $\sigma\in \{0.1, 0.2, 0.4, 0.5\}$ \\
		$(p50n100\rho0.2)$ & Correlation misspecified & $\rho\in \{-0.01, 0,\allowbreak 0.1,\allowbreak 0.21,\allowbreak 0.22,\allowbreak \dots,\allowbreak 0.9\}$,
		
		$\rho\in \{(-0.9)^{|i-j|},\allowbreak \dots,\allowbreak (-0.1)^{|i-j|},\allowbreak 0.1^{|i-j|},\allowbreak 0.2^{|i-j|},\allowbreak 0.21^{|i-j|},\allowbreak \dots,\allowbreak  0.99^{|i-j|}\}$  (five blocks)\\  
		$(p50n100\rho0.2)$ & Coefficients and correlation misspecified & $\beta_I$ and $\rho= 0.5$ \\
		$(p50n100\rho0.2)$ & Error sd and correlation misspecified & $\sigma= 0.4$ and $\rho= 0.5$\\
		$(p50n100\rho0.2)$ & Expectation of second half of features misspecified & $\mu\in\{0.05,\allowbreak 0.1,\allowbreak \dots,\allowbreak 5\}$\\
		$(p50n100\rho0.2)$ & Expectation of all features misspecified & $\mu = 1$\\
		$(p50n100\rho0.2)$ & Variance of second half of features misspecified & $\sigma^2\in\{0.1,\allowbreak 0.11,\allowbreak \dots,\allowbreak 0.99,\allowbreak 1.05,\allowbreak 1.1,\allowbreak \dots,\allowbreak 10,\allowbreak 10.1,\allowbreak \dots,\allowbreak 20,\allowbreak 32,\allowbreak 64,\allowbreak 128,\allowbreak 256,\allowbreak 512,\allowbreak 1024,\allowbreak 2048,\allowbreak 4096,\allowbreak 8192,\allowbreak 16384,\allowbreak 32768,\allowbreak 65536,\allowbreak 131072\}$\\
		$(p50n100\rho0.2)$ & Distribution of second half of features misspecified as Gaussian mixture with $N(0,10)$ & $\alpha \in \{0.01, 0.02,\dots, 0.99\}$\\
		$(p50n100\rho0.2)$ & Distribution of second half of features misspecified as Gaussian mixture with $N(3,1)$ & $\alpha \in \{0.01, 0.02,\dots, 0.99\}$\\
		$(p50n100\rho0.2)$ & Distribution of second half of features misspecified as log-normal & $logN(0,1)$\\
		$(p50n100\rho0.2)$ & Distribution of second half of features misspecified as Bernoulli & $\pi\in\{0.3, 0.35, 0.4, 0.45, 0.5\}$\\
		$(p50n100\rho0.2)$ & Error distribution misspecified & $t(10)$, $t(3)$, $\chi^2(1)$, $\chi^2(5)$\\
		$(p50n100\rho0.2^{|i-j|})$ & Error sd misspecified & $\sigma\in \{0.1, 0.2, 0.4, 0.5\}$ \\
		$(p50n100\rho0.2^{|i-j|})$ & Correlation misspecified & $\rho\in \{(-0.9)^{|i-j|},\allowbreak \dots,\allowbreak (-0.1)^{|i-j|},\allowbreak 0.1^{|i-j|},\allowbreak 0.21^{|i-j|},\allowbreak 0.22^{|i-j|},\allowbreak \dots,\allowbreak  0.99^{|i-j|}\}$  (five blocks),
		
		$\rho\in \{0, 0.2,\allowbreak 0.21,\allowbreak \dots,\allowbreak 0.9\}$ (no blocks)\\  
		$(p50n100\rho0.2^{|i-j|})$ & Coefficients and correlation misspecified & $\beta_I$ and $\rho= 0$ \\
		$(p50n100\rho0.2^{|i-j|})$ & Error sd and correlation misspecified & $\sigma= 0.4$ and $\rho= 0$\\
		$(p50n100\rho0.2^{|i-j|})$ & Expectation of second half of features misspecified & $\mu\in\{0.05,\allowbreak 0.1,\allowbreak \dots,\allowbreak 5\}$\\
		$(p50n100\rho0.2^{|i-j|})$ & Variance of second half of features misspecified & $\sigma^2\in\{0.1,\allowbreak 0.11,\allowbreak \dots,\allowbreak 0.99,\allowbreak 1.05,\allowbreak 1.1,\allowbreak \dots,\allowbreak 10,\allowbreak 10.1,\allowbreak \dots,\allowbreak 20\}$\\
		$(p50n100\rho0.2^{|i-j|})$ & Distribution of second half of features misspecified as Gaussian mixture with $N(0,10)$ & $\alpha \in \{0.01, 0.02,\dots, 0.8\}$\\
		$(p50n100\rho0.2^{|i-j|})$ & Distribution of second half of features misspecified as Gaussian mixture with $N(3,1)$ & $\alpha \in \{0.01, 0.02,\dots, 0.8\}$\\
		$(p50n100\rho0.2^{|i-j|})$ & Distribution of second half of features misspecified as log-normal & $logN(0,1)$\\
		$(p50n100\rho0.2^{|i-j|})$ & Distribution of second half of features misspecified as Bernoulli & $\pi\in\{0.3, 0.35, 0.4, 0.45, 0.5\}$\\
		$(p50n100\rho0.2^{|i-j|})$ & Error distribution misspecified & $t(10)$, $t(3)$, $\chi^2(1)$, $\chi^2(5)$\\
		$(p50n100\rho0.5^{|i-j|})$ & Error sd misspecified & $\sigma\in \{0.1, 0.2, 0.4, 0.5\}$ \\
		$(p50n100\rho0.5^{|i-j|})$ & Correlation misspecified & $\rho\in \{(-0.9)^{|i-j|},\allowbreak (-0.8)^{|i-j|},\allowbreak \dots,\allowbreak (-0.1)^{|i-j|},\allowbreak 0.1^{|i-j|},\allowbreak 0.2^{|i-j|},\allowbreak \dots,\allowbreak 0.4^{|i-j|},\allowbreak 0.51^{|i-j|},\allowbreak 0.52^{|i-j|},\allowbreak \dots,\allowbreak  0.99^{|i-j|}\}$  (five blocks),
		
		$\rho\in \{0, 0.1,\allowbreak 0.2,\allowbreak \dots,\allowbreak 0.4,\allowbreak 0.51,\allowbreak,\allowbreak 0.52 \dots,\allowbreak 0.9\}$ (no blocks)\\  
		$(p50n100\rho0.5^{|i-j|})$ & Coefficients and correlation misspecified & $\beta_I$ and $\rho= 0$ \\
		$(p50n100\rho0.5^{|i-j|})$ & Error sd and correlation misspecified & $\sigma= 0.4$ and $\rho= 0$\\
		$(p50n100\rho0.5^{|i-j|})$ & Expectation of second half of features misspecified & $\mu\in\{0.05,\allowbreak 0.1,\allowbreak \dots,\allowbreak 5\}$\\
		$(p50n100\rho0.5^{|i-j|})$ & Variance of second half of features misspecified & $\sigma^2\in\{0.1,\allowbreak 0.11,\allowbreak \dots,\allowbreak 0.99,\allowbreak 1.05,\allowbreak 1.1,\allowbreak \dots,\allowbreak 10,\allowbreak 10.1,\allowbreak \dots,\allowbreak 20\}$\\
		$(p50n100\rho0.5^{|i-j|})$ & Distribution of second half of features misspecified as Gaussian mixture with $N(0,10)$ & $\alpha \in \{0.01, 0.02,\dots, 0.8\}$\\
		$(p50n100\rho0.5^{|i-j|})$ & Distribution of second half of features misspecified as Gaussian mixture with $N(3,1)$ & $\alpha \in \{0.01, 0.02,\dots, 0.8\}$\\
		$(p50n100\rho0.5^{|i-j|})$ & Distribution of second half of features misspecified as log-normal & $logN(0,1)$\\
		$(p50n100\rho0.5^{|i-j|})$ & Distribution of second half of features misspecified as Bernoulli & $\pi\in\{0.3, 0.35, 0.4, 0.45, 0.5\}$\\
		$(p50n100\rho0.5^{|i-j|})$ & Error distribution misspecified & $t(10)$, $t(3)$, $\chi^2(1)$, $\chi^2(5)$\\
		(quake) & Error sd misspecified & $\sigma\in \{0.1, 0.2, 0.4, 0.5\}$ \\
		(quake) & Correlation misspecified & $\rho\in \{(-0.9)^{|i-j|},\allowbreak (-0.8)^{|i-j|},\allowbreak \dots,\allowbreak (-0.1)^{|i-j|},\allowbreak 0.01^{|i-j|},\allowbreak 0.02^{|i-j|},\allowbreak \dots,\allowbreak 0.4^{|i-j|},\allowbreak 0.5^{|i-j|}, \dots,\allowbreak  0.9^{|i-j|}\}$  (one block),
		
		$\rho\in \{0, 0.01,\allowbreak 0.02,\allowbreak \dots,\allowbreak 0.4,\allowbreak 0.5,\allowbreak,\allowbreak 0.6 \dots,\allowbreak 0.9\}$ (no blocks)\\  
		(quake) & Coefficients and correlation misspecified & $\beta_I$ and $\rho= 0$ \\
		(quake) & Error sd and correlation misspecified & $\sigma= 0.4$ and $\rho= 0$\\
		(quake) & Expectation of second half of features misspecified & $\mu\in\{0.05,\allowbreak 0.1,\allowbreak \dots,\allowbreak 1, 2\}$\\
		(quake) & Variance of second half of features misspecified & $\sigma^2\in\{0.1,\allowbreak 0.11,\allowbreak \dots,\allowbreak 0.99,\allowbreak 1.05,\allowbreak 1.1,\allowbreak \dots,\allowbreak 1.5,\allowbreak 2,\allowbreak 5\}$\\
		(quake) & Distribution of second half of features misspecified as Gaussian mixture with $N(0,10)$ & $\alpha \in \{0.01, 0.02,\dots, 0.05\}$\\
		(quake) & Distribution of second half of features misspecified as Gaussian mixture with $N(3,1)$ & $\alpha \in \{0.01, 0.02,\dots, 0.05\}$\\
		(quake) & Distribution of second half of features misspecified as log-normal & $logN(0,1)$\\
		(quake) & Distribution of second half of features misspecified as Bernoulli & $\pi\in\{0.3, 0.35, 0.4, 0.45, 0.5\}$\\
		(quake) & Error distribution misspecified & $t(10)$, $t(3)$, $\chi^2(1)$, $\chi^2(5)$\\
		
		(wine\_quality), (pol), (Yolanda) & Error sd misspecified & $\sigma\in \{0.1, 0.2, 0.4, 0.5\}$ \\
		(wine\_quality), (pol) & Correlation misspecified & $\rho\in \{(-0.9)^{|i-j|},\allowbreak (-0.8)^{|i-j|},\allowbreak \dots,\allowbreak  0.9^{|i-j|}\}$  (two blocks),
		
		$\rho\in \{0, 0.1,\allowbreak 0.2,\allowbreak \dots,\allowbreak 0.9\}$ (no blocks)\\  
		(Yolanda) & Correlation misspecified & $\rho\in \{(-0.9)^{|i-j|},\allowbreak (-0.8)^{|i-j|},\allowbreak \dots,\allowbreak  0.9^{|i-j|}\}$  (ten blocks),
		
		$\rho\in \{0, 0.1,\allowbreak 0.2,\allowbreak \dots,\allowbreak 0.9\}$ (no blocks)\\ 
		(wine\_quality), (pol), (Yolanda) & Coefficients and correlation misspecified & $\beta_I$ and $\rho= 0$ \\
		(wine\_quality), (pol), (Yolanda) & Error sd and correlation misspecified & $\sigma= 0.4$ and $\rho= 0$\\
		(wine\_quality), (pol), (Yolanda) & Expectation of second half of features misspecified & $\mu\in\{0.05,\allowbreak 0.1,\allowbreak \dots,\allowbreak 5\}$\\
		(wine\_quality), (pol), (Yolanda) & Variance of second half of features misspecified & $\sigma^2\in\{0.1,\allowbreak 0.11,\allowbreak \dots,\allowbreak 0.99,\allowbreak 2,\allowbreak 5\}$\\
		(wine\_quality), (pol), (Yolanda) & Distribution of second half of features misspecified as Gaussian mixture with $N(0,10)$ & $\alpha \in \{0.01, 0.02,\dots, 0.99\}$\\
		(wine\_quality), (pol), (Yolanda) & Distribution of second half of features misspecified as Gaussian mixture with $N(3,1)$ & $\alpha \in \{0.01, 0.02,\dots, 0.99\}$\\
		(wine\_quality), (pol), (Yolanda) & Distribution of second half of features misspecified as log-normal & $logN(0,1)$\\
		(wine\_quality), (pol), (Yolanda) & Distribution of second half of features misspecified as Bernoulli & $\pi\in\{0.3, 0.35, 0.4, 0.45, 0.5\}$\\
		(wine\_quality), (pol), (Yolanda) & Error distribution misspecified & $t(10)$, $t(3)$, $\chi^2(1)$, $\chi^2(5)$\\
		
	\end{xltabular}
	
	\newpage
	\section{Simulation of Bernoulli, log normal, and Gaussian mixture variables with fixed correlations}\label{app:calc.cor}
	In each case, we first draw data from a multivariate normal and then transform some of the variables. To get $Ber(\pi)$-distributed data, we dichotomize at the $\pi$-quantile $u_{\pi}$ of the respective normal distribution, i.e. values of the normal variable that are smaller than the $\pi$-quantile of this normal are set to one, values larger than the quantile are set to zero. To get to log-normal data, we use the exponential function on the respective normal variables. To get to Gaussian mixture variables, we generate two normal variables according to the two distributions to be mixed and then take the observations from the first variable with a probability of $\alpha$ and otherwise the observations of the second variable. \\
	The main challenge is to ensure that the resulting variables fulfill a given covariance structure. For this, we have to find a matching covariance matrix for the underlying multivariate normal variables such that the transformed variables have the given covariance structure. The solutions to this problem for each type of transformation will be presented in the following. Since we want to be able to mix each type of variable with normal variables for our simulation, we need the underlying covariances for pairs of transformed variables as well as for pairs where one variable is transformed while the other variable stays normal. Note that positive definiteness of the desired covariance matrix does not ensure positive definiteness of the underlying covariance matrix. In cases where the determined covariance matrix of the underlying multivariate normal distribution is not positive definite, we still use it for sampling since we checked the distributions of the resulting variables and they looked good even in that case.
	\subsection{Bernoulli}
	Let $X_1\sim N(\mu_1, \sigma_1^2)$ be a normal variable that will not be transformed, $X_2, X_3\sim N(0,1)$ standard normal variables used to generate two Bernoulli variables and denote by $\sigma_{ij} = \mathbb{C}\text{ov}(X_i, X_j)$ the covariances of the original variables. By setting $Y_1 = \mathbbm{1}(X_2\le u_{\pi_1})$ and $Y_2 = \mathbbm{1}(X_3\le u_{\pi_2})$ we get $Y_1\sim Ber(\pi_1)$ and $Y_2\sim Ber(\pi_2)$ where $\pi_i\in(0,1), i = 1,2$.\\
	\textcite{emrich_method_1991} derived the correlation between two Bernoulli variables generated from standard normals as above as 
	\[
	\mathbb{C}\text{or}(Y_1, Y_2) = \frac{F_{X_2, X_3}(u_{\pi_1}, u_{\pi_2}) - \pi_1\pi_2}{\sqrt{\pi_1(1-\pi_1)\cdot \pi_2(1-\pi_2)}},
	\]
	where $F_{X_2, X_3}$ denotes the multivariate normal distribution function of $X_2$ and $X_3$ that is of $N\left(\begin{pmatrix}
		0\\0
	\end{pmatrix}, \begin{pmatrix}
		1 & \sigma_{23}\\
		\sigma_{23} & 1
	\end{pmatrix}\right)$. This means that if we want to set the correlation of the transformed variables to $\rho$, we have to solve 
	\[
	F_{X_2, X_3}(u_{\pi_1}, u_{\pi_2}) \stackrel{!}{=} \rho \cdot \sqrt{\pi_1(1-\pi_1)\cdot \pi_2(1-\pi_2)} + \pi_1\pi_2
	\]
	for $\sigma_{23}$. Since there is no closed form for $F_{X_2, X_3}$ we solve the equation numerically by performing a grid search over $\sigma_{23}\in[-1, 1]$ in steps of $10^{-4}$ and taking the value for which the resulting value of the left-hand side is closest to the required value of the right-hand side.\\
	The correlation between a Bernoulli variable generated from a standard normal and another untransformed normal variable is given as follows. 
	
	\begin{align*} 
		\mathbb{C}\text{ov}(X_1, Y_1) & = \mathbb{C}\text{ov}(X_1, \mathbbm{1}(X_2\le u_{\pi_1}))\\
		&= \E(X_1\cdot \mathbbm{1}(X_2\le u_{\pi_1})) - \E(X_1) \E(\mathbbm{1}(X_2\le u_{\pi_1}))\\
		&= \int\int_\mathbb{R} x_1 \cdot \mathbbm{1}(x_2\le u_{\pi_1}) f_{X_1, X_2}(x_1, x_2) \dif x_1 \dif x_2 - \mu_1\pi_1 \\
		&= \int_{-\infty}^{u_{\pi_1}}\int_\mathbb{R} x_1 f_{X_1, X_2}(x_1, x_2) \dif x_1 \dif x_2 - \mu_1\pi_1 \\
		&= \int_{-\infty}^{u_{\pi_1}}\left[\int_\mathbb{R} x_1 f_{X_1|X_2}(x_1|x_2) \dif x_1\right] f_{X_2}(x_2) \dif x_2 - \mu_1\pi_1 \\
		&= \int_{-\infty}^{u_{\pi_1}}\E\left[X_1|X_2=x_2\right] f_{X_2}(x_2) \dif x_2 - \mu_1\pi_1 \\
		&= \int_{-\infty}^{u_{\pi_1}} \left[\mu_1 + \sigma_{12} x_2\right] f_{X_2}(x_2) \dif x_2 - \mu_1\pi_1 \\
		&= \int_{-\infty}^{u_{\pi_1}} \mu_1 f_{X_2}(x_2) \dif x_2 + \int_{-\infty}^{u_{\pi_1}} \left[\sigma_{12} x_2\right] f_{X_2}(x_2) \dif x_2 - \mu_1\pi_1 \\
		&= \mu_1 \mathbb{P}(X_2 \le u_{\pi_1}) + \sigma_{12} \underbrace{\E(X_2|X_2\le u_{\pi_1})}_{\text{Expectation of truncated normal}} P(X_2\le u_{\pi_1}) - \mu_1\pi_1 \\
		&= \mu_1 \pi_1 + \sigma_{12} \left[\E(X_2) - \sigma_2 \varphi(u_{\pi_1}) / \Phi(u_{\pi_1})\right] \pi_1 - \mu_1\pi_1 \\
		&= -\sigma_{12}\varphi(u_{\pi_1}), 
	\end{align*}
	where $\varphi$ and $\Phi$ denote the density and cumulative distribution function of a standard normal distribution
	
	\begin{equation*} 
		\Rightarrow\mathbb{C}\text{or}(X_1, Y_1)  = -\frac{\sigma_{12}\varphi(u_{\pi_1})}{\sqrt{\sigma_1^2\pi_1(1-\pi_1)}},
	\end{equation*}
	so we get the solution
	\[
	\sigma_{12} \stackrel{!}{=} -\frac{\rho\sqrt{\sigma_1^2\pi_1(1-\pi_1)}}{\varphi(u_{\pi_1})}.
	\]

	\subsection{Log normal}
	Let $X_i\sim N(\mu_i, \sigma_i^2), i = 1, 2, 3$ normal variable and denote by $\sigma_{ij} = \mathbb{C}\text{ov}(X_i, X_j)$ the covariances of the original variables. By setting $Y_1 = \exp(X_2)$ and $Y_2 = \exp(X_3)$ we get $Y_1\sim N(\mu_2, \sigma^2_2)$ and $Y_2\sim N(\mu_3, \sigma^2_3)$.\\
	For the correlations between two transformed log-normal variables, we obtain 
	\[
	\mathbb{C}\text{or}(Y_1, Y_2)  = \frac{\exp(\sigma_{23})-1}{\sqrt{(\exp(\sigma_2^2)-1)(\exp(\sigma_3^2)-1)}}
	\]
	analogously to the results of \textcite{astivia_population_2017} for log normals generated from standard normals. Therefore, we have to set 
	\[
	\sigma_{23} \stackrel{!}{=} \log\left(\rho \sqrt{(\exp(\sigma_2^2)-1)(\exp(\sigma_3^2)-1)} + 1\right).
	\]
	Note, that the lower bound of the possible correlations between $Y_1$ and $Y_2$ might be larger than $-1$ depending on $\sigma_2^2$ and $\sigma_3^2$, e.g. for standard normals ($\sigma_2^2 = \sigma_3^2$), the lower bound for $\rho$ that can be reached is $\approx -0.632$.\\
	The correlation between a log-normal variable generated from a normal and another untransformed normal variable is given as follows. 
	\begin{align*}
		\mathbb{C}\text{ov}(X_1, Y_1) & = \mathbb{C}\text{ov}(X_1, \exp(X_2))\\
		&= \E(X_1\cdot \exp(X_2)) - \E(X_1) \E(\exp(X_2))\\
		&= \E\left[\E\left(X_1\exp(X_2)|X_2\right)\right] - \mu_1\exp\left(\mu_2+\frac{\sigma^2_2}{2}\right)\\
		&= \E\left[\exp(X_2)\E\left(X_1|X_2\right)\right] - \mu_1\exp\left(\mu_2+\frac{\sigma^2_2}{2}\right)\\
		&= \E\left[\exp(X_2)\left(\mu_1 + \sigma_{12}/\sigma_2^2 \left(X_2 - \mu_2\right)\right)\right] - \mu_1\exp\left(\mu_2+\frac{\sigma^2_2}{2}\right)\\
		&= \E\left[\exp(X_2)\mu_1\right] + \sigma_{12}/\sigma_2^2 \E\left[\exp(X_2) X_2\right] - \sigma_{12}/\sigma_2^2  \mu_2\E\left[\exp(X_2)\right]\\
		& \quad - \mu_1\exp\left(\mu_2+\frac{\sigma^2_2}{2}\right)\\
		&= \mu_1 \exp\left(\mu_2+\frac{\sigma^2_2}{2}\right) + \sigma_{12}/\sigma_2^2 \E\left[\exp(X_2) X_2\right] - \sigma_{12}/\sigma_2^2  \mu_2\exp\left(\mu_2+\frac{\sigma^2_2}{2}\right)\\
		& \quad - \mu_1\exp\left(\mu_2+\frac{\sigma^2_2}{2}\right)\\
		&= \sigma_{12}/\sigma_2^2 \int_{\mathbb{R}} x_2\exp{x_2} f_{X_2}(x_2) \dif x_2 - \sigma_{12}/\sigma_2^2  \mu_2\exp\left(\mu_2+\frac{\sigma^2_2}{2}\right)\\
		&= \sigma_{12}/\sigma_2^2 \exp\left(\mu_2+\frac{\sigma^2_2}{2}\right) (\mu_2 + \sigma_2^2) - \sigma_{12}/\sigma_2^2  \mu_2\exp\left(\mu_2+\frac{\sigma^2_2}{2}\right)\\
		&= \sigma_{12} \exp\left(\mu_2+\frac{\sigma^2_2}{2}\right) \\
		\Rightarrow \mathbb{C}\text{or}(X_1, Y_1) & = \frac{\sigma_{12} \exp\left(\mu_2+\frac{\sigma^2_2}{2}\right)}{\sqrt{\sigma_1^2 \cdot \left[\exp(\sigma_2^2) - 1\right] \exp(2\mu_2 + \sigma^2)}}\\
	\end{align*}
	So our solution is given as 
	\[
	\sigma_{12} \stackrel{!}{=} \frac{\rho \sqrt{\sigma_1^2 \cdot \left[\exp(\sigma_2^2) - 1\right] \exp(2\mu_2 + \sigma^2)}}{\exp\left(\mu_2+\frac{\sigma^2_2}{2}\right)}
	\]
	
	\subsection{Gaussian mixture}
	Let $X_i\sim N(\mu_i, \sigma_i^2), i = 1, \dots, 5$ normal variable and denote by $\sigma_{ij} = \mathbb{C}\text{ov}(X_i, X_j)$ the covariances of the original variables. Additionally let $\Delta_i\sim Ber(\alpha_i), i = 1,2$ be independent of all $X_i$. By setting $Y_1 = \Delta_1 X_2 + (1-\Delta_1) X_3$ and $Y_2 = \Delta_2 X_4 + (1-\Delta_2) X_5$ we get $Y_1\sim \alpha_1 N(\mu_2, \sigma^2_2) + (1 -\alpha_1) N(\mu_3, \sigma^2_3)$ and $Y_2\sim \alpha_2 N(\mu_4, \sigma^2_4) + (1 -\alpha_2) N(\mu_5, \sigma^2_5)$.\\
	For the transformed variables, it holds 
	\begin{align*}
		\E(Y_1) &= \alpha_1 \mu_2 + (1-\alpha_1) \mu_3\\
		\E(Y_2) &= \alpha_2 \mu_4 + (1-\alpha_2) \mu_5\\
		\mathbb{V}\text{ar}(Y_1) &= \alpha_1 \sigma_2^2 + (1-\alpha_1) \sigma_3^2 + \alpha_1 (1-\alpha_1^2)(\mu_2 - \mu_3)^2\\
		\mathbb{V}\text{ar}(Y_2) &= \alpha_2 \sigma_4^2 + (1-\alpha_2) \sigma_5^2 + \alpha_2 (1-\alpha_2^2)(\mu_4 - \mu_5)^2\\
		\E(Y_1 Y_2) &= \alpha_1 \alpha_2 (\sigma_{24} + \mu_2\mu_4) + (1-\alpha_1)\alpha_2(\sigma_{34} + \mu_3\mu_4) + \alpha_1(1-\alpha_2)(\sigma_{25} + \mu_2\mu5) \\
		&\quad + (1-\alpha_1)(1-\alpha_2)(\sigma_{35} + \mu_3\mu5).
	\end{align*}
	\[
	\Rightarrow \mathbb{C}\text{ov}(Y_1, Y_2) = \alpha_1\alpha_2\sigma_{24} + (1-\alpha_1)\alpha_2\sigma_{34} + \alpha_1(1-\alpha_2)\sigma_{25} + (1-\alpha_1)(1-\alpha_2)\sigma_{35}
	\]
	
	\[
	\resizebox{.9\linewidth}{!}{$\Rightarrow \mathbb{C}\text{or}(Y_1, Y_2) = \frac{\alpha_1\alpha_2\sigma_{24} + (1-\alpha_1)\alpha_2\sigma_{34} + \alpha_1(1-\alpha_2)\sigma_{25} + (1-\alpha_1)(1-\alpha_2)\sigma_{35}}{\sqrt{(\alpha_1 \sigma_2^2 + (1-\alpha_1) \sigma_3^2 + \alpha_1 (1-\alpha_1^2)(\mu_2 - \mu_3)^2)(\alpha_2 \sigma_4^2 + (1-\alpha_2) \sigma_5^2 + \alpha_2 (1-\alpha_2^2)(\mu_4 - \mu_5)^2)}}$}
	\]
	
	The correlation between a Gaussian Mixture variable generated from two normals and another untransformed normal variable is given as follows. 
	\begin{align*}
		\E(Y_1 X_1) &= \alpha_1\sigma_{12} + \alpha_1\mu_1\mu_2 + (1-\alpha_1)\sigma_{13} + (1-\alpha_1)\mu_1\mu_3\\
		\Rightarrow \mathbb{C}\text{ov}(Y_1, X_1) &= \alpha_1\sigma_{12} + (1-\alpha_1)\sigma_{13}\\
		\Rightarrow \mathbb{C}\text{or}(Y_1, X_1) &= \frac{\alpha_1\sigma_{12} + (1-\alpha_1)\sigma_{13}}{\sqrt{\sigma_1^2(\alpha_1\sigma_2^2 + (1-\alpha_1)\sigma_3^2 + \alpha_1(1-\alpha_1^2)(\mu_2-\mu_3)^2)}}.
	\end{align*}
	
	For the Gaussian mixture variables, there is no unique solution. For symmetry reasons, we set
	\begin{align*}
		\sigma_{12} &= \sigma_{13} = \rho \sqrt{\sigma_1^2[\alpha_1\sigma_2^2 + (1-\alpha_1)\sigma_3^2 + \alpha_1(1-\alpha_1^2)(\mu_2 - \mu_3)^2]}\\
		\sigma_{14} &= \sigma_{15} = \rho \sqrt{\sigma_1^2[\alpha_2\sigma_4^2 + (1-\alpha_2)\sigma_5^2 + \alpha_2(1-\alpha_2^2)(\mu_4 - \mu_5)^2]}\\
		\sigma_{14} &= \sigma_{15} = 0\\
		\sigma_{24} &= \sigma_{35} = \sigma_{34} = \sigma_{25} \\
		&= \rho\frac{\sqrt{\left[\alpha_1\sigma_2^2+(1-\alpha_1)\sigma_3^2+\alpha_1(1-\alpha_1^2)(\mu_2-\mu_3)^2\right]\left[\alpha_2\sigma_4^2+(1-\alpha_2)\sigma_5^2+\alpha_2(1-\alpha_2^2)(\mu_4-\mu_5)^2\right]}}{\alpha_1\alpha_2 + (1-\alpha_1)\alpha_2 + \alpha_1(1-\alpha_2) + (1-\alpha_1)(1-\alpha_2)}.\\
	\end{align*}
\end{document}